\documentclass[journal,comsoc]{IEEEtran}
\pdfoutput=1
\usepackage[T1]{fontenc}

\usepackage[switch,columnwise]{lineno}
\usepackage{amsmath}
\usepackage{color}
\usepackage{url}
\usepackage[cmintegrals]{newtxmath}
\usepackage{graphicx}
\usepackage{caption}
\usepackage{multirow}
\usepackage{cite}
\usepackage{subfig}
\usepackage{float}
\usepackage{array}
\usepackage{flushend}
\usepackage{color}
\newcommand{\beq}{\begin{equation}}
\newcommand{\eeq}{\end{equation}}
\newcommand{\beqa}{\begin{eqnarray}}
\newcommand{\eeqa}{\end{eqnarray}}
\DeclareMathOperator*{\argmax}{arg\,max}
\begin{document}

\title{An experimental evaluation and characterization of VoIP over an LTE-A network}

\author{Mario~Di~Mauro,~\IEEEmembership{Member,~IEEE,}
	Antonio~Liotta,~\IEEEmembership{Senior Member,~IEEE}
	\IEEEcompsocitemizethanks{\IEEEcompsocthanksitem M. Di Mauro is with the Department of Information and Electrical Engineering and Applied Mathematics (DIEM), University of Salerno, 84084, Fisciano, Italy (E-mail: mdimauro@unisa.it).
   \newline
   
  A. Liotta is with the School of Computing, Edinburgh Napier University, Edinburgh EH10 5DT , U.K. (E-mail: a.liotta@napier.ac.uk)
  
	}
}

\maketitle

\begin{abstract}

Mobile telecommunications are converging towards all-IP solutions. This is the case of the Long Term Evolution (LTE) technology that, having no circuit-switched bearer to support voice traffic, needs a dedicated VoIP infrastructure, which often relies on the IP Multimedia Subsystem architecture. Most telecom operators implement LTE-A, an advanced version of LTE often marketed as $4G+$, which achieves data rate peaks of 300 Mbps. Yet, although such novel technology boosts the access to advanced multimedia contents and services, telco operators continue to consider the VoIP market as the major revenue for their business. In this work, the authors propose a detailed performance assessment of VoIP traffic by carrying out experimental trials across a real LTE-A environment. The experimental campaign consists of two stages. First, we characterize VoIP calls between fixed and mobile terminals, based on a data-set that includes more than 750,000 data-voice packets. We analyze quality-of-service metrics such as round-trip time (RTT) and jitter, to capture the influence of uncontrolled factors that typically appear in real-world settings. In the second stage, we further consider VoIP flows across a range of codecs, looking at the trade-offs between quality and bandwidth consumption. Moreover, we propose a statistical characterization of jitter and RTT (representing the most critical parameters), identifying the optimal approximating distribution, namely the Generalized Extreme Value (GEV). The estimation of parameters through the Maximum Likelihood criterion, leads us to reveal both the short- and long-tail behaviour for jitter and RTT, respectively.

\end{abstract}


\IEEEpeerreviewmaketitle

\section{Introduction}

\IEEEPARstart{N}{owadays}, the high demand for low-cost, voice-based communication services has a great impact on the growth of the mobile VoIP (Voice over IP) market. According to a recent Ericsson research \cite{eri2017}, VoLTE (Voice over LTE) subscriptions will reach 5.5 billion by the end of 2023. This unprecedented growth is sustained by an ever decreasing cost of internet-based services, including VoIP, and flat-rate packages. However, by contrast to conventional IP services, which operate on a best-effort model, VoIP sessions pose real-time constraints on the network Quality of Service (QoS) metrics, aiming at acceptable levels of user Quality of Experience (QoE). VoIP sessions typically require a one-way latency between the callers of no more than 150 milliseconds, average jitter (that is the delay variation) under 30 msec, and packet loss rates not exceeding 1\% \cite{cisco}. 

However, the relationship between such QoS parameters and the achievable VoIP service QoE is highly non linear, and several compensation mechanisms exist in the network and terminals to address jitter or packet loss. When the audio codecs work in tandem with packet buffering and just-on-time re-transmission, it is possible to tackle to a certain extent the variability of mobile network conditions (e.g. in poor reception areas of during handover). Yet, determining how individual QoS parameters affect QoE under a broad range of conditions and codecs is a hard problem. 
With the present study, we aim to better evaluate the actual performance of mobile VoIP across a real-world LTE-Advanced (LTE-A) domain, under a range of operative conditions. We embark on an empirical analysis which allows us to capture the actual end-to-end VoIP quality achieved in a pilot study, for different codecs and network conditions. 
Two original contributions emerge in this work. 

The first one, is an experimental performance assessment based on a real-world LTE measurement campaign. The end-to-end VoIP traffic has been collected in both mobile-to-mobile and mobile-to-fixed settings. Several voice flows are included, considering a range of audio codecs. In addition to the network parameters, we consider diverse metrics that are not typically taken into account in the literature, including the R-factor,  the voice signal levels, the call setup delay, and the session disconnect delay.

The second contribution pertains the statistical modeling of the two most critical parameters in VoIP traffic: jitter and RTT. These metrics are treated as random variables and characterized by means of Generalized Extreme Value (GEV) distribution, along the estimation of their parameters (shape, scale, location) via the Maximum Likelihood (ML) criterion. This analysis  reveals that jitter and RTT exhibit short and long-tail behaviors, respectively. It provides numerical results that play a crucial role in creating close-to-reality models. With this work, we make a step forward in the quest to better understand how the mismatch between real-time VoIP services (QoE) and non-real-time (best-effort) networks (QoS) affects end-to-end quality in mobile cellular networks.

The paper is organized as follows: Section \ref{sec:rw} presents a review of the most significant literature in relation to this article. In Section \ref{sec:expset}, we provide a detailed description of the experimental setting used to carry out the measurement campaign. Section \ref{sec:perfeval} provides performance evaluation results, with the dual purpose of providing scientific insights, but also serving as a benchmark for an operator to improve the efficiency of VoIP services. In Section \ref{sec:statchar}, we propose a statistical modeling of jitter and RTT, where some basic concepts about extreme value distributions are recalled. Finally, Section \ref{sec:concl} draws conclusions and explores possible research directions.

\section{Related Research}
\label{sec:rw}

The characterization of mobile voice traffic continues to be an important research theme, since the network infrastructures are continuously evolving. It is particularly useful to evaluate the suitability of mobile networks as these get ready to support truly advanced services such as HD Video, VoLTE, HD Voice, to mention but a few. These have a great impact both on QoE (a subjective measure of perceived quality) \cite{murroni,atzori1,atzori2,torres18_2}, and on network QoS metrics. QoS can benefit from the new high-speed networks but is negatively affected by uncontrolled phenomena that typically arise in a radio environment (weather conditions, obstacles, time-varying load of network nodes). 

Most of works proposed in the literature are focused on characterizing such QoS metrics by performing simulations in ``artificial" environments where some parameters (e.g. the packet loss) can be tuned. A simulated scenario obtained by means of the Vienna LTE-A downlink simulator has been considered in \cite{olaifa16}, where authors evaluate some throughput-based performance metrics, considering various spatial deployments. A simulator that accounts for the physical layer of the LTE infrastructure is instead proposed in \cite{xu2013}, where the radio channel is reproduced by combining deterministic wave propagation structures with stochastic channel models. Interesting also the RTP performance study over a simulated LTE environment presented in \cite{sarker2014}, where a mobility scenario characterized by movement at 3 km/hour between two radio cells has been taken into account.

Many other works rely on the popular NS-2/NS-3 simulator equipped with LTE modules. Among such works we cite: \cite{carullo16} that proposes a performance evaluation of WebRTC traffic by using a mixed simulated/emulated environment with an LTE customized module; \cite{ahmad15} where delay and throughput measures are derived for TCP traffic in an LTE setting; \cite{abidi14} where authors analyze the effect of mobility at high speeds with and without handover procedures; \cite{bermudez17} that  presents a performance evaluation of a streaming video service based on the Real-Time Messaging Protocol across an LTE domain, and where different QoS parameters have been considered.

Other works, focused on characterizing the performance of data traffic in a real LTE environment, have some limitations. A semi-realistic testbed is proposed in \cite{kassim17}, where a limited set of performance figures are derived for 3G and 4G networks using Jperf, a software used to analyze and measure network traffic. In \cite{wylie14} a more realistic, yet still limited, campaign has been carried out to evaluate some traffic metrics in an LTE environment but no codecs are considered to vary across traffic sessions. Again, a throughput analysis has been assessed in \cite{buenestado14} in a live LTE setting, where some key indicators (jitter, delay, R-Factor) are not be considered.

Inspired by this literature, we propose a detailed performance assessment of real mobile VoIP traffic across an LTE-A environment with the double aim of: \textit{i)} going beyond the obvious limitations encountered when dealing with simulation tools, which typically cannot carry out an end-to-end quality analysis; and \textit{ii)} offering a richer characterization of mobile VoIP traffic by considering also metrics that are typically not available in other studies.

Results of our study can also be applied to the 5G domain, whose architecture is quite close to LTE-A. In this sense, interesting findings appear in \cite{murroni} where, by means of an extensive simulation campaign carried out through the OMNet++ network simulator,  the performance of an algorithm called TYDER has been evaluated in terms of metrics such as throughput and delay. Furthermore, results presented in our paper might also be conveyed in 5G-compliant procedures such as VLM+ proposed in \cite{trivisonno2015}, where novel QoS-based concepts (e.g. Virtual link end-to-end QoS, profit driven QoS class prioritization) are taken into account. Finally, some attempts in characterizing QoE-related metrics for 5G scenarios have been performed in \cite{5gqoe}. Although carried out in a controlled environment, where no external phenomena are present, authors evaluate metrics related to the subjective users' perception of streamed videos. 

\section{Experimental Setting}
\label{sec:expset}

\begin{figure*}[t!]
	\centering
	\captionsetup{justification=centering}
	\includegraphics[scale=0.58,angle=90]{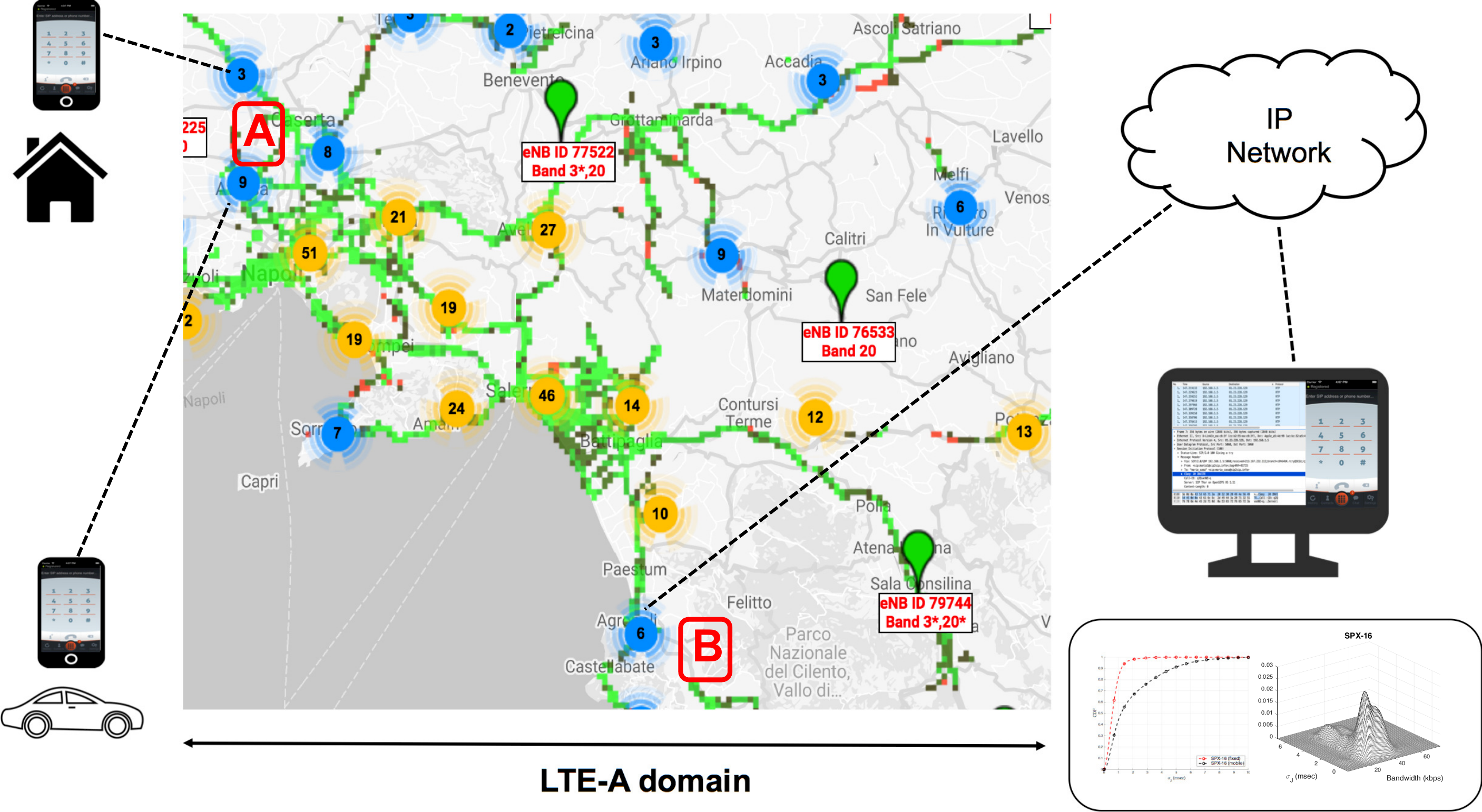}
	\caption{Pictorial representation of the considered experimental setting that includes: \textit{i)} user device (point $A$) equipped with a VoIP softphone and a 4G-plus Sim Card across two scenarios: mobile (car) and fixed; \textit{ii)} real LTE-A domain of Vodafone telco operator (in the middle); \textit{iii)} a standard pc equipped with a softphone and a network analyzer to extract VoIP parameters during the call (point $B$).}
	\label{fig:architecture}
\end{figure*} 

This section is devoted at describing the scenario and the setting that have been considered during the experimental campaign (performed across the city of Salerno, Italy) pictorially illustrated in Fig. \ref{fig:architecture}. All the described trials have been carried out by exploiting a real LTE network that, typically, is based on the following key nodes: the eNB (enhanced Node B) designated to supervise the overall radio communication, the MME (Mobility Management Entity) in charge of managing the VoIP sessions and the authentication procedures together with HSS (Home Subscriber Server), the SGW (Serving Gateway) having routing functions, the PGW (Packet Data Gateway) playing as a proxy server for external networks or domains (e.g. IP Multimedia Subsystem), the PCRF (Policy and Charging Rules Function) devoted at managing billing and charging procedures. 

Two endpoints (say, $A$ and $B$) have been considered in the experiment. The first one ($A$) is an iOS-based smartphone equipped with Linphone, a popular open-source VoIP software that allows to make/receive phone calls by means of two protocols: the Session Initiation Protocol (SIP) in charge of managing the signalling part of the communication; and the Real-time Transport Protocol (RTP) in charge of carrying the call content. Linphone natively includes some audio codecs, even if some extra licensed codecs have been purchased to enrich the experiment (see Table \ref{tab:codec} for a complete list of used codecs along with their main features). The smartphone is equipped with a 4G-Plus Sim Card that allows to exploit the LTE-A services offered by the telecom operator (Vodafone) during voice calls. The second endpoint ($B$) is a personal computer equipped with the same VoIP application and connected to a network sniffer used to capture the entire VoIP session during the call. 
The two endpoints are synchronized by means of the Network Time Protocol (NTP), and are both attached to the NTP primary server of the Italian Institute on Metrologic Research (thus, the NTP inaccuracy is negligible). All measurements are performed at the second endpoint where, through the network sniffer, we can access packets sent and received from both ends ($A \rightarrow B$, $B \rightarrow A$) providing consistent information, thanks to the common clock reference.

\subsection{Mobile and fixed scenarios}

\begin{table}[t!]
	\caption {VoIP Codecs considered in the experiment} \label{tab:codec}
	\resizebox{.48\textwidth}{!}{
		\begin{tabular}{|c|c|c|c|}
			\hline
			Codec & Algorithm Type & Bit Rate (Kbps) & Sampling Rate (KHz) \\
			\hline
			\\[-8pt]
			G.711 (a-law) & PCM & 64 Kbps & 8 KHz \\ \hline
			G.722 & ADPCM & 64 Kbps & 16 KHz \\ \hline
			G.729 & CS-ACELP & 8 Kbps & 8 KHz \\ \hline
			GSM & RPE-LTP & 8 Kbps & 8 KHz \\ \hline
			Speex-8 (Narrow Band) & CELP & 8 Kbps & 8 KHz \\ \hline
			Speex-16 (Wide Band) & CELP & 16 Kbps & 16 KHz \\ \hline
			OPUS & LP-MDTC & VBR (6 to 128 Kbps)& 48 KHz \\ \hline
			(AAC) MPEG4-16 & CELP & 16 Kbps & 16 KHz \\ 
			\hline
		\end{tabular}}
	\end{table}
	
The whole experiment can be divided in two settings: the first scenery is a urban mobility environment where a driver speaks using a hands-free connection to a smartphone, driving at an average speed of 60 Km/h.\footnote{The actual speed ranges between approximately 20 Km/h (average urban speed) and 100 Km/h (average highway speed).}. The experiments start from point $A$ of Fig. \ref{fig:architecture}. The other communication end-point is labeled as point $B$ and involves a second user with a personal computer. When the user located in $A$ initiates the VoIP call towards user located in $B$, there is a preliminary phase (managed by the SIP protocol) during which the devices exchange information (e.g. the SIP addresses) and negotiate some capabilities (e.g. codecs that must be the same for the two peers). During this experiment, a total of $359,768$ RTP packets (corresponding approximately to $1$ hour and $10$ minutes of conversations) between the two endpoints have been exchanged.

The second scenery is set as for the first one, except that the driver remains stationary at point $A$ for the whole experiment. To avoid ambiguity, we refer to the first scenario as \textit{mobile}, and to the second scenario as \textit{fixed}. The approximate distance between the fixed caller equipment and (fixed) callee equipment is 100 km. The urban territory between the two users has an approximate density of 2,000 people per square km and is surrounded by more than 100 e-nodeB installations \cite{cellmapper} of the operator considered for the test. In Fig. \ref{fig:architecture}, the figures in the blue and yellow circles indicate the number of eNBs across the traversed territory. On the other hand, the distance between mobile caller equipment and fixed callee equipment varies between 80 and 30 km. During the fixed experiment, a total amount of $410,150$ RTP packets (corresponding approximately to $1$ hour and $20$ minutes of conversations) between the two endpoints have been exchanged.
Table \ref{table:gen_par} summarizes the main values of experiment parameters.
	\begin{table}[t!]
		\caption{Experiment Parameters}
		\resizebox{.48\textwidth}{!}{
			\begin{tabular}{|c|c|c|c}
				\hline
				Number of eNBs surrounding the covered area & $\sim 100$\\ \cline{1-2}
				Mobile Technology  & LTE-A (Vodafone operator) \\ \cline{1-2}
				Type of territory  & Urban ($\sim 2,000$ people sqKm) \\ \cline{1-2}
				Distance between caller (fixed) and callee &  $\sim 100$ km \\ \cline{1-2}
				Distance between caller (mobile) and callee & $\sim 30-80$ km \\ \cline{1-2}
				Average speed of mobile user (car) & $\sim 60$ Km/h \\ \cline{1-2}
				Average duration per call & $\sim 500$ seconds \\ \cline{1-2}
				Total number of RTP packets exchanged & 769,918 \\
				\hline 
			\end{tabular}}
			\label{table:gen_par}
		\end{table}
When operating in a real network environment (as opposed to an emulated/simulated framework), there are two fundamental aspects to take into account. The first one concerns the high variability of the parameters at stake, which could not be governed under simulations. These include: interference phenomena (as consequence of propagation of real radio signals) introduced by buildings and vehicles; radio cell switching due to the car mobility scenario; and signal degradation due to atmospheric conditions. Such unpredictable events obviously affect the parameters of a VoIP session (throughput, jitter, RTT, etc.) whereas specific codecs introduce their own variability. 

The second aspect concerns the difficulty of interpreting real results that, in some cases, can be counter-intuitive. For example, in a simulated environment, one would expect a throughput that decreases as packet loss grows, whereas in a real environment some negative effects could compensate each other, thus producing an unexpected improvement. 

Before delving into results, some clarifications about adopted metrics are needed. As concerns jitter, since it is strongly affected by the unpredictability of real conditions (time-variance of moving objects, micro-climate conditions, traffic conditions, external interferences), a characterization in terms of jitter deviation from average values will allow us to better grasp the ``stability" of a VoIP communication. This is, in fact, an approach that is often adopted in literature \cite{voip-handbook,chang2012,jou2003,icufn}. 
To confirm this, we have carried out a range of measurements (unreported herein, for brevity), which confirm that in both the mobile and the fixed scenarios (and for each of the codecs under scrutiny), the average jitter values are in the range of tens of millisecond, with standard deviations falling within the same order of magnitude. Hence, the deviation of jitter can provide useful information about its ``stability", as external conditions become increasingly critical and may lead to a frequent violation of QoS thresholds.	Such considerations led us to treat the jitter's deviation as a random variable, as discussed in the following.

In order to capture fine dynamics of jitter \cite{garch}, we derive $\sigma_J$ as a moving standard deviation, useful to prevent the {\em homoskedasticity} constraint implicit in the use of the sample standard deviation. Such an approach calls for the introduction of a moving window of a specific size which, in our case, amounts to $1$. In fact, window size should be chosen to be small enough compared to the data size and reflects a tradeoff between resolution and estimation error. This choice allows to characterize the standard deviation as a random variable exhibiting different micro-behaviors in each window. Thus, a more granular analysis is possible. 
Similarly, in the case of consumed bandwidth metric (having the same significance of the achieved throughput for a data session) we apply a moving average technique so that time variability could emerge across the whole analysis.

\begin{figure}[t]
	\centering
	\captionsetup{justification=centering}
	\includegraphics[scale=0.31,angle=90]{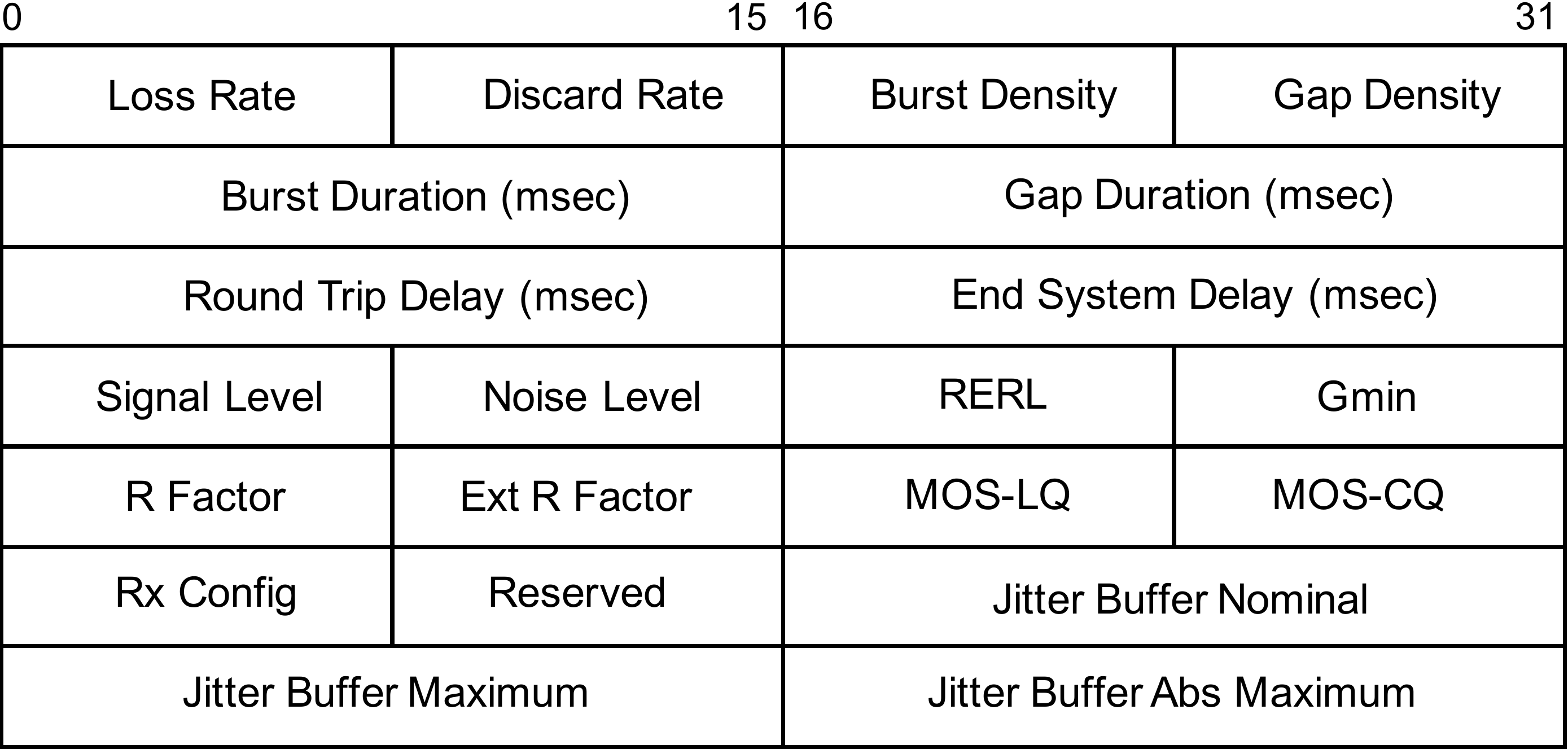}
	\caption{VoIP metrics report block (RTCP XR).}
	\label{fig:rtcp-xr}
\end{figure} 

\begin{figure*}[ht!]
	\centering
	\begin{minipage}[t]{0.26\textwidth}
		\includegraphics[width=\textwidth]{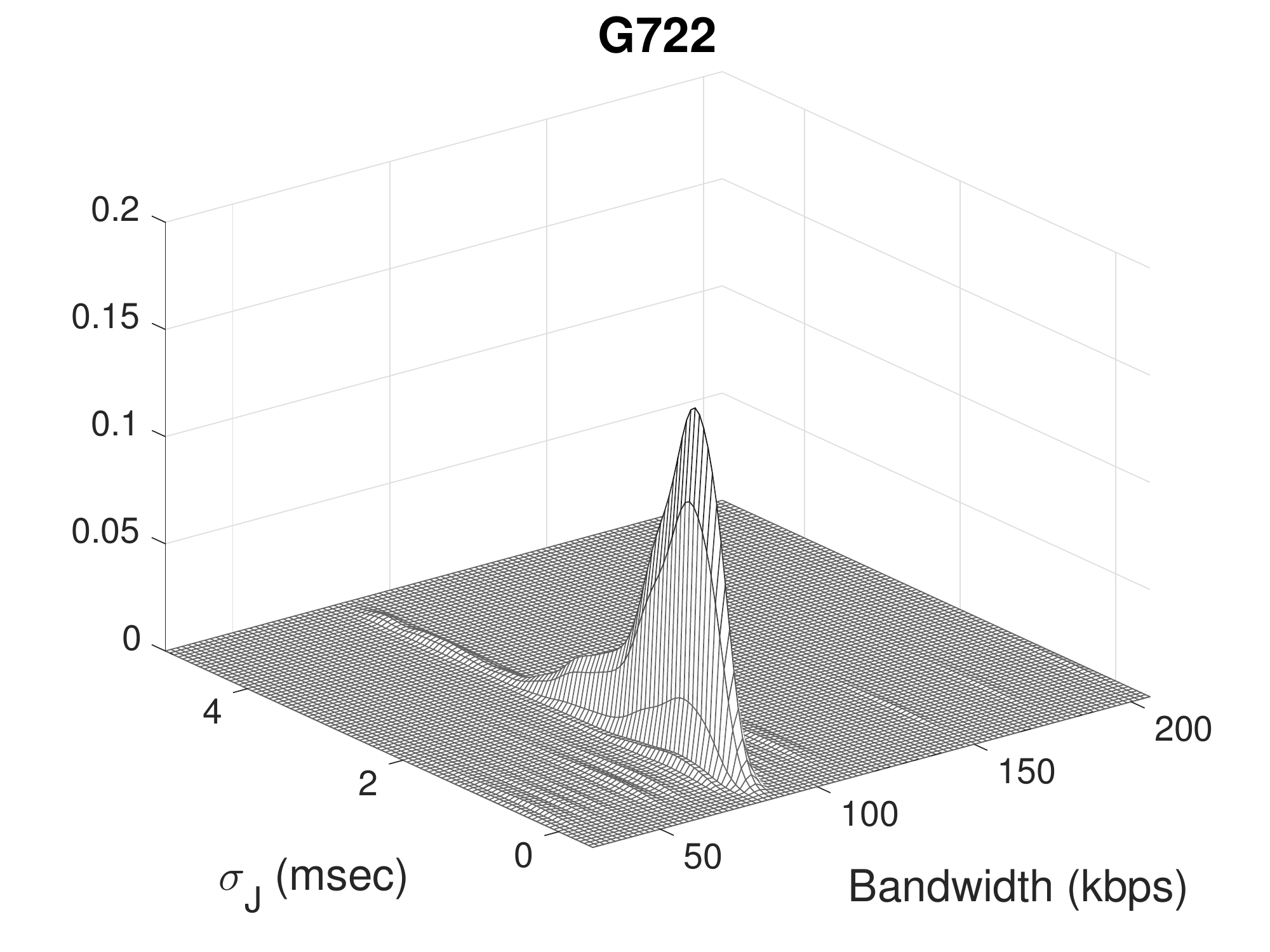}
	\end{minipage} \hspace{2pt}
		\begin{minipage}[t]{0.26\textwidth}
			\includegraphics[width=\textwidth]{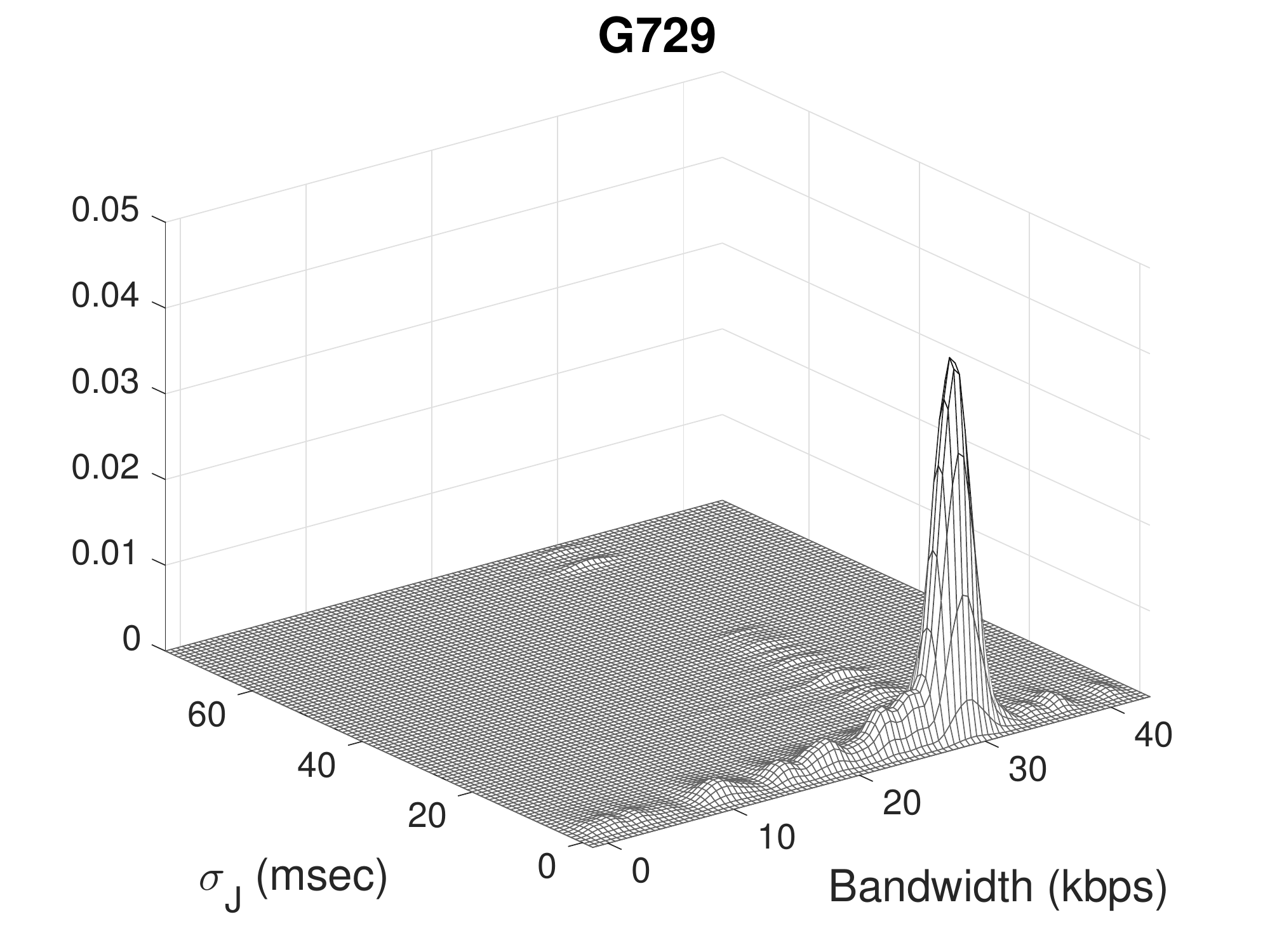}
		\end{minipage} \hspace{2pt}
	\begin{minipage}[t]{0.26\textwidth}
		\includegraphics[width=\textwidth]{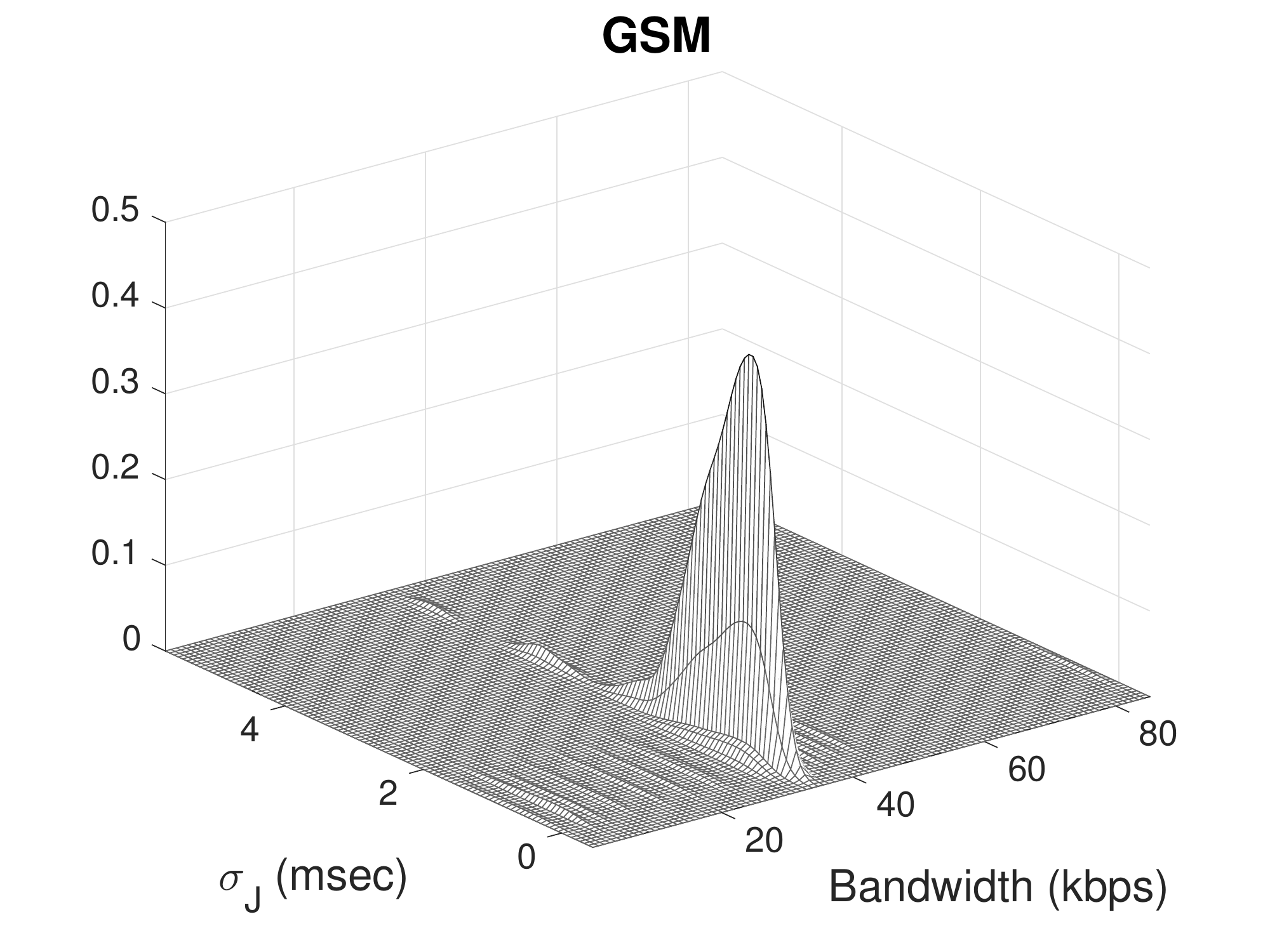}
	\end{minipage}	\hspace{2pt}
	\begin{minipage}[t]{0.26\textwidth}
		\includegraphics[width=\textwidth]{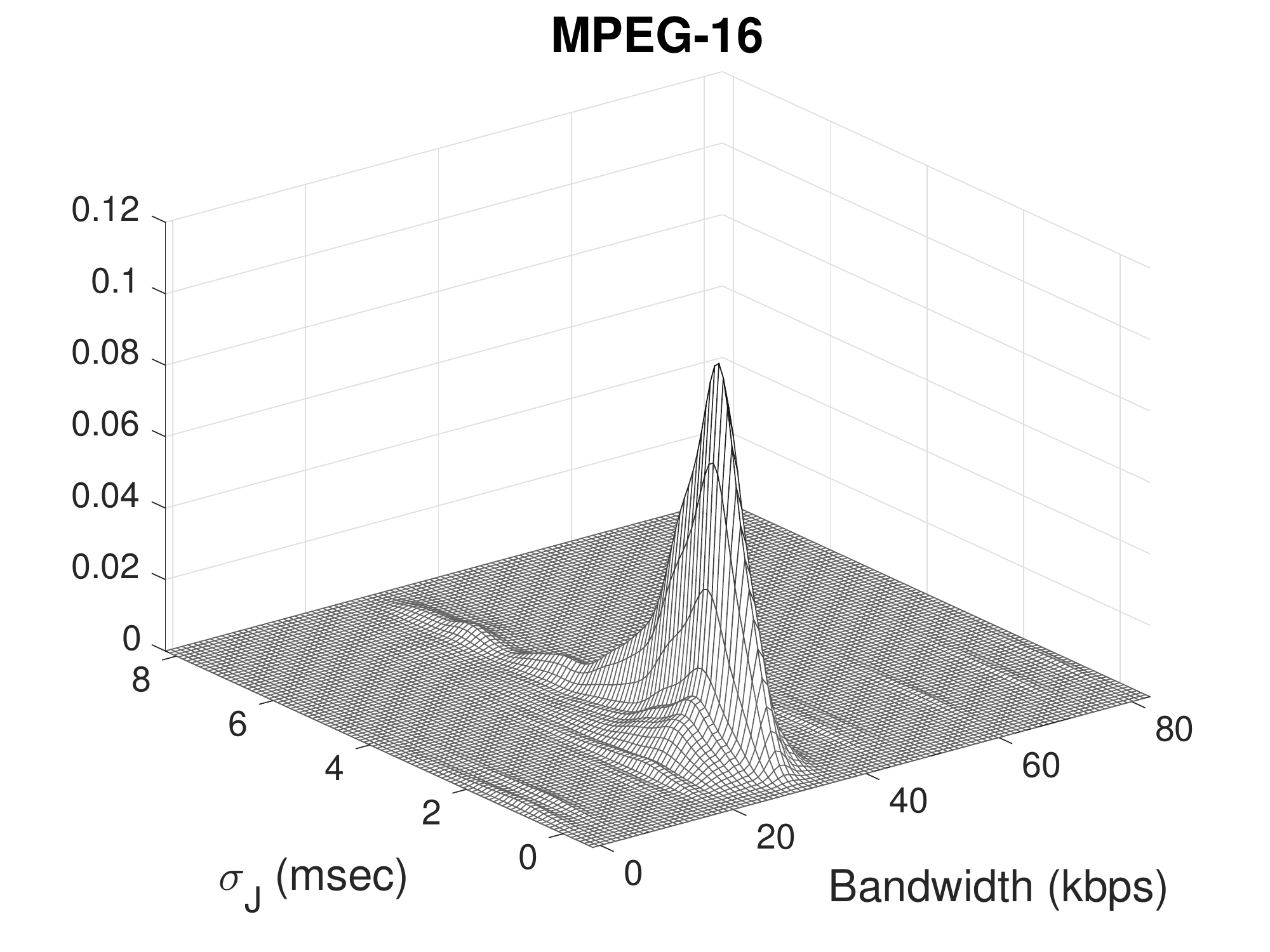}
	\end{minipage}	\hspace{2pt}
	\begin{minipage}[t]{0.26\textwidth}
		\includegraphics[width=\textwidth]{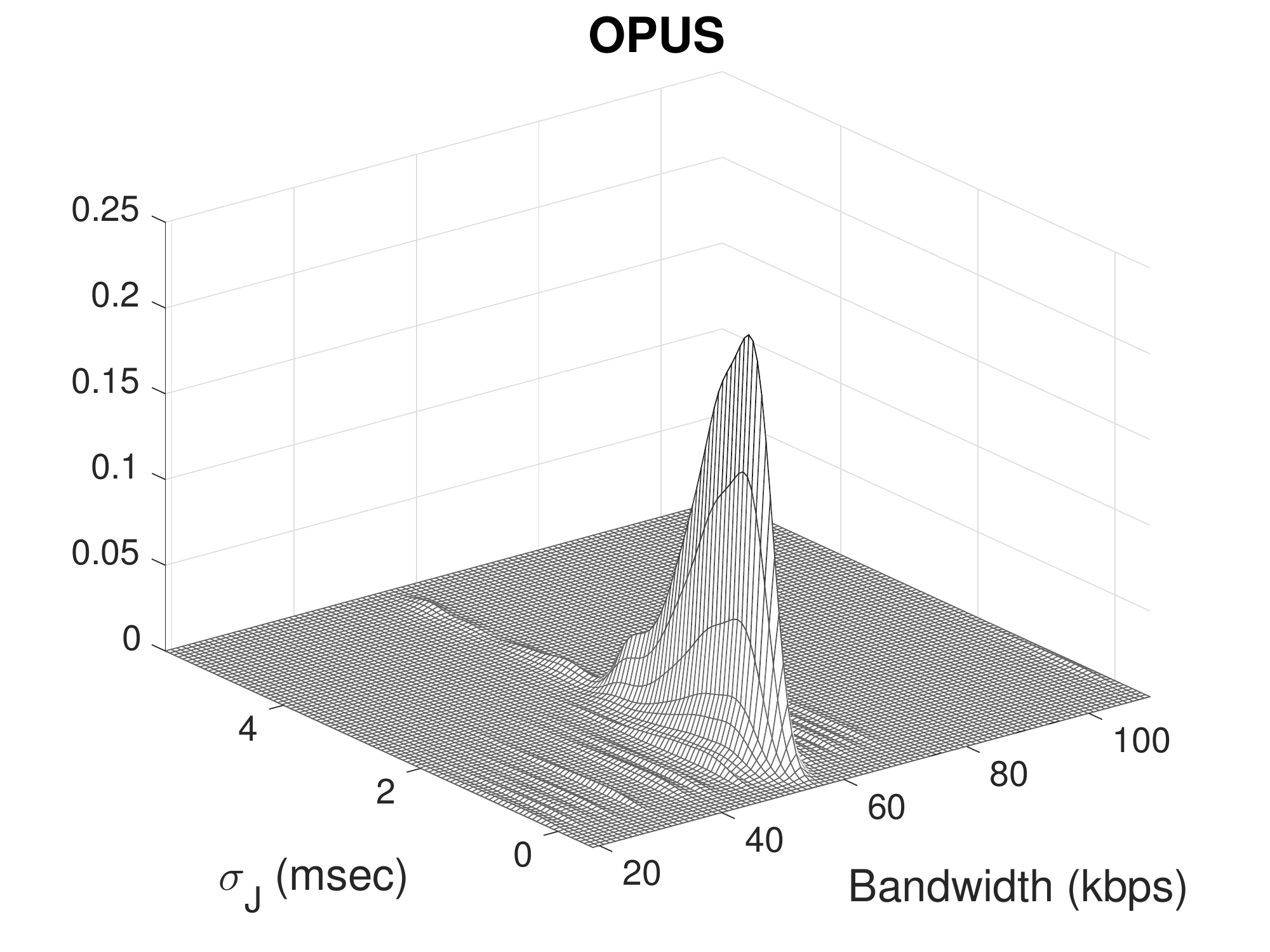}
	\end{minipage}	\hspace{2pt}	
	\begin{minipage}[t]{0.26\textwidth}
		\includegraphics[width=\textwidth]{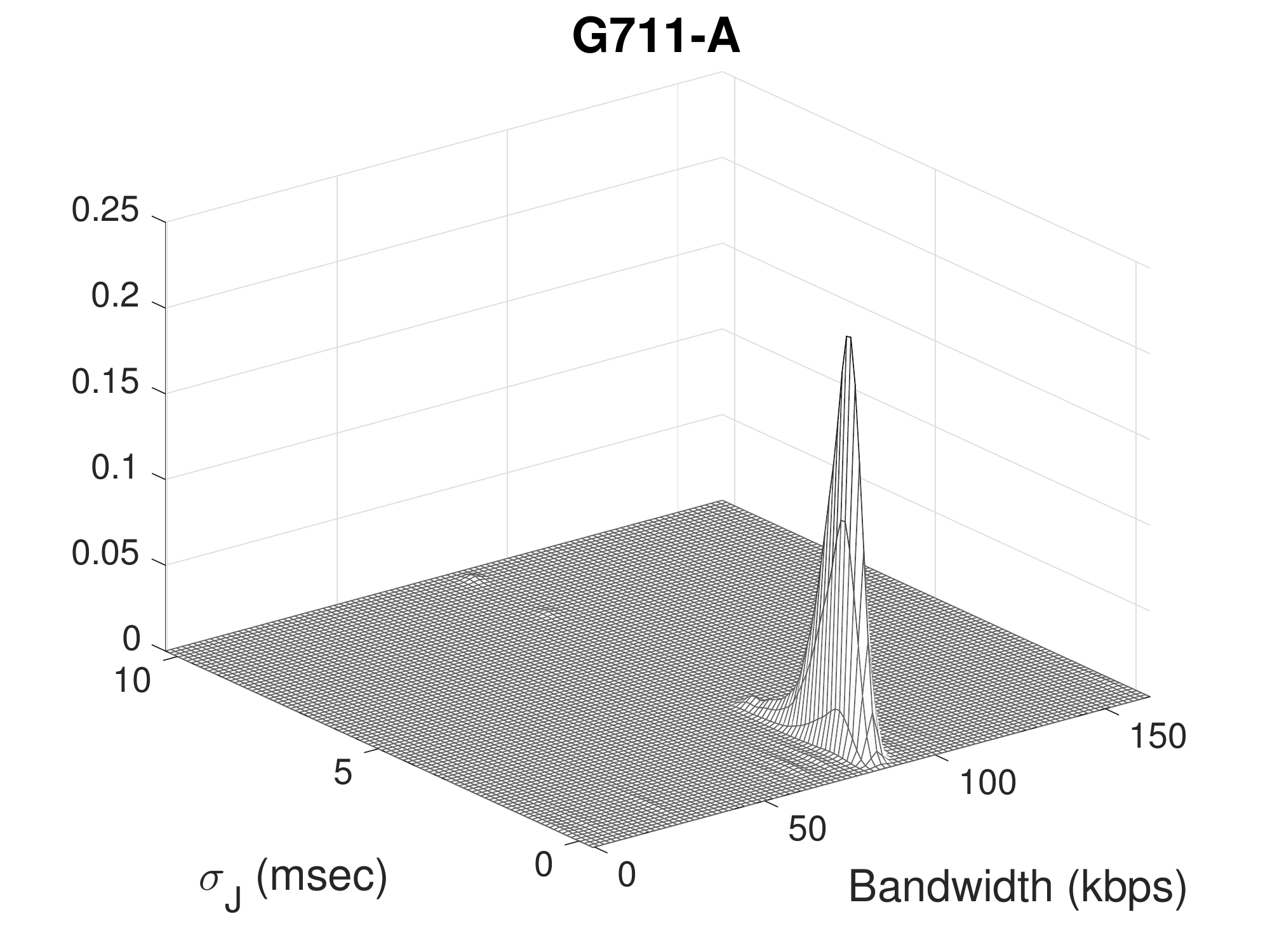}
	\end{minipage}	\hspace{2pt}
	\begin{minipage}[t]{0.26\textwidth}
		\includegraphics[width=\textwidth]{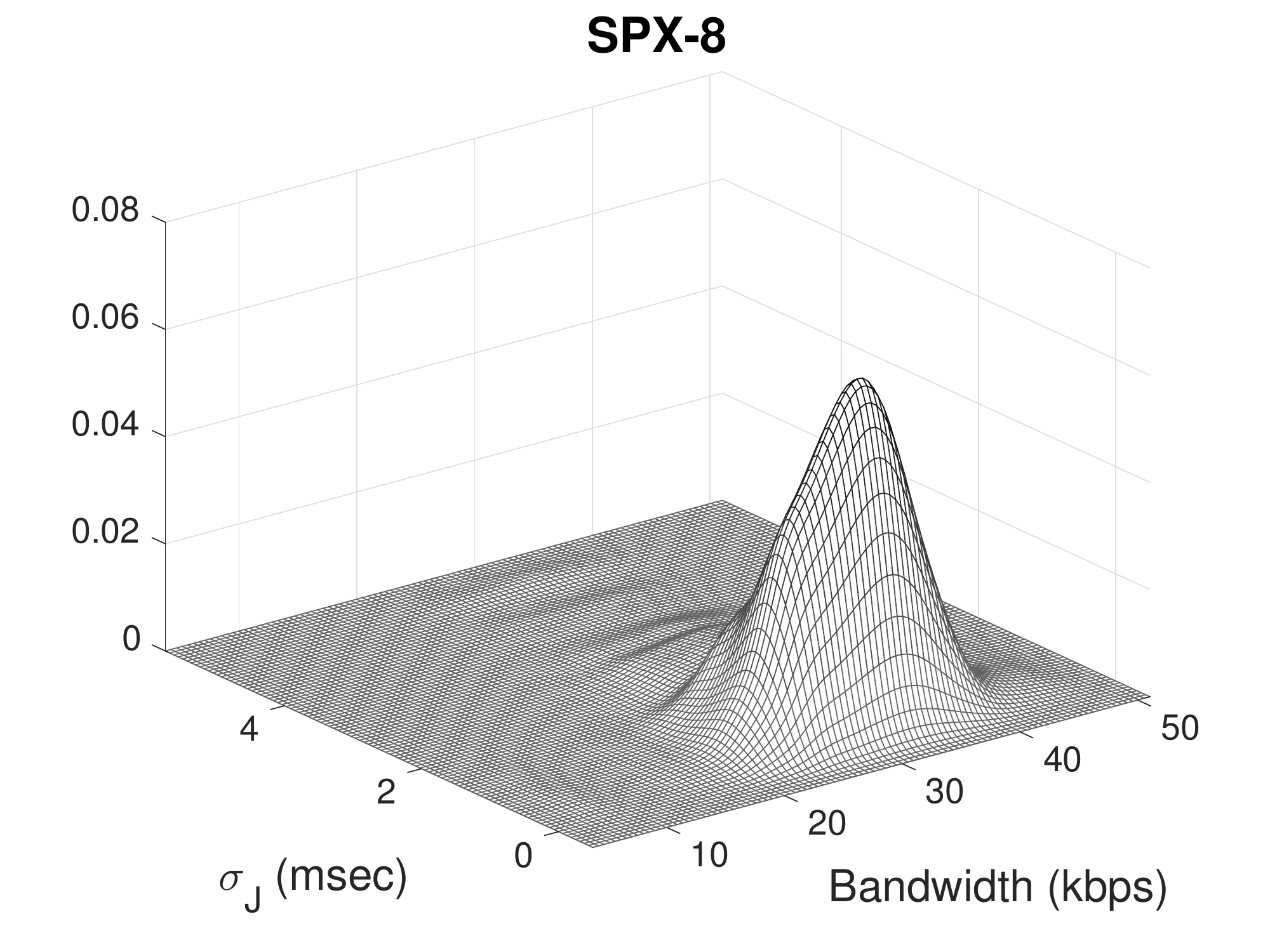}
	\end{minipage}		\hspace{2pt}
	\begin{minipage}[t]{0.26\textwidth}
		\includegraphics[width=\textwidth]{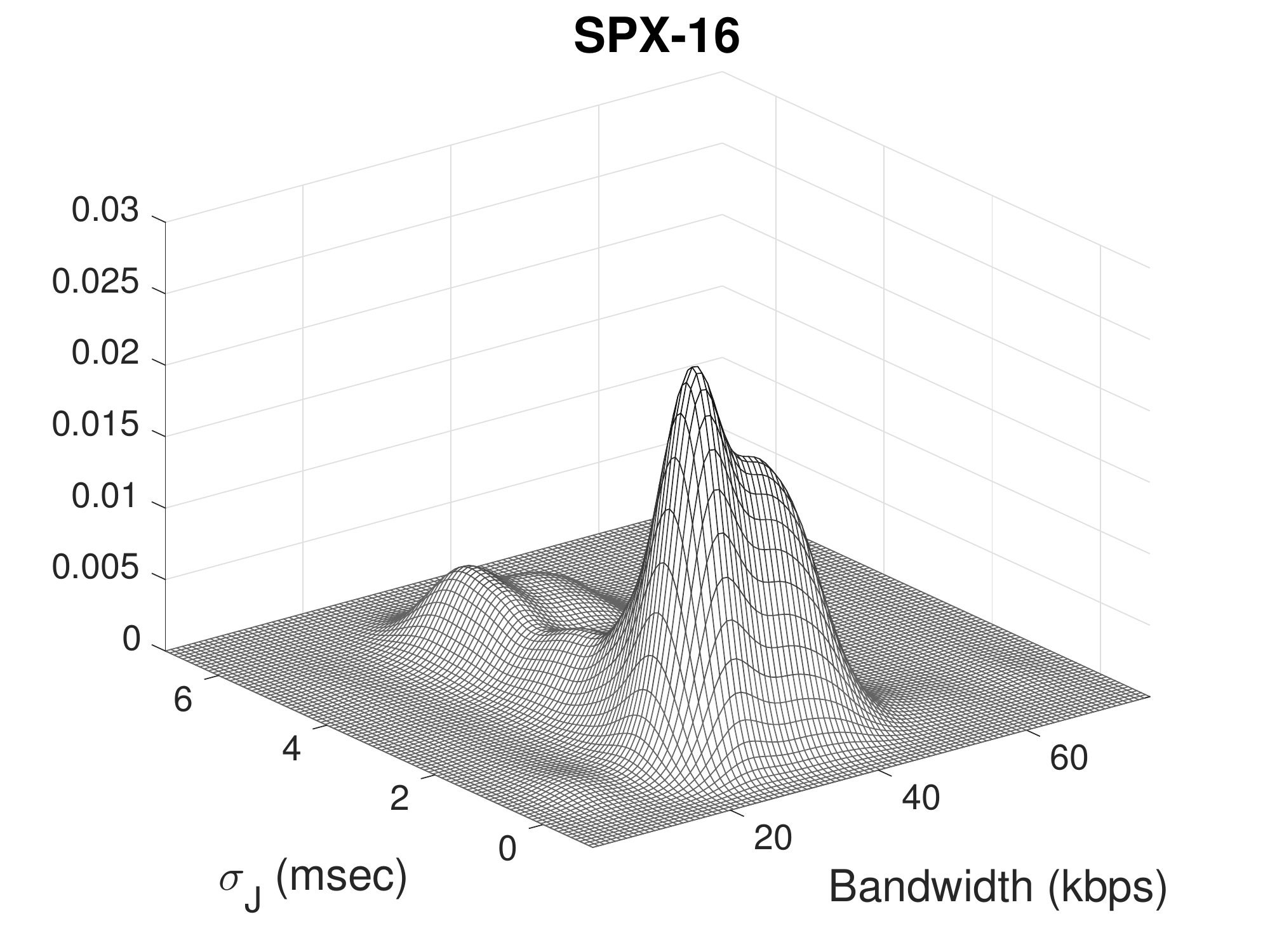}
	\end{minipage}		
	\caption{Bivariate distribution of consumed bandwidth (kbps) and jitter standard deviation $\sigma_J$ (msec) for various VoIP sessions with different codecs.}
	\label{fig:tputdevjit}
\end{figure*}

\subsection{Metrics available through RTCP Extended Reports (XR)}
The RTP Control Protocol (RTCP) \cite{rtcp} has the main objective of providing some out-of-band statistics and control information for an RTP flow. A slightly more recent version of this protocol, called RTP Control Protocol - Extendend Reports (RTCP-XR) and defined in \cite{rtcp-xr}, carries additional QoS-related metrics in a dedicated block called VoIP metrics Report Block, encoded as seven $32$-bit words and shown in Fig. \ref{fig:rtcp-xr}. Because the implementation of RTCP-XR depends also on the specific VoIP softphone, it justifies the choice of Linphone to manage our trials. 

We actually exploit a combination of classic and extended metrics to assess our performance analysis. Among classic metrics we consider: \textit{i)} jitter, calculated by the formula  $J_n=|({t_r}_{(n)} - {t_t}_{(n)}) - ({t_r}_{(n-1)} - {t_t}_{(n-1)}  )|$, representing the jitter of $n$-th packet that depends on the transmitting time of the $n$-th packet (${t_t}_{(n)}$) and on the receiving time of the $n$-th packet (${t_r}_{(n)}$){\footnote{This quantity has been measured at the entry of endpoint $B$ jitter buffer.}; \textit{ii)} bandwidth consumption that is strongly related to the codec type (poor compression implies more consumed bandwidth). Among additional metrics available through RTCP XR we consider: \textit{i)} round-trip delay (coinciding with RTT), a 16-bit field in RTCP XR (see Fig. \ref{fig:rtcp-xr}), defined as the time interval starting when a packet is sent from a source and completed as soon as the ack is received from the destination. For RTP traffic, the timestamp value is used to approximate the RTT value;  \textit{ii)} R-Factor (rating factor) \cite{itutg107}, a 8-bit field in RTCP XR (see Fig. \ref{fig:rtcp-xr}) defined as $R=R_0 - I_s - I_d - I_{eff} + A$ where: $R_0$ represents the signal-to-noise ratio (including sources such as circuit noise and room noise), $I_s$ includes a combination of various impairments occurring simultaneously with the voice signal, $I_d$ expresses delay impairments, $I_{eff}$ represents impairments caused by low bit-rate codecs, and $A$ is an advantage factor to compensate various impairment factors; \textit{iii)} voice signal level, an 8-bit field in RTCP XR (see Fig. \ref{fig:rtcp-xr}) specified only for packets containing speech energy, defined as the ratio of the signal level to a 0 dBm0 reference \cite{eiatia2000}.

\section{Performance Evaluation}
\label{sec:perfeval}

Aimed at providing a detailed characterization and a useful performance assessment, in this section we consider a set of experiments and measurements that include: \textit{i)} a set of bivariate analyses useful to catch the mutual influences between some key factors (consumed bandwidth, RTT, R-Factor) and the jitter variation within the mobile scenario for various codecs; \textit{ii)} a set of CDF-based comparative performance analyses amenable to characterize the behavior of some VoIP parameters when acting in the mobile and fixed scenario, and for various codecs; \textit{iii)} a set of measurements aimed at characterizing some high-level parameters that intervene during the signalling phase of a VoIP session.

\subsection{Bivariate analysis}

Let us first consider the relationships among the different parameters in the mobile scenario. The first set of results is expressed in terms of joint probability density functions, and, more precisely, in terms of bivariate distributions giving the probability that a couple of random variables falls in a specific range of values\footnote{Some alternative representations are possible such as PCA. Please refer to Appendix $A$ where an exemplary PCA is reported.}. In Fig. \ref{fig:tputdevjit} we analyze the mutual influence of consumed bandwidth (expressed in kbps) and standard  deviation of jitter ($\sigma_J$ expressed in msec) in a mobile environment, being both responsible for voice call quality.
From results two main things emerge. First, the range of values assumed by the consumed bandwidth is clearly codec-dependent. This was to be expected and is in line with technical literature (see \cite{cisco-codec}). For example, codecs G711-A and G722 result to be the less efficient ones, exhibiting a bandwidth consumption (on average) of about $85$ kbps  (see also Table \ref{tab:codec} for Bit Rate values of each  codec). It follows OPUS, with a bandwidth consumption of $54.8$ kbps. For all the other codecs, the bandwidth consumption is less than $35$ kbps, thus guaranteeing a good efficiency. Secondly, the peaks of bandwidth are concentrated around small values of  $\sigma_J$. Such values, according to scientific literature \cite{sheluhin-book}, are in the order of few milliseconds. More specifically, the used bandwidth is spanned around values of $\sigma_J$ not greater than $8$ msec. 
\begin{figure*}[t!]
	\centering
	\begin{minipage}[t]{0.26\textwidth}
		\includegraphics[width=\textwidth]{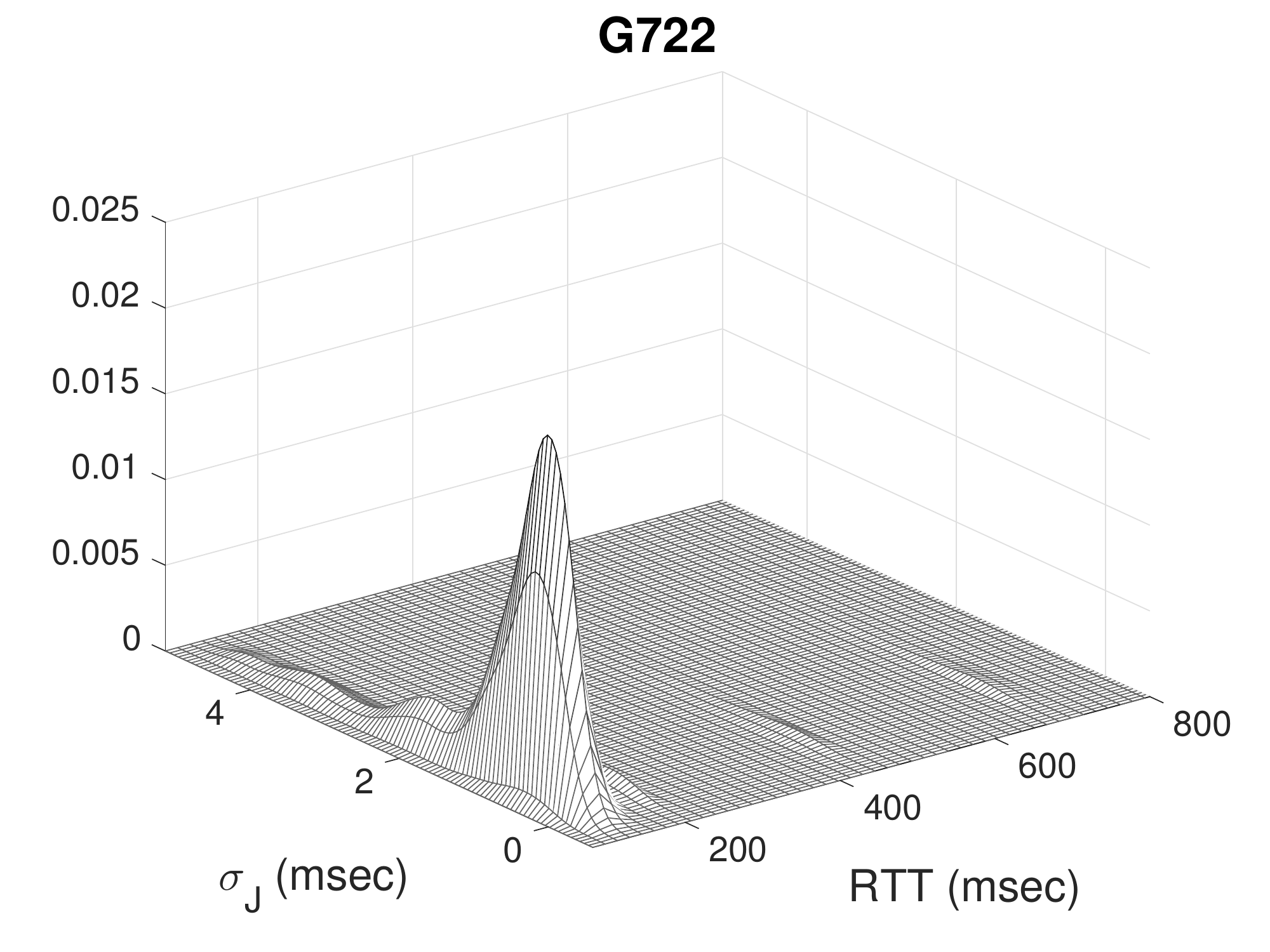}
	\end{minipage} \hspace{2pt}
	\begin{minipage}[t]{0.26\textwidth}
		\includegraphics[width=\textwidth]{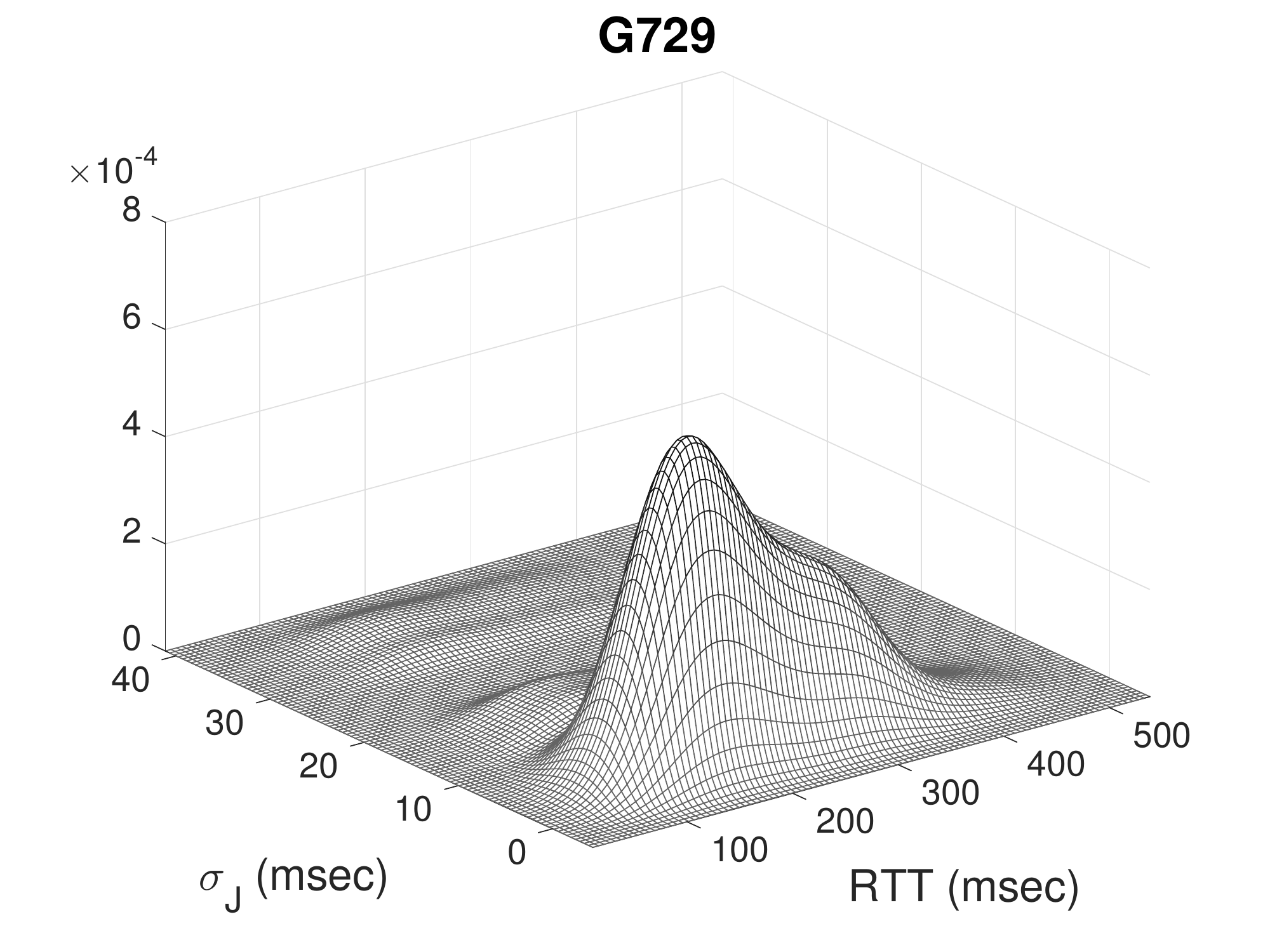}
	\end{minipage} \hspace{2pt}
	\begin{minipage}[t]{0.26\textwidth}
		\includegraphics[width=\textwidth]{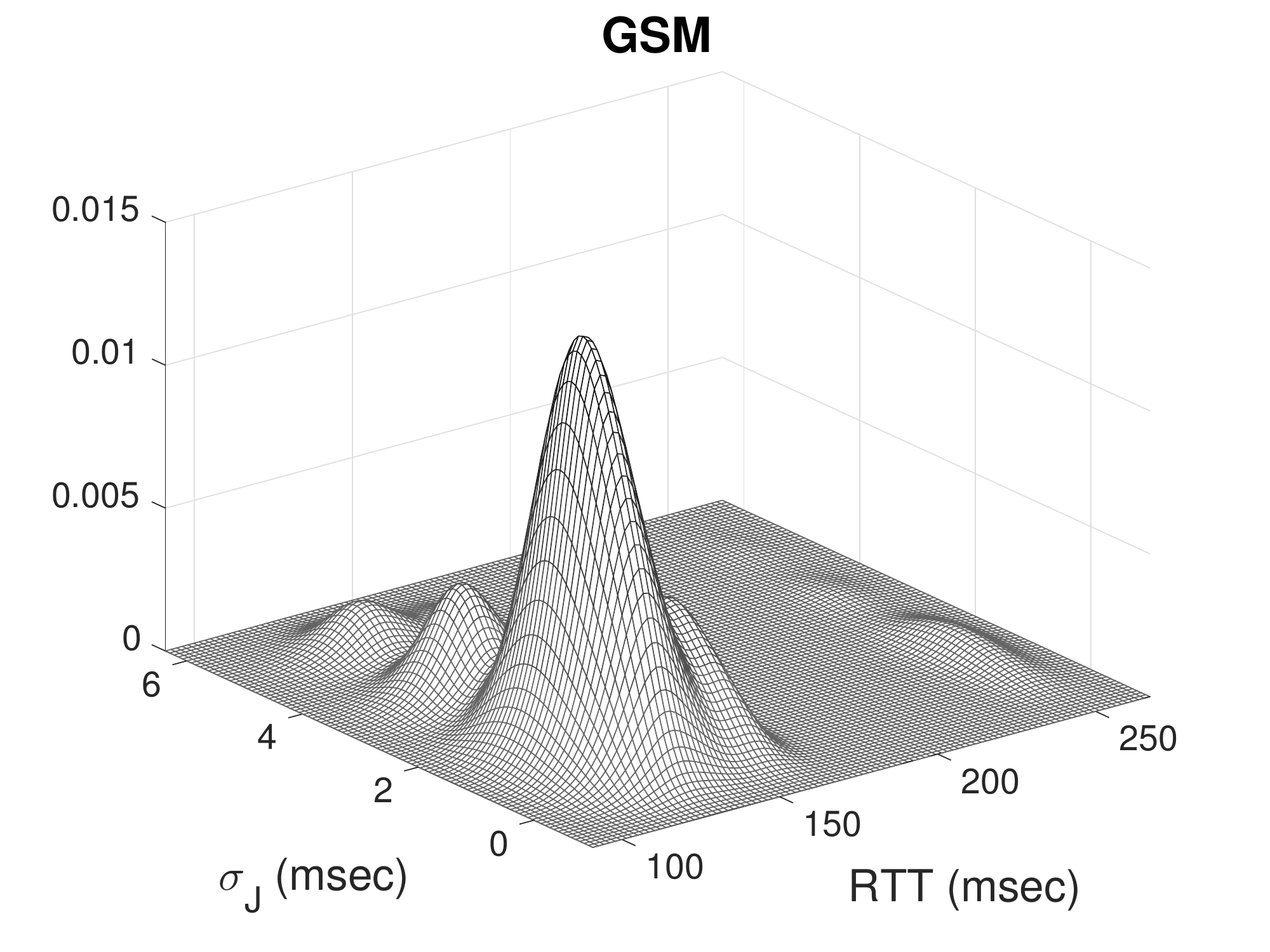}
	\end{minipage}	\hspace{2pt}
	\begin{minipage}[t]{0.26\textwidth}
		\includegraphics[width=\textwidth]{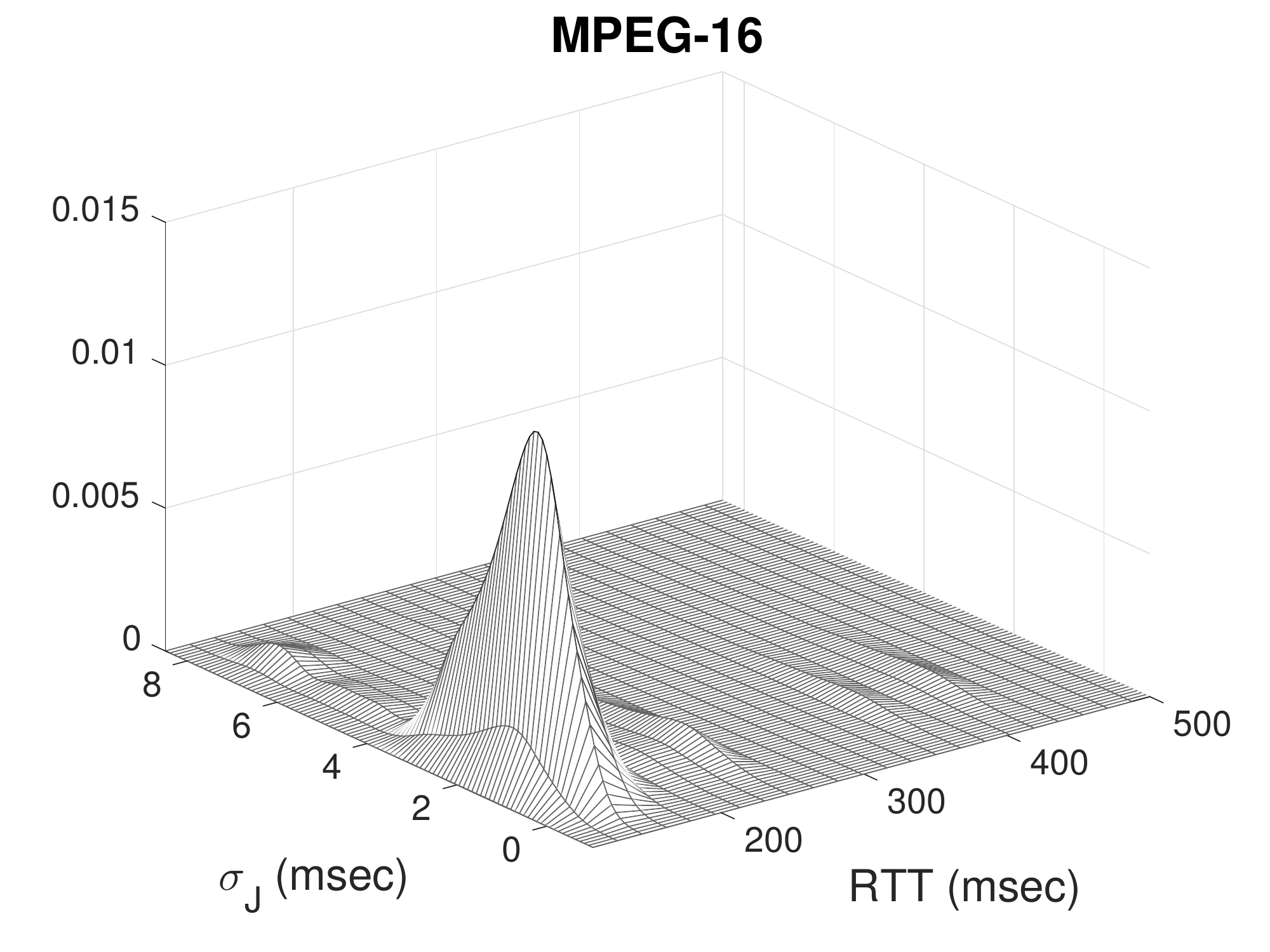}
	\end{minipage}	\hspace{2pt}
	\begin{minipage}[t]{0.26\textwidth}
		\includegraphics[width=\textwidth]{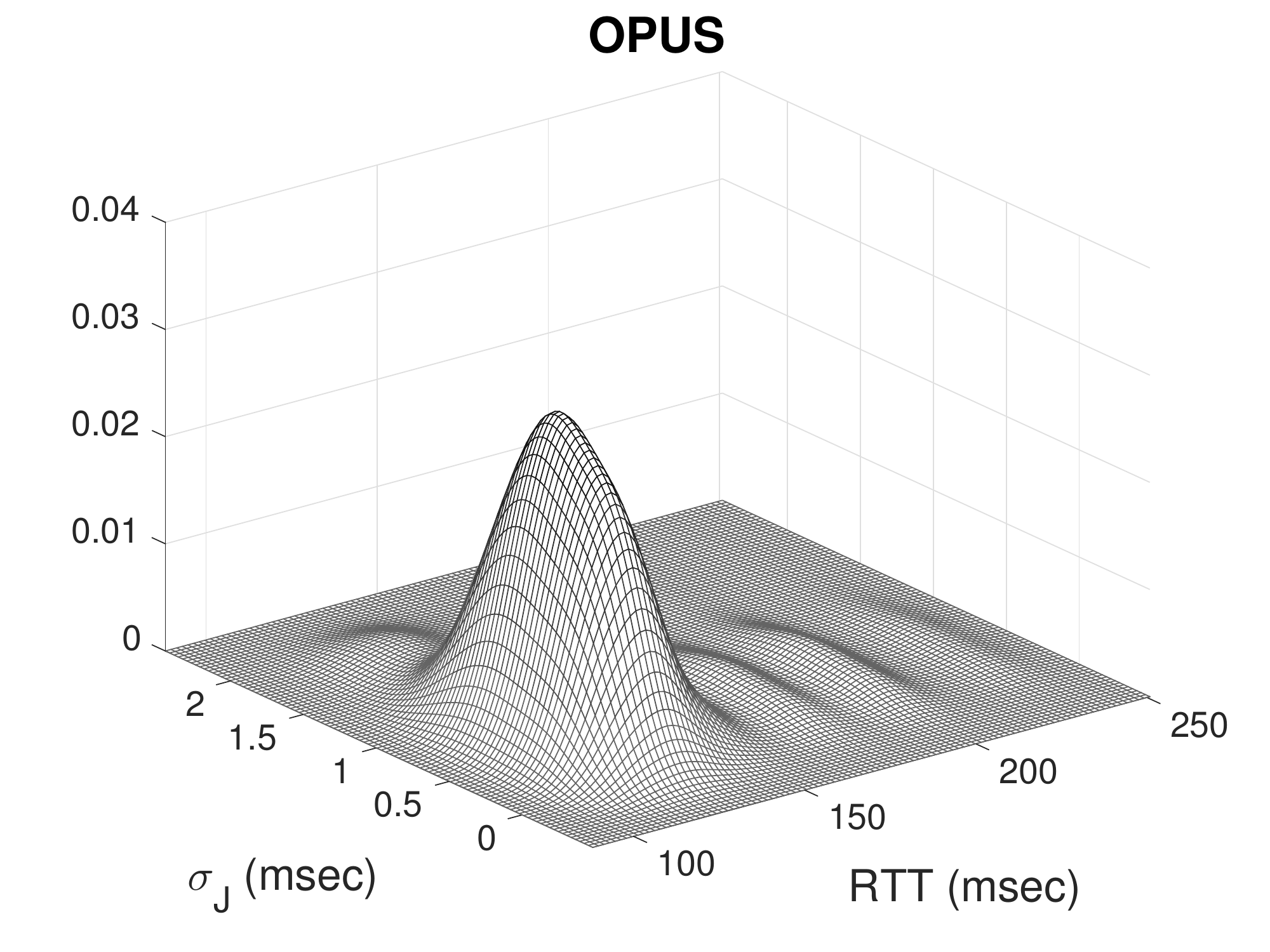}
	\end{minipage}	\hspace{2pt}	
	\begin{minipage}[t]{0.26\textwidth}
		\includegraphics[width=\textwidth]{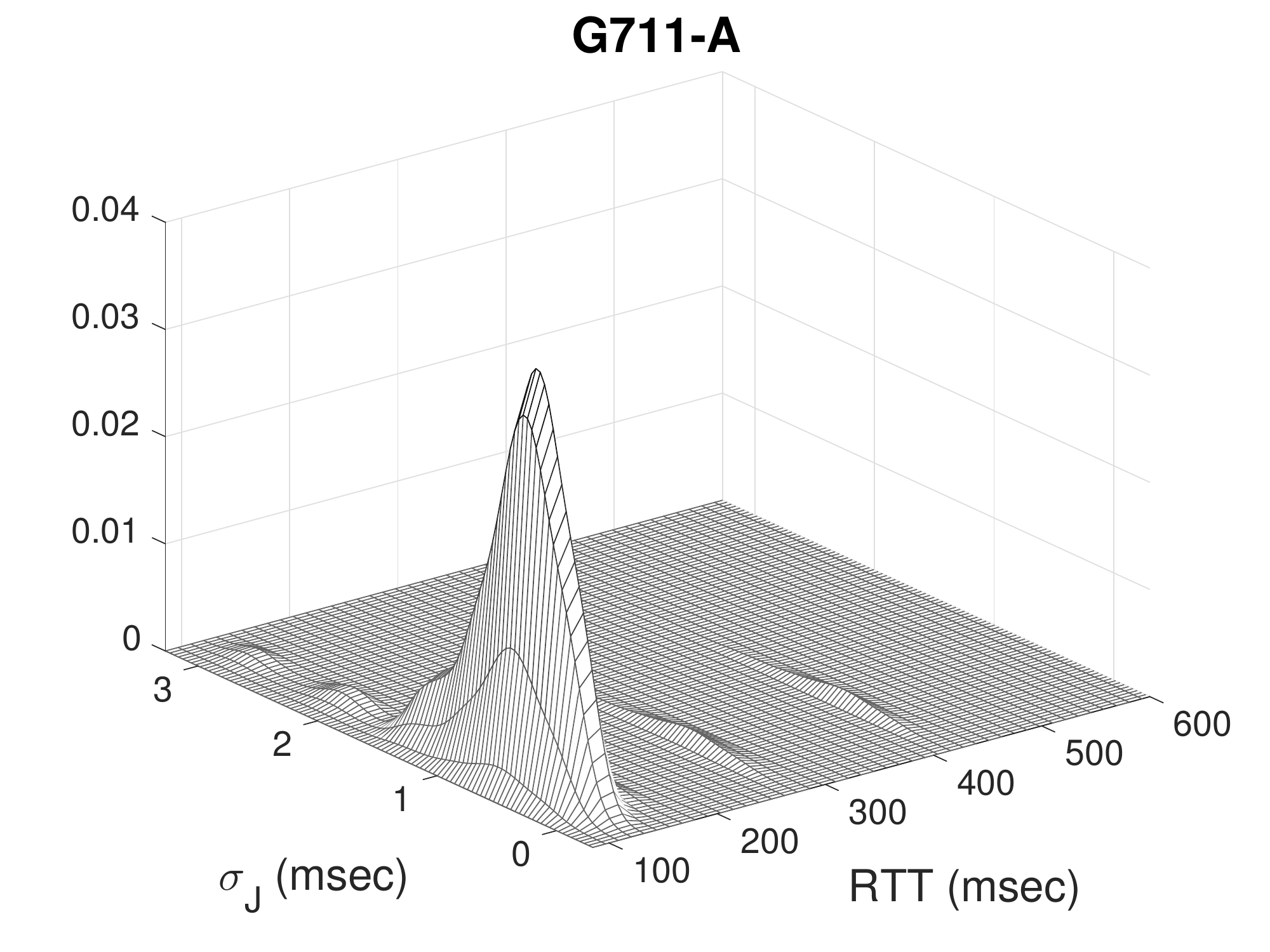}
	\end{minipage}	\hspace{2pt}
	\begin{minipage}[t]{0.26\textwidth}
		\includegraphics[width=\textwidth]{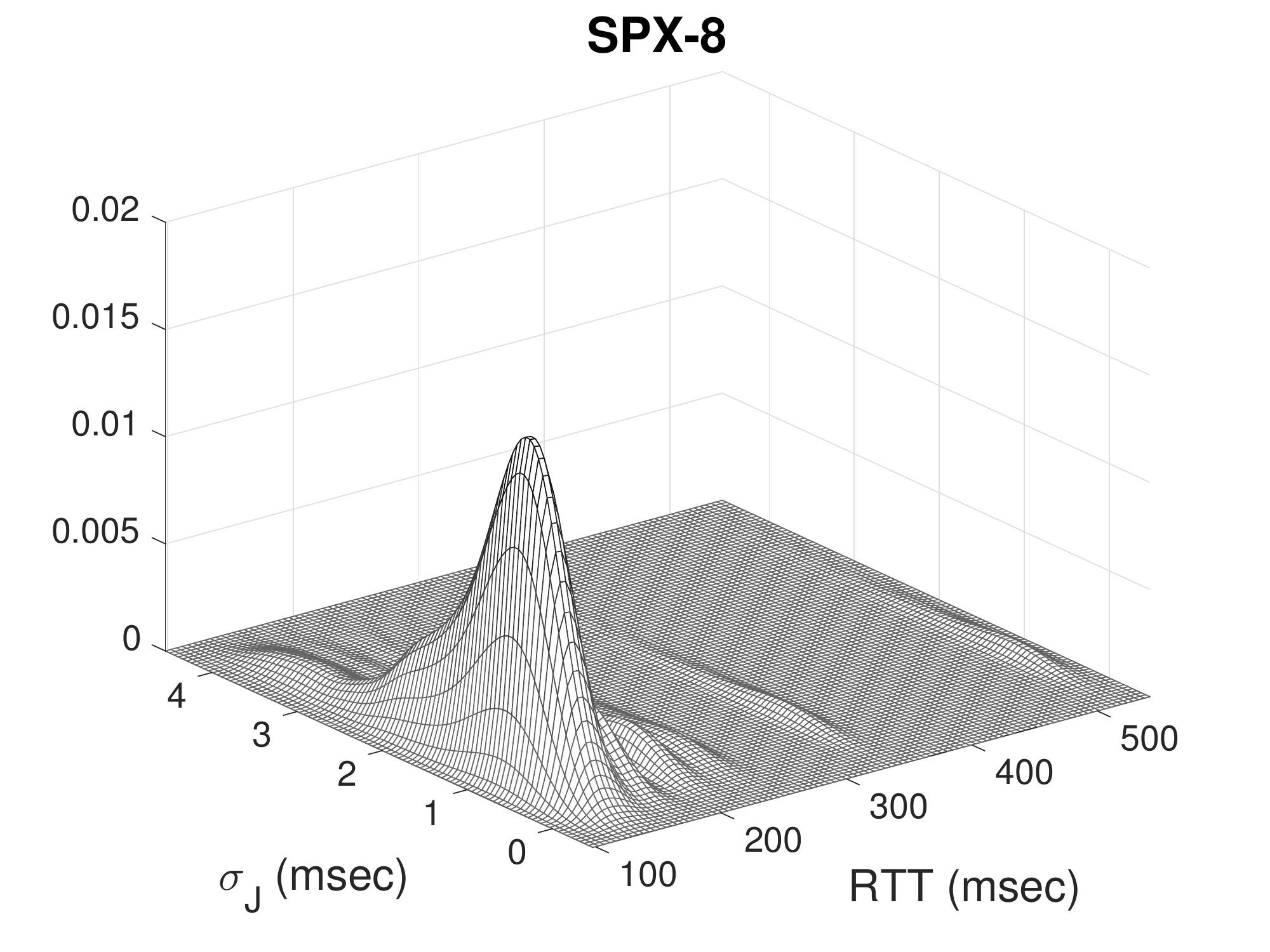}
	\end{minipage}		\hspace{2pt}
	\begin{minipage}[t]{0.26\textwidth}
		\includegraphics[width=\textwidth]{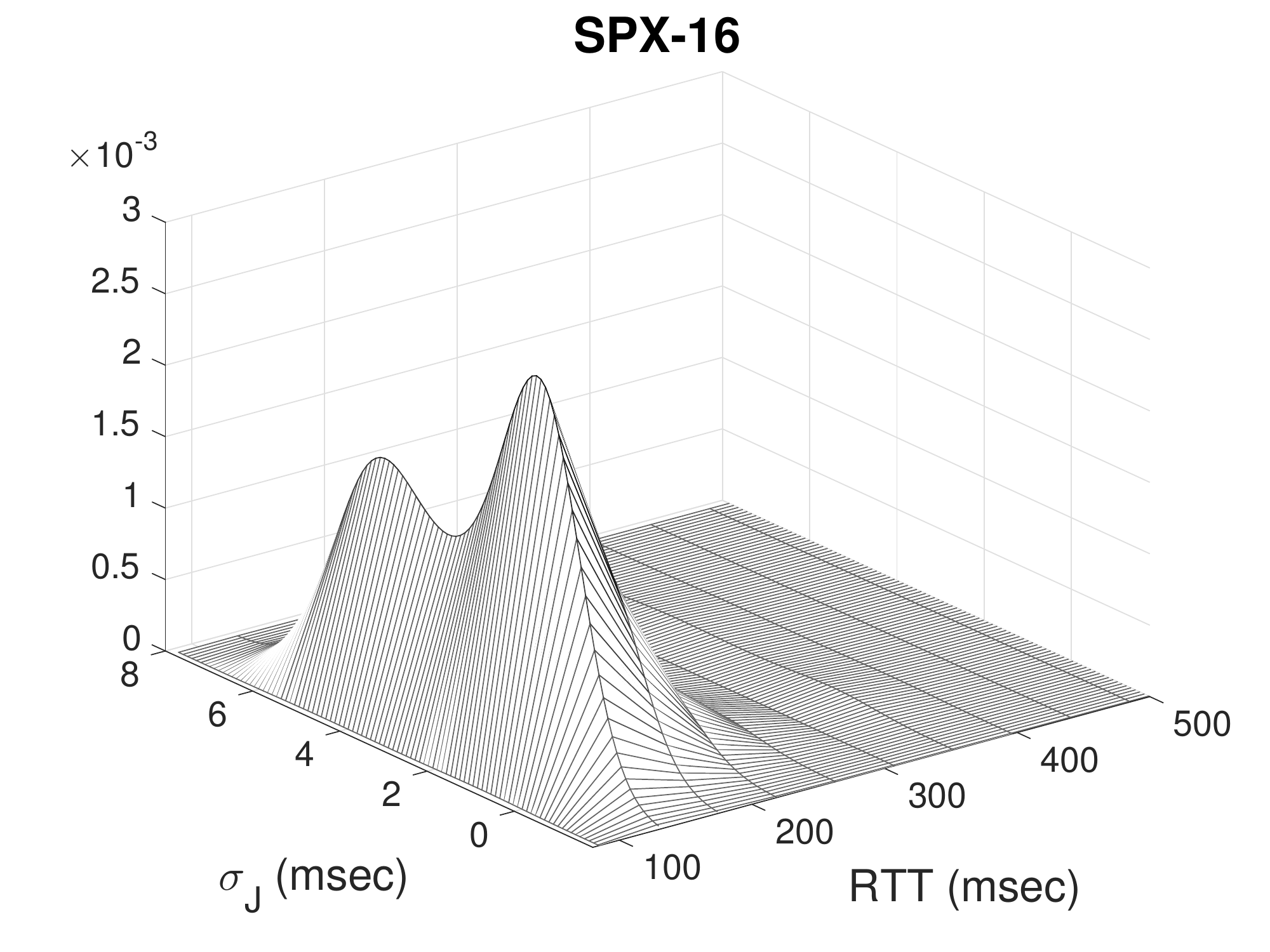}
	\end{minipage}		
	\caption{Bivariate distribution of RTT (msec) and jitter standard deviation $\sigma_J$ (msec) for various VoIP sessions with different codecs.}
	\label{fig:rttdevjit}
\end{figure*}
A second set of performance results is expressed in terms of bivariate distributions of RTT with respect to $\sigma_J$, as shown in Fig. \ref{fig:rttdevjit}. RTT can be approximately considered as the one-way latency in the forward and backward directions of a communication, so its value is expected to be less than $300$ msec. Figure \ref{fig:rttdevjit} shows that, for almost all cases, the peak of RTT lies between $100$ and $150$ msec, in correspondence of a $\sigma_J$ value lying between $0$ and $3$ msec. A slight exception refers to SPX-16 sessions where $\sigma_J$ reaches the value of $6$ msec, probably due to particular conditions of the mobile environment that occurred during the experiment.

\begin{figure*}[t!]
	\centering
	\begin{minipage}[t]{0.26\textwidth}
		\includegraphics[width=\textwidth]{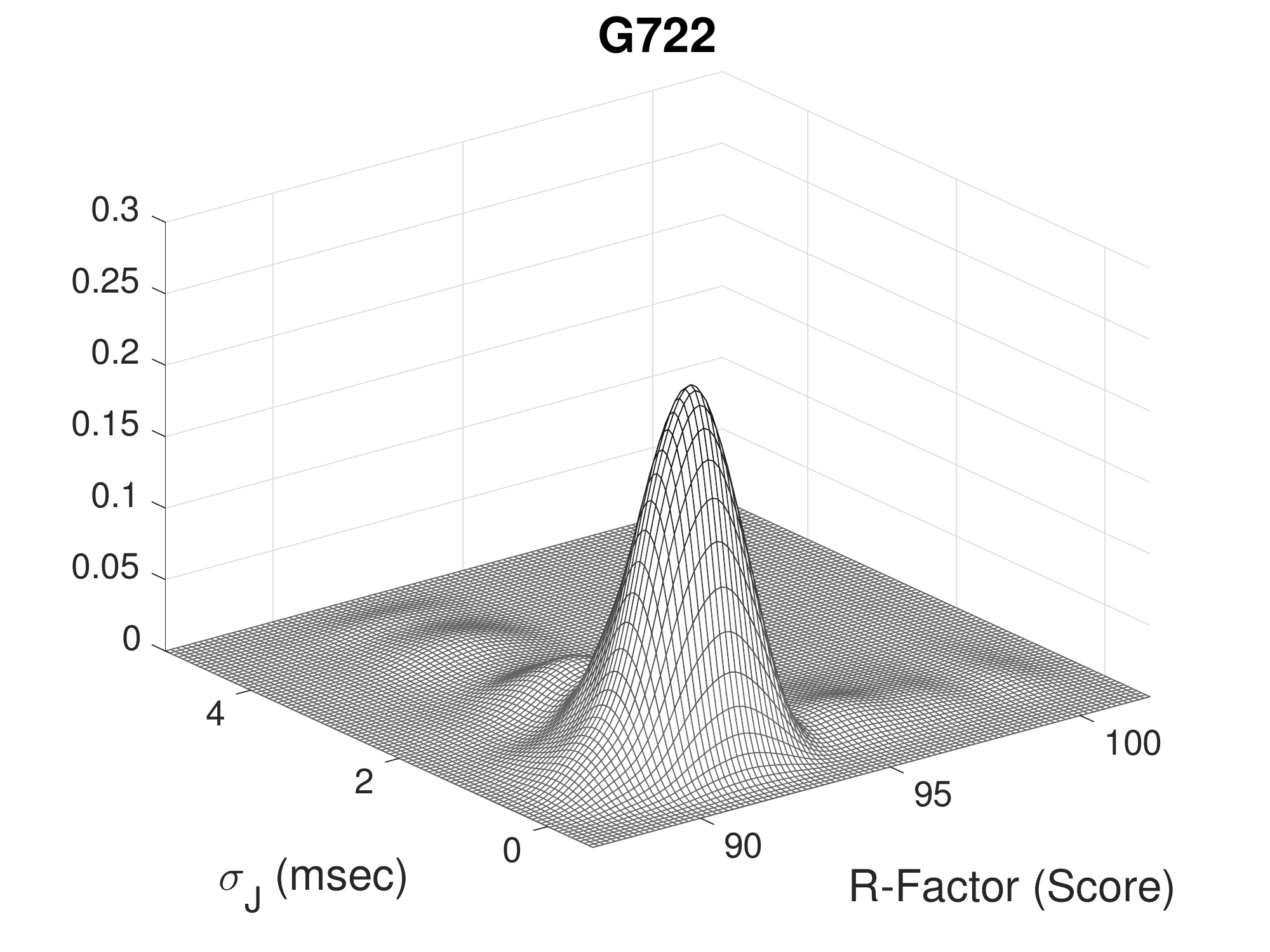}
	\end{minipage} \hspace{2pt}
	\begin{minipage}[t]{0.26\textwidth}
		\includegraphics[width=\textwidth]{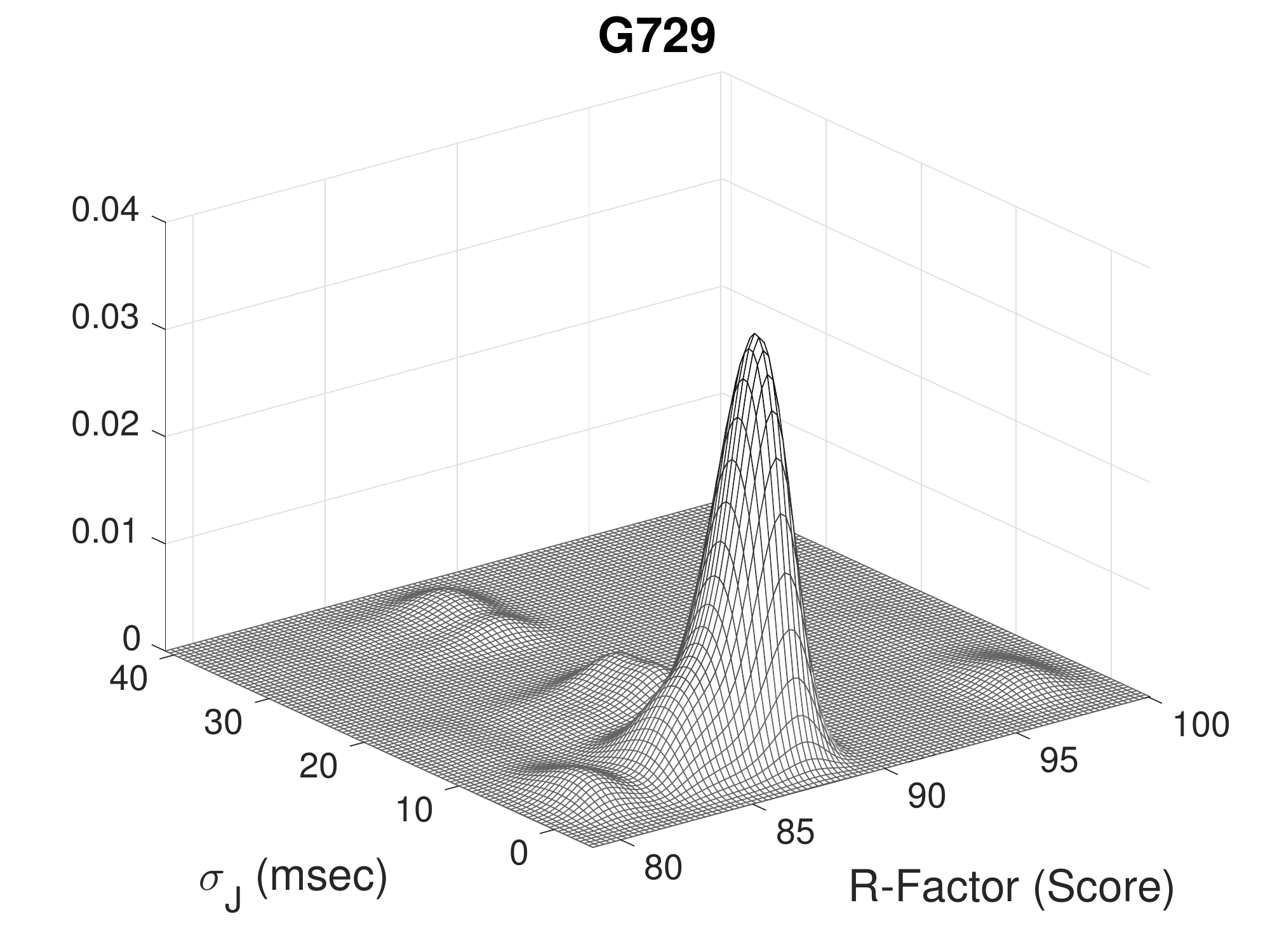}
	\end{minipage} \hspace{2pt}
	\begin{minipage}[t]{0.26\textwidth}
		\includegraphics[width=\textwidth]{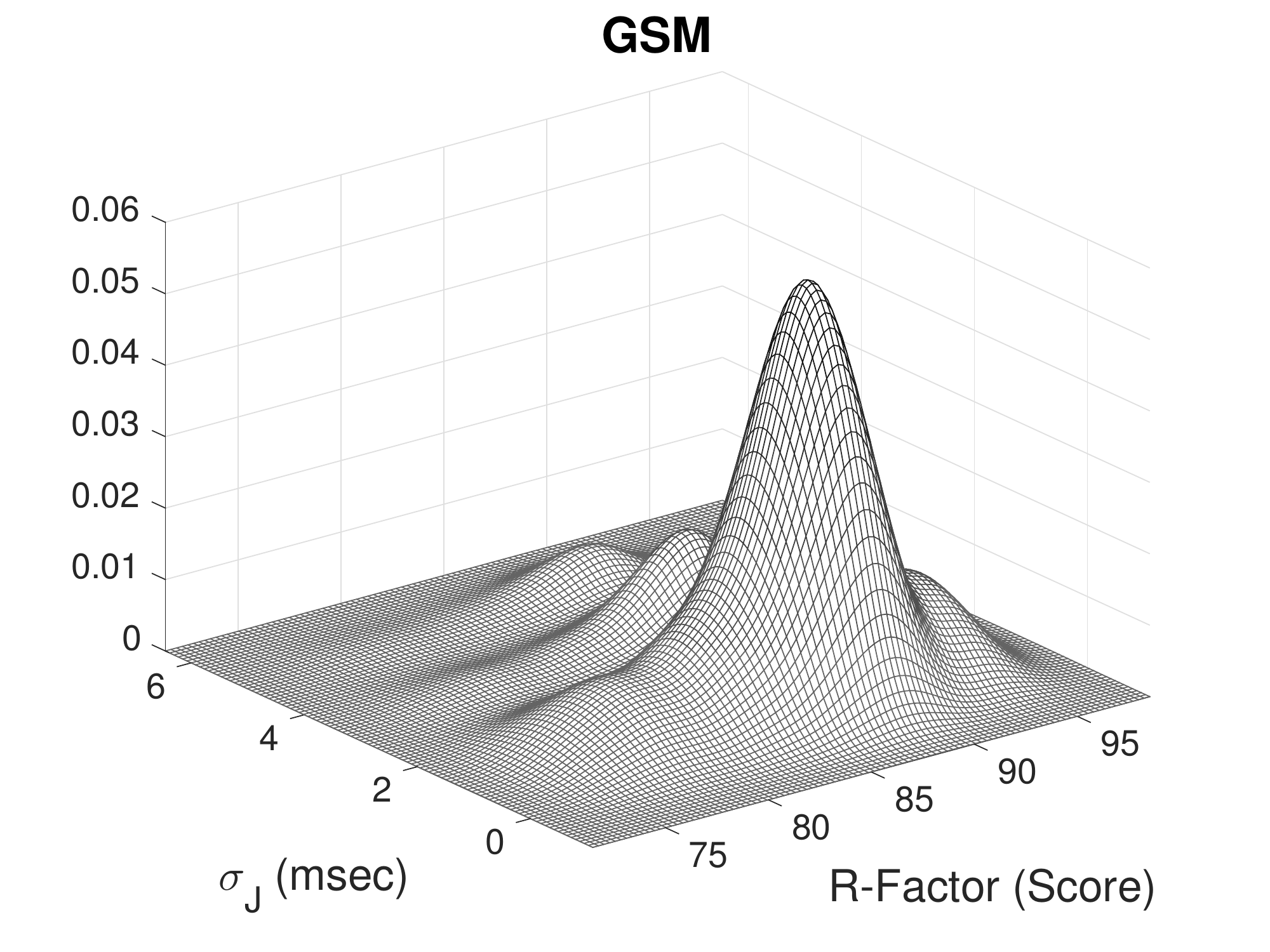}
	\end{minipage}	\hspace{2pt}
	\begin{minipage}[t]{0.26\textwidth}
		\includegraphics[width=\textwidth]{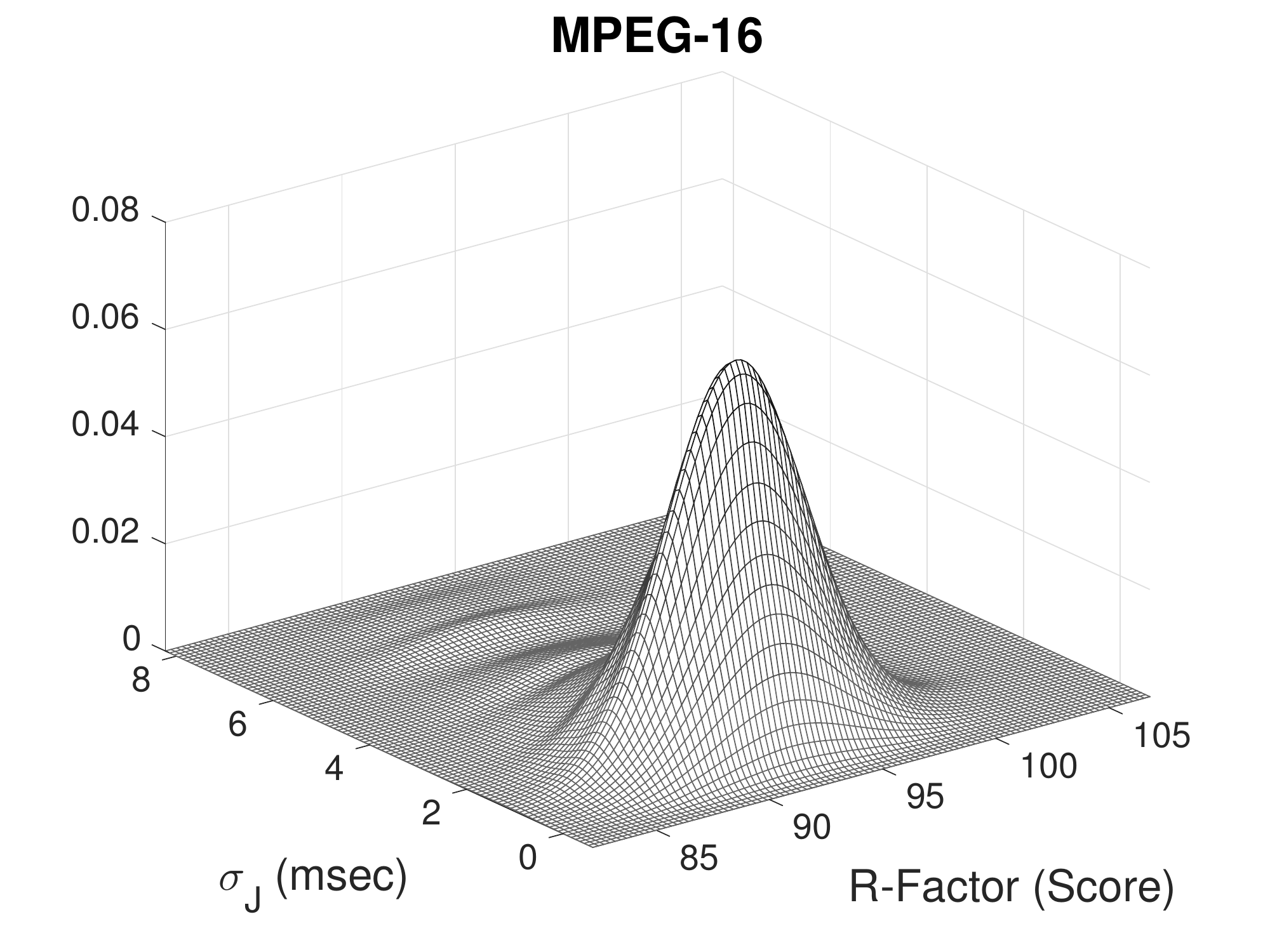}
	\end{minipage}	\hspace{2pt}
	\begin{minipage}[t]{0.26\textwidth}
		\includegraphics[width=\textwidth]{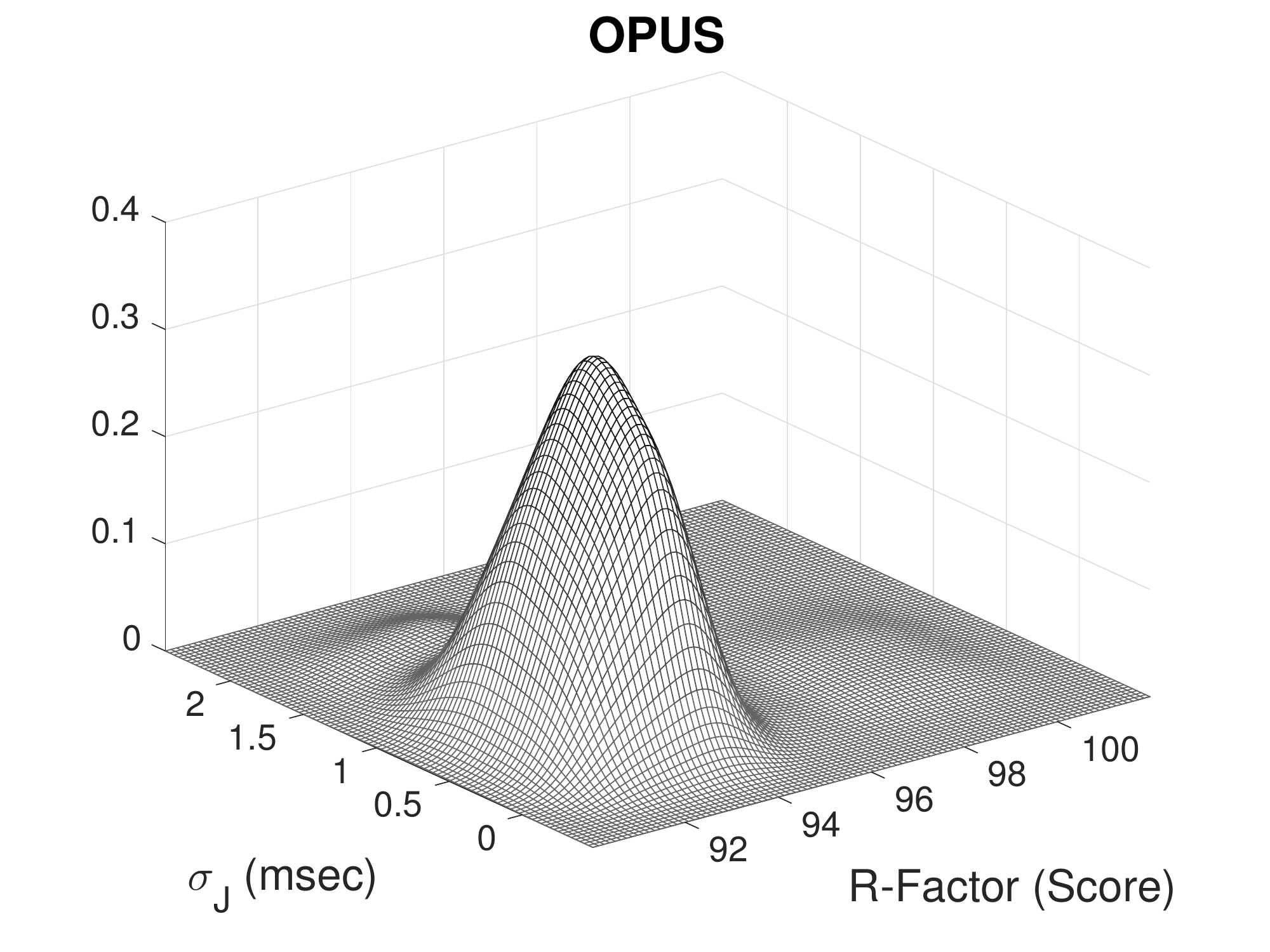}
	\end{minipage}	\hspace{2pt}	
	\begin{minipage}[t]{0.26\textwidth}
		\includegraphics[width=\textwidth]{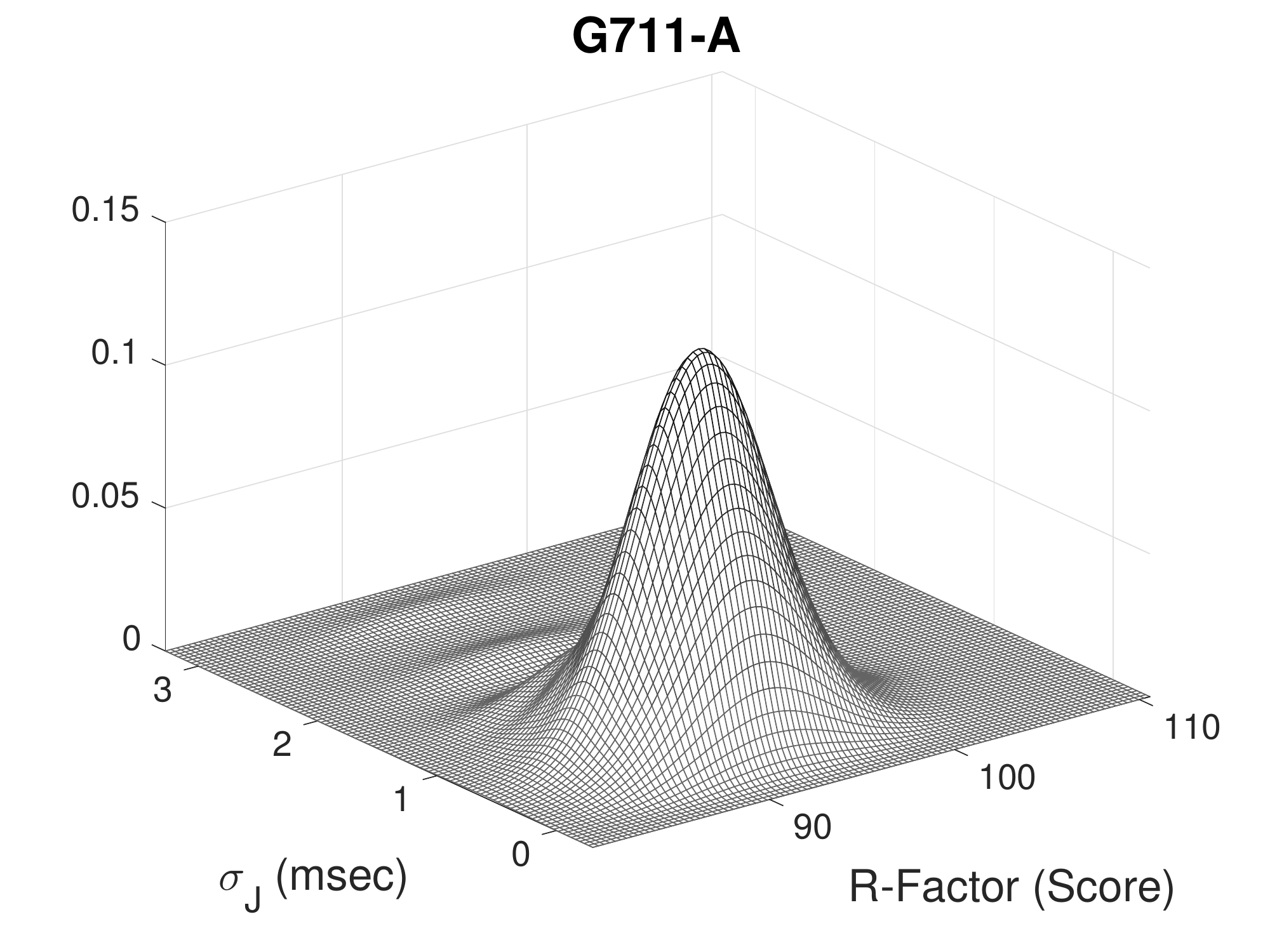}
	\end{minipage}	\hspace{2pt}
	\begin{minipage}[t]{0.26\textwidth}
		\includegraphics[width=\textwidth]{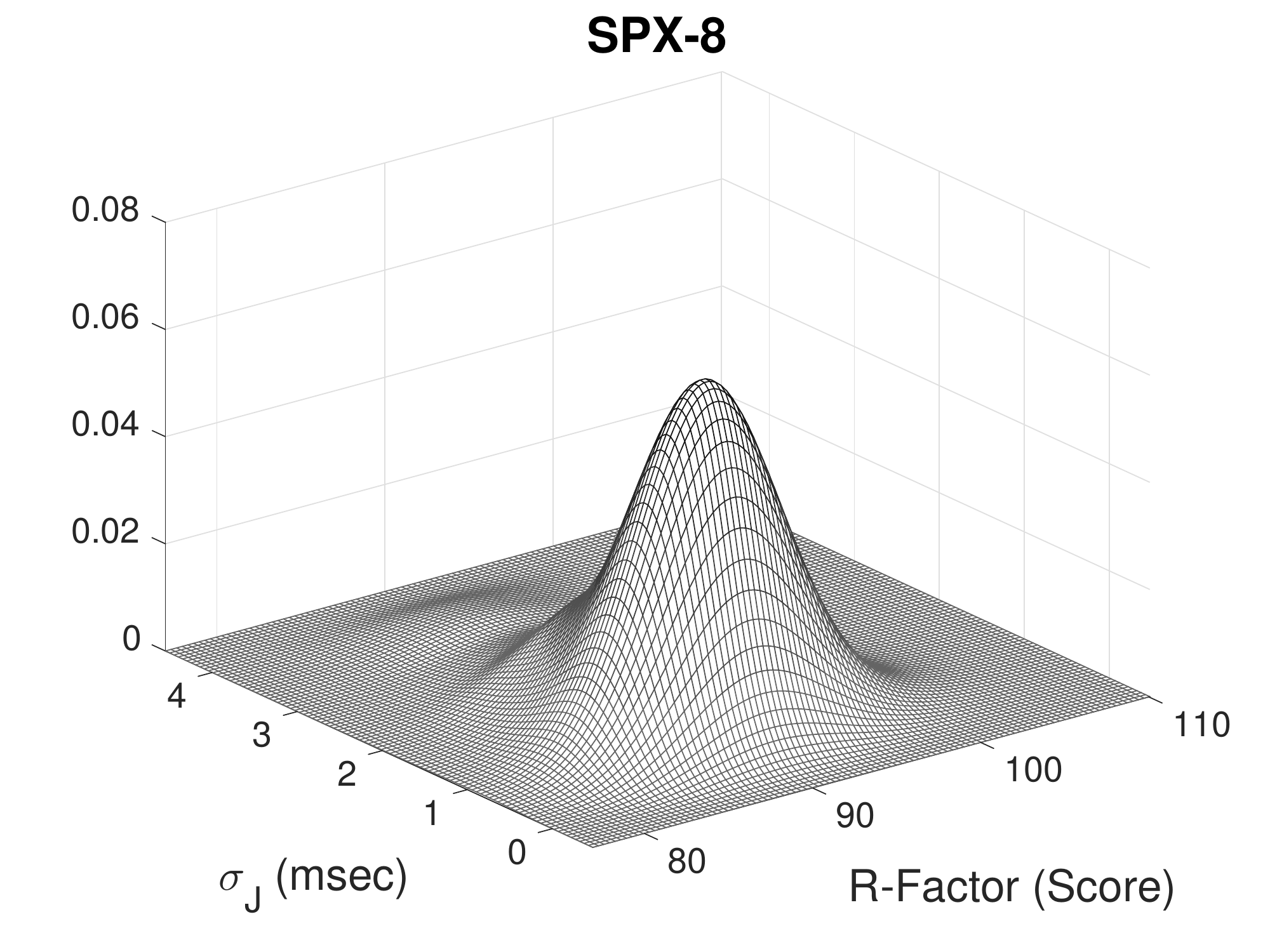}
	\end{minipage}		\hspace{2pt}
	\begin{minipage}[t]{0.26\textwidth}
		\includegraphics[width=\textwidth]{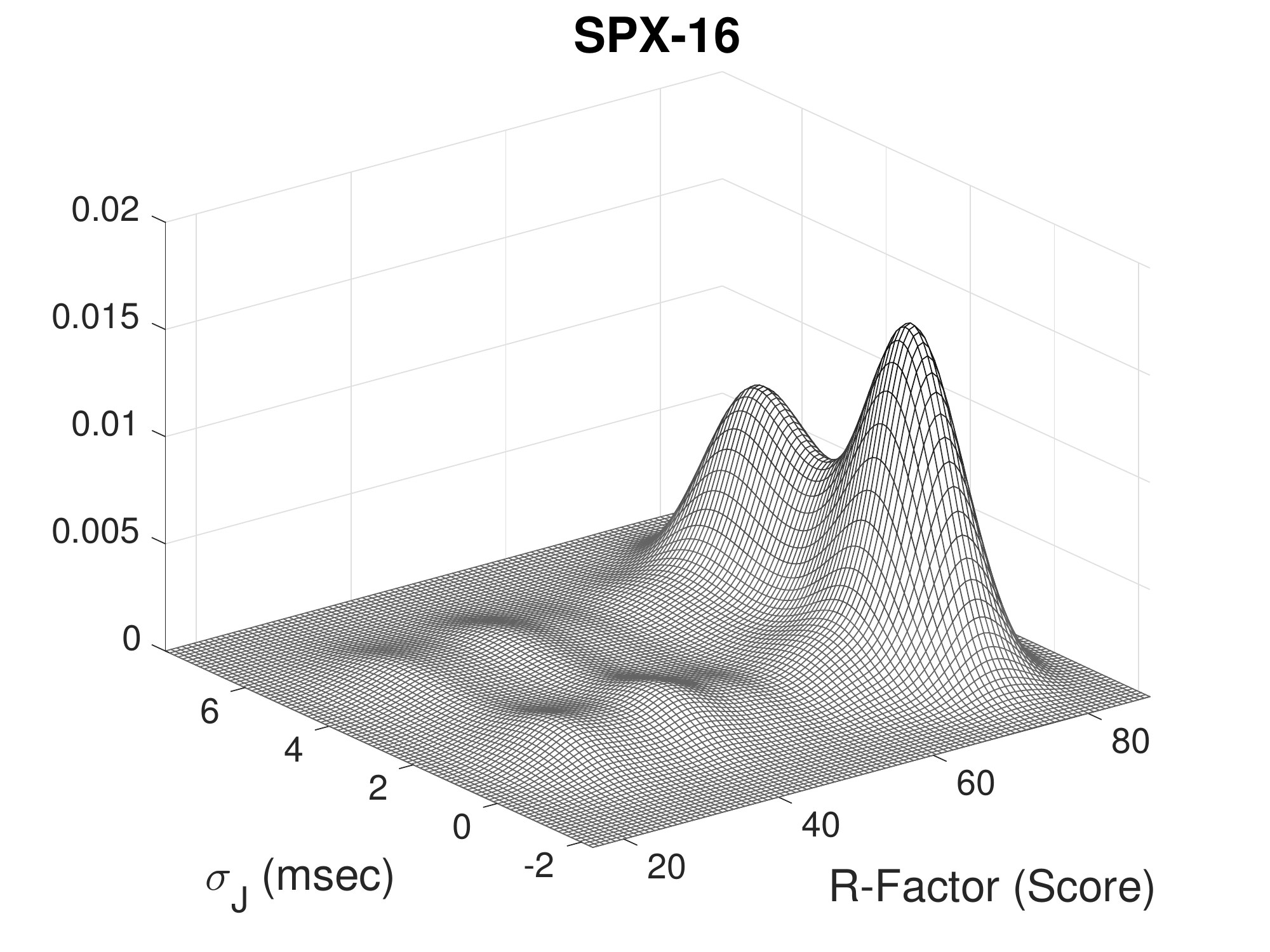}
	\end{minipage}		
	\caption{Bivariate distribution of R-factor (score) and jitter standard deviation $\sigma_J$ (msec) for various VoIP sessions with different codecs.}
	\label{fig:rfacdevjit}
\end{figure*}
\begin{figure*}[t!] 
	\centering
	\begin{tabular}{cccc}
		\subfloat{\includegraphics[scale=0.25]{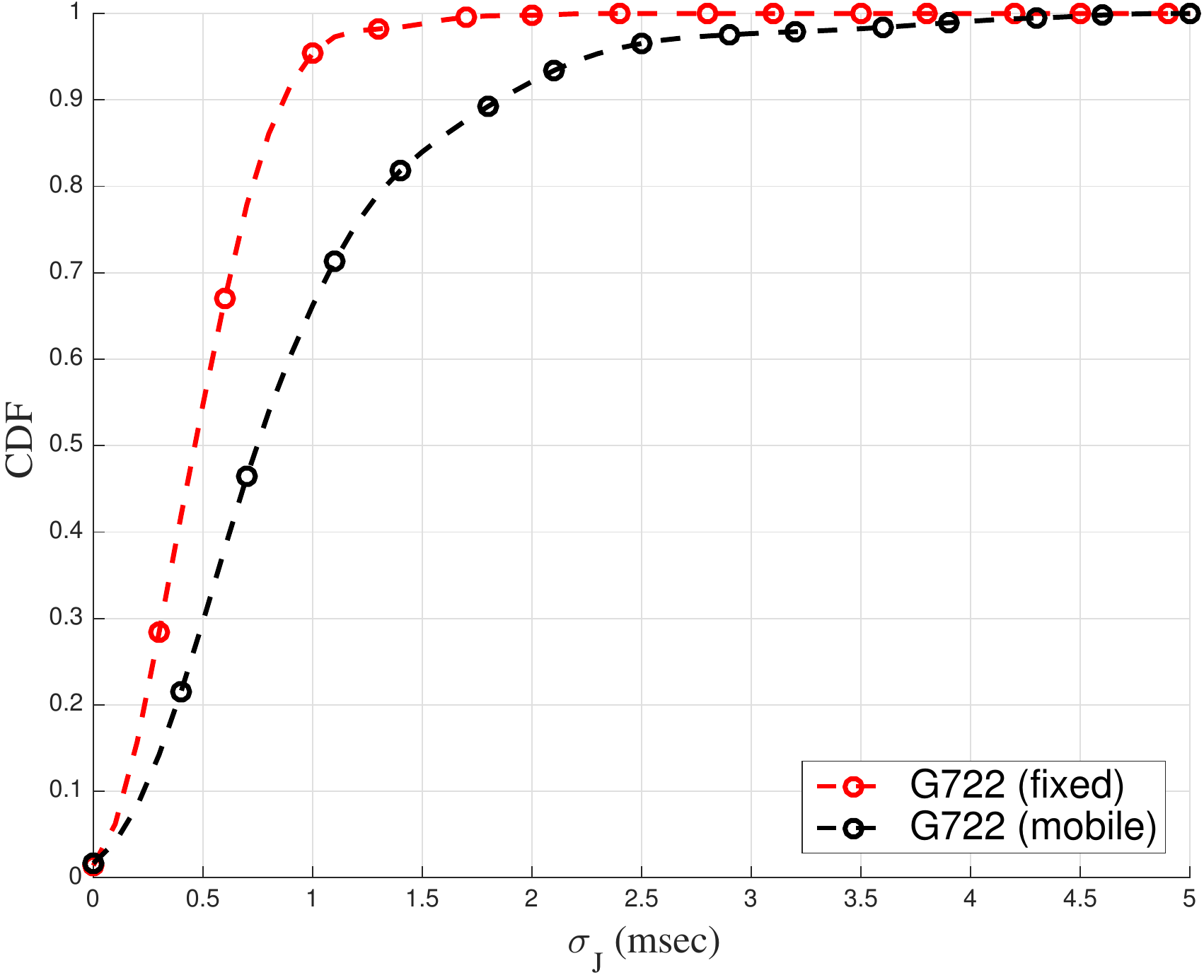}}  \hspace{2mm}
		\subfloat{\includegraphics[scale=0.25]{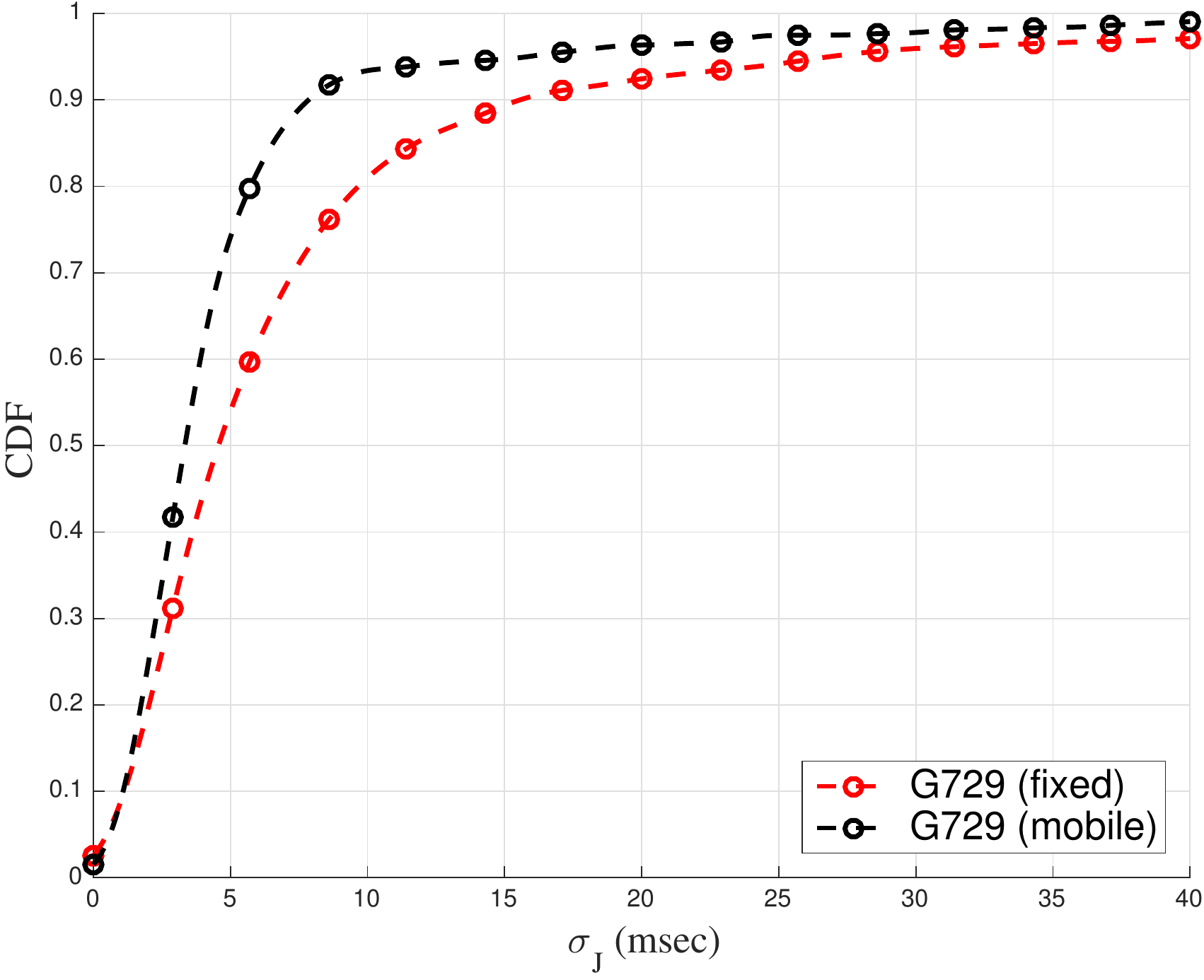}}  \hspace{2mm}
		\subfloat{\includegraphics[scale=0.25]{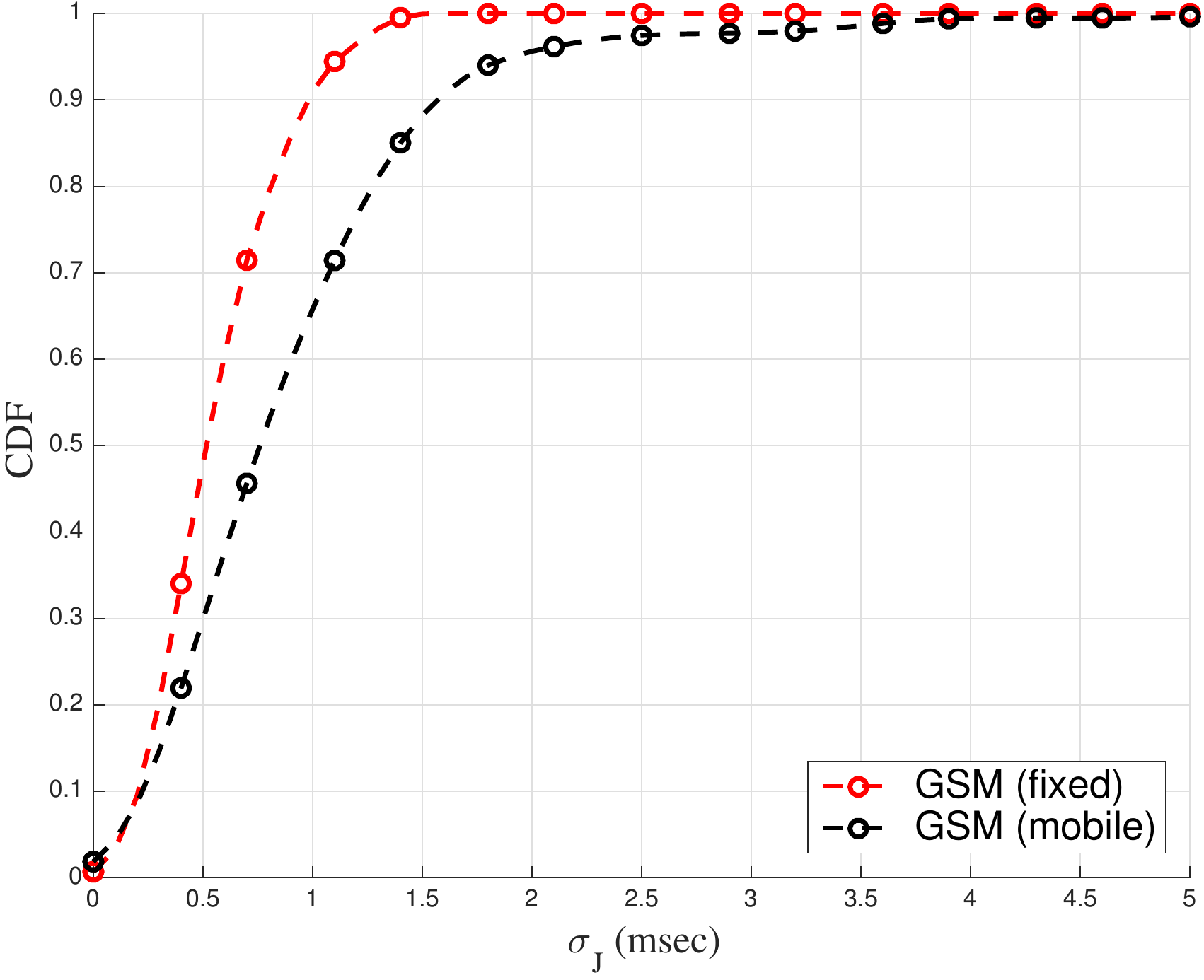}}  \hspace{2mm}
		\subfloat{\includegraphics[scale=0.25]{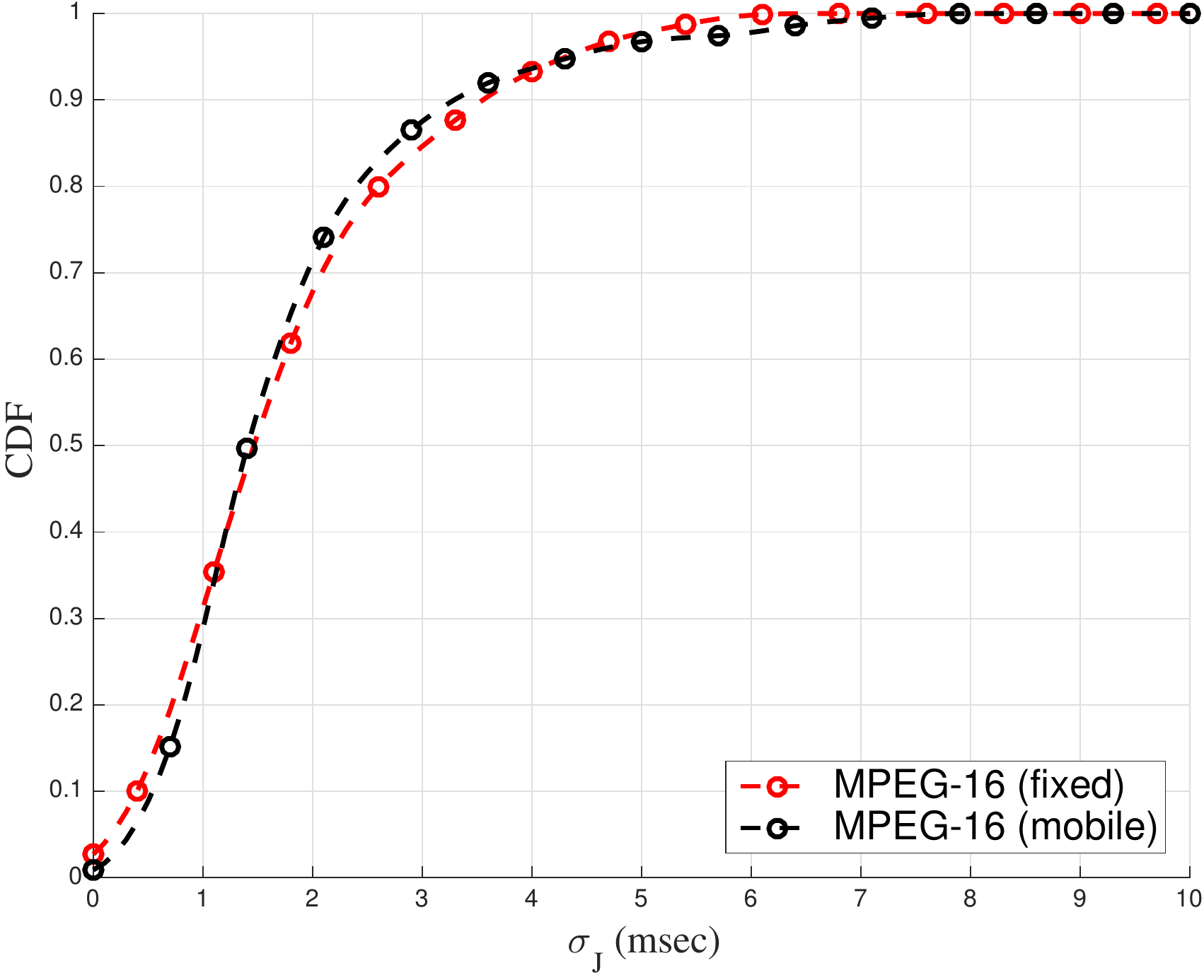}} \\
		\subfloat{\includegraphics[scale=0.25]{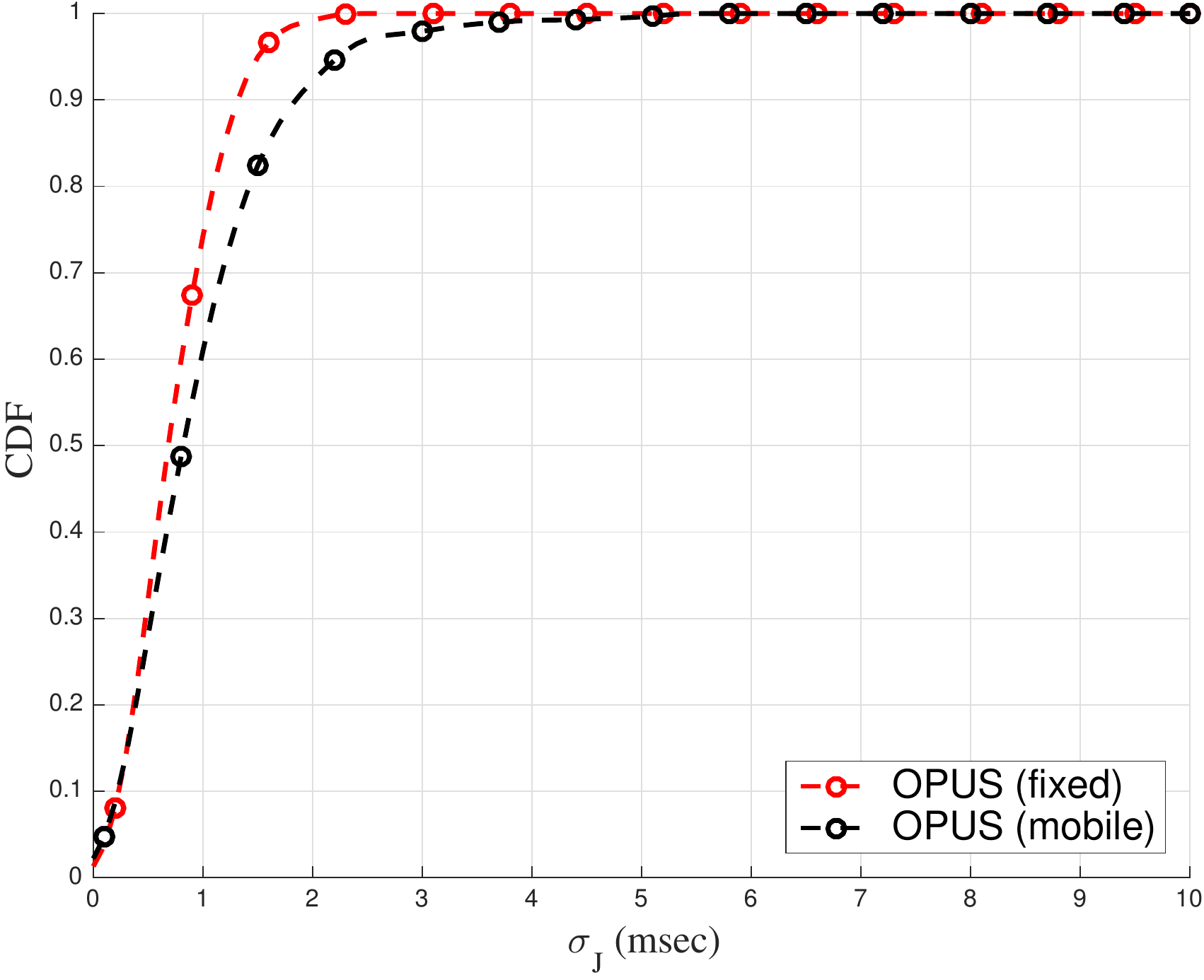}}  \hspace{2mm}
		\subfloat{\includegraphics[scale=0.25]{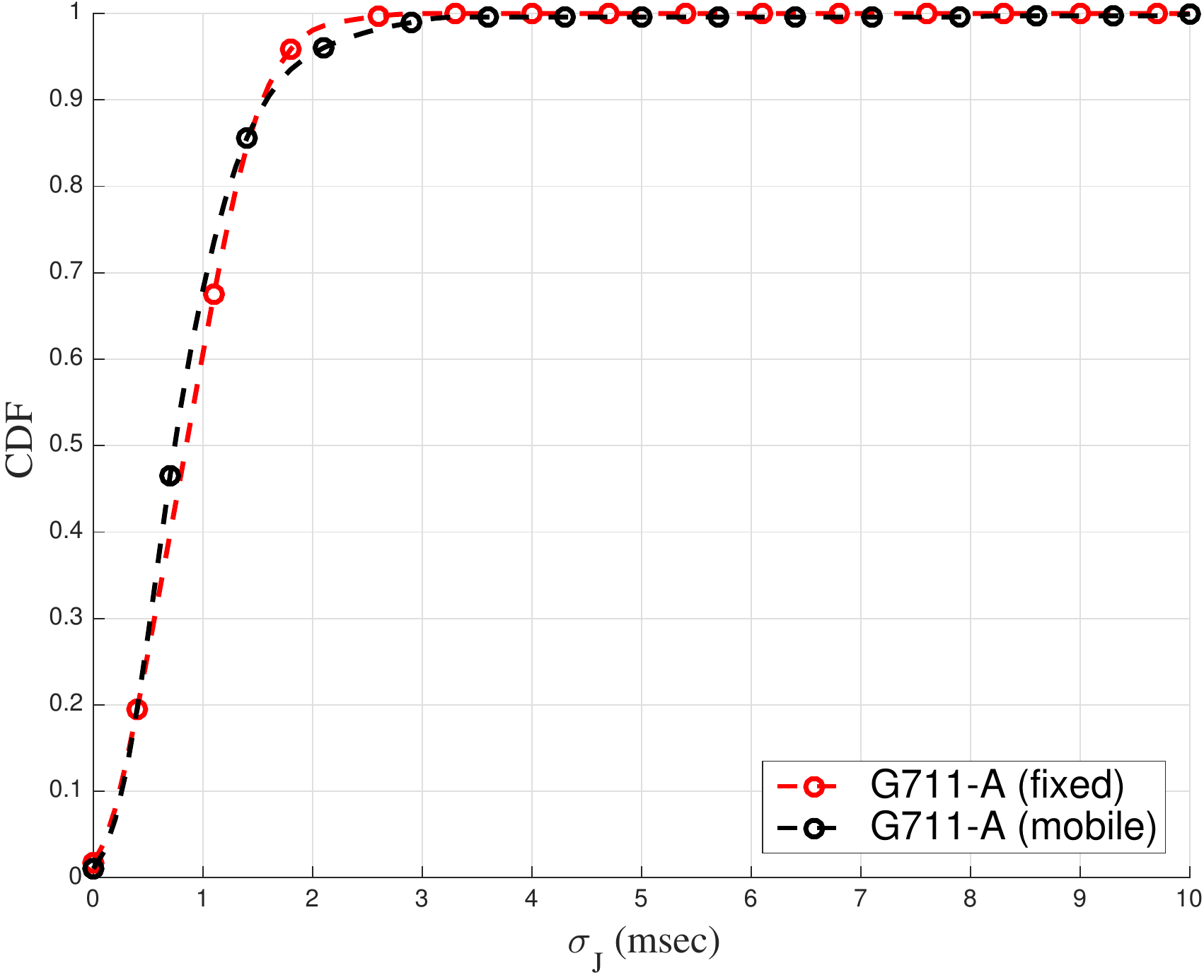}} \hspace{2mm}
		\subfloat{\includegraphics[scale=0.25]{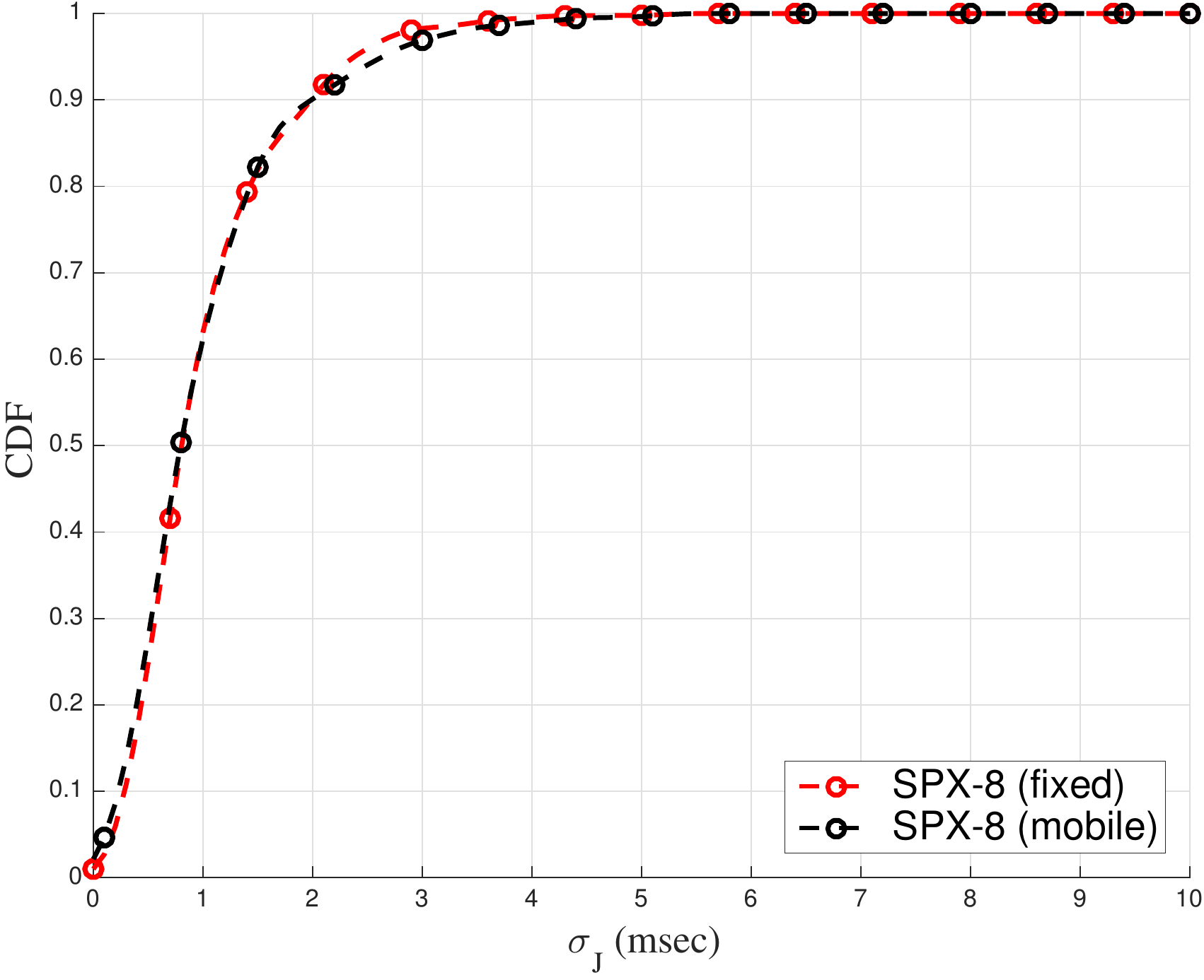}} \hspace{2mm}
		\subfloat{\includegraphics[scale=0.25]{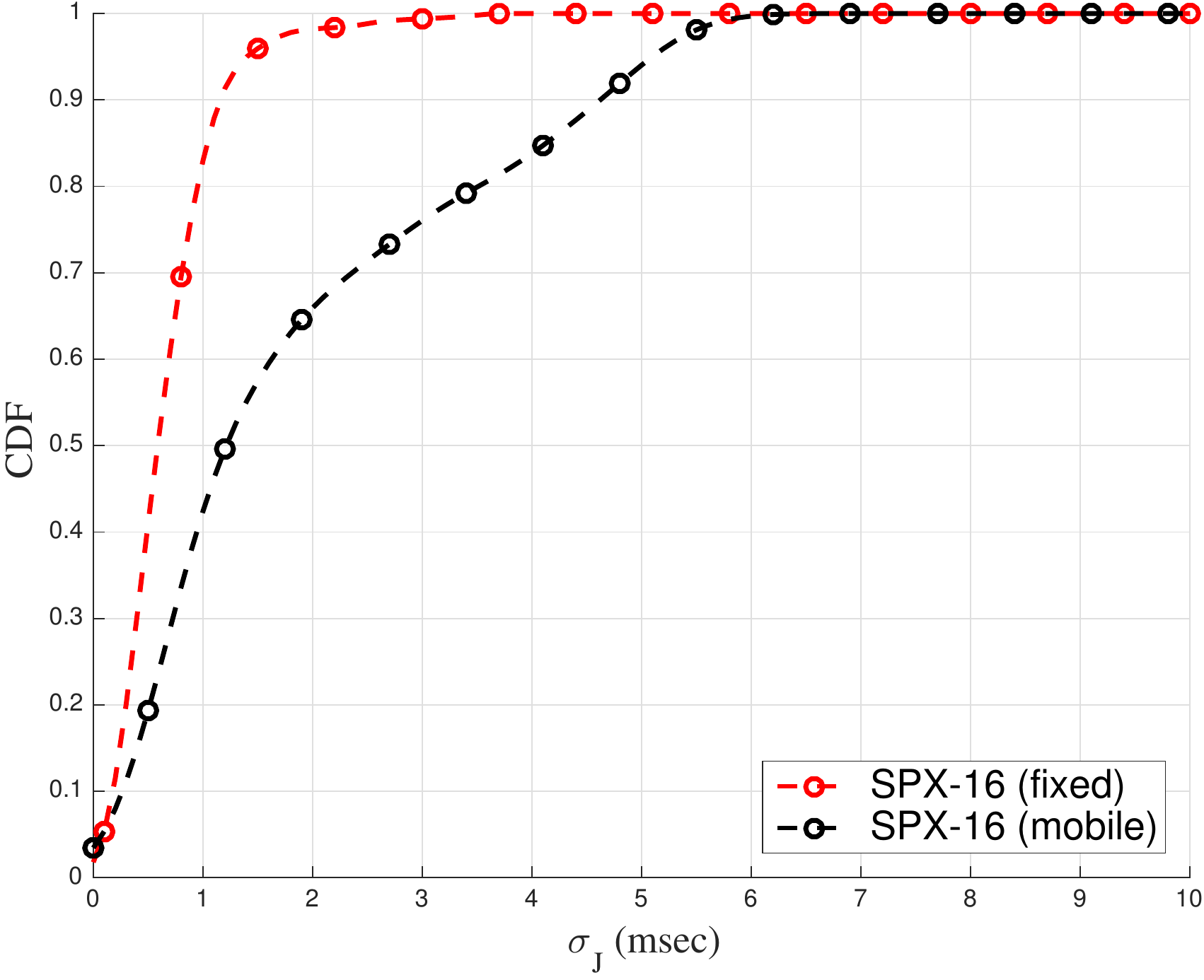}} 
	\end{tabular}
	\caption{Cumulative Distribution Functions (CDFs) of jitter standard deviation $\sigma_{J}$ (msec) in two cases: fixed scenario (red curve) and mobile scenario (black curve) for various VoIP sessions with different codecs.}
	\label{fig:cdfdevjit}
\end{figure*}

A third set of results concerns the analysis of relationships between R-Factor and $\sigma_J$, as shown in Fig. \ref{fig:rfacdevjit}. It is useful to recall that the R-Factor is a voice quality metric expressed as an integer in the interval $[0,100]$ with a value of $94$ indicating the best communication quality, whereas values below $50$ are typically considered unacceptable. Such a metric is directly connected to the Mean Opinion Score (MOS), but, it is considered more precise due to its finer granularity. 
For almost all cases, the peak of the R-Factor score distribution is around $90$ in correspondence of small values of $\sigma_J$ (around $3$ msec). An exception is given for the case of codec SPX-16 where the R-Factor reaches its maximum value at $80$. 

This behavior can be explained by invoking the dependence of MOS (and then of the R-Factor) from some time-variant phenomena such as the Peak Signal-to-Noise Ratio (PSRN), value that is directly related to the temporary quality of channel. In this case, a transitory degradation of channel quality determined a decay of the R-Factor.

\begin{figure*}[t!] 
	\centering
	\begin{tabular}{cccc}
		\subfloat{\includegraphics[scale=0.25]{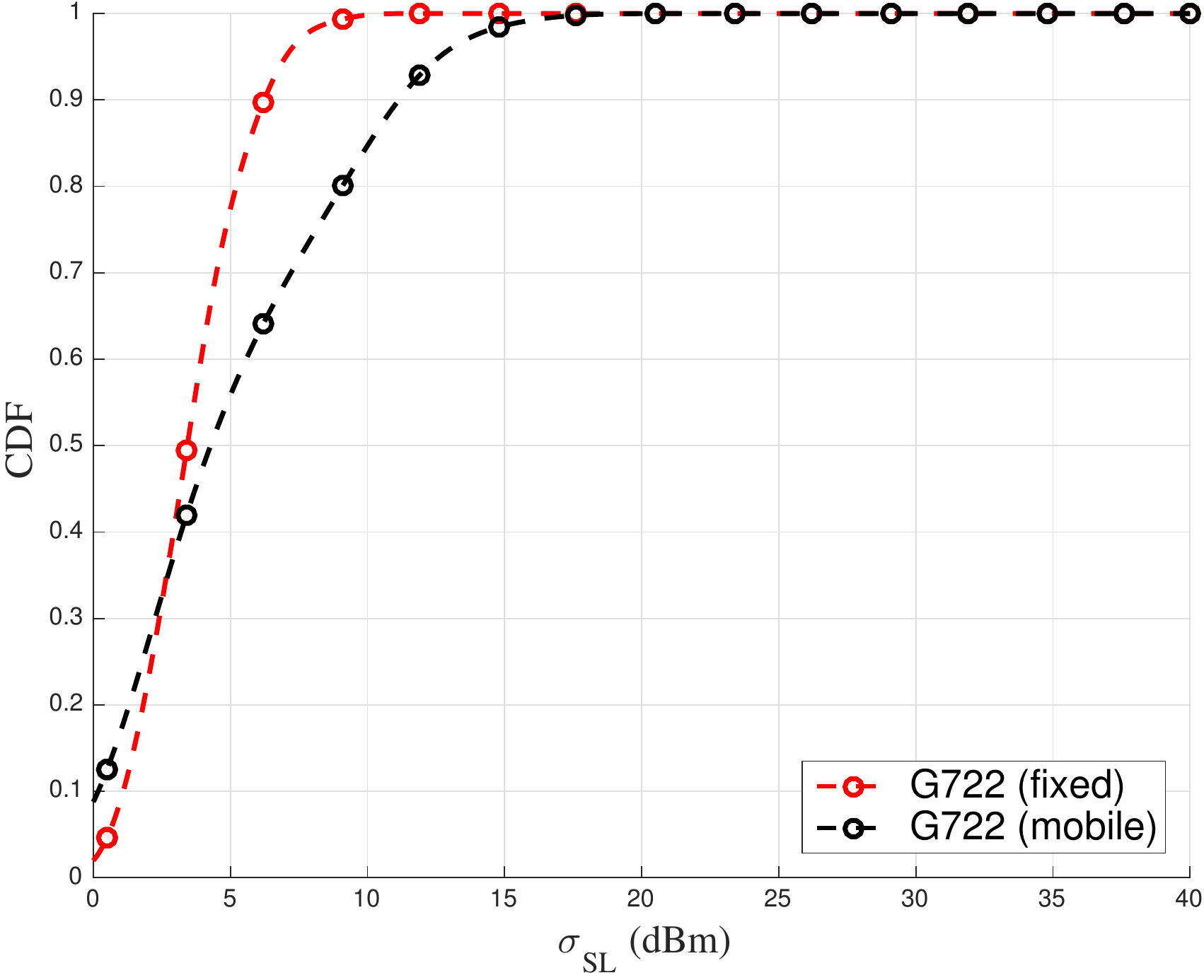}}  \hspace{2mm}
		\subfloat{\includegraphics[scale=0.25]{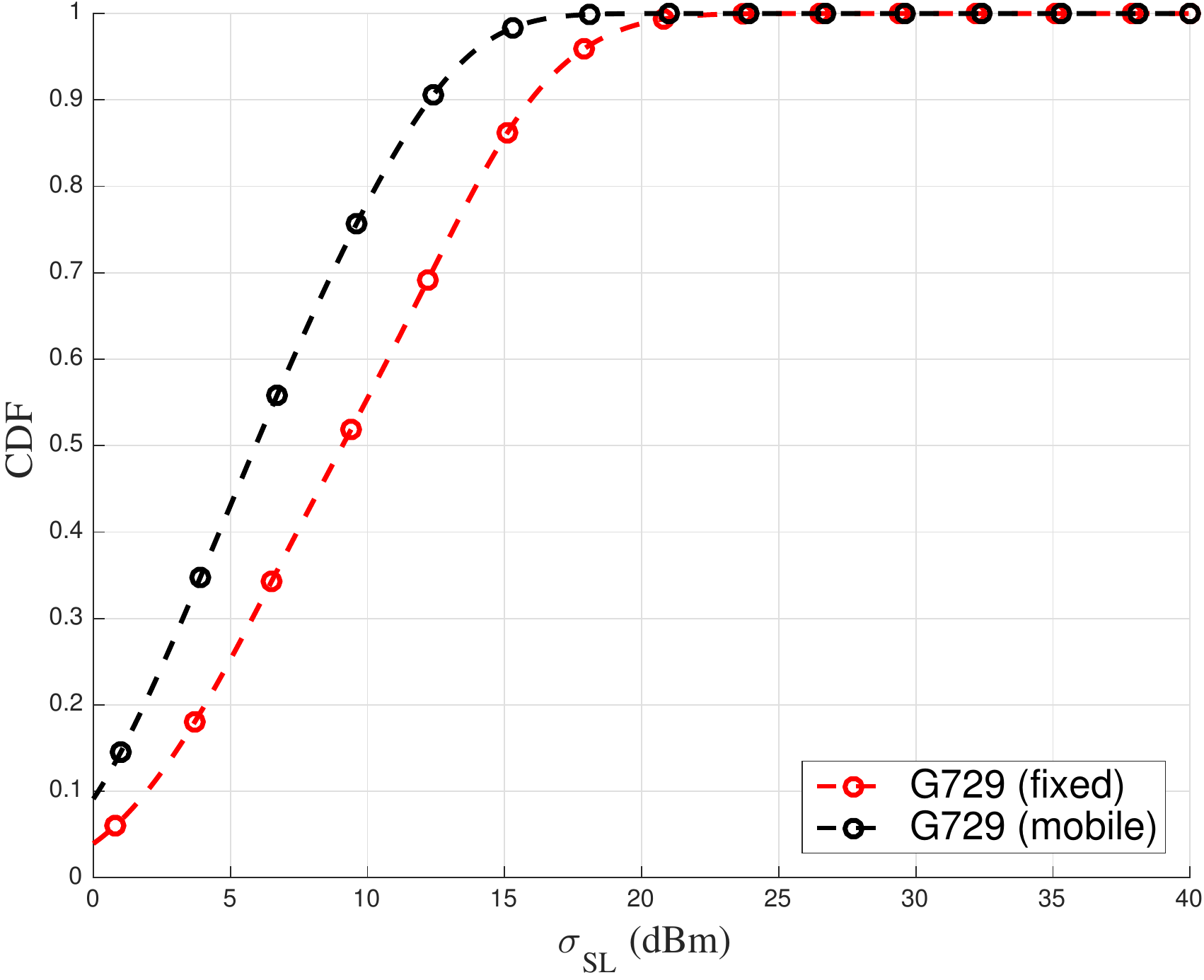}}  \hspace{2mm}
		\subfloat{\includegraphics[scale=0.25]{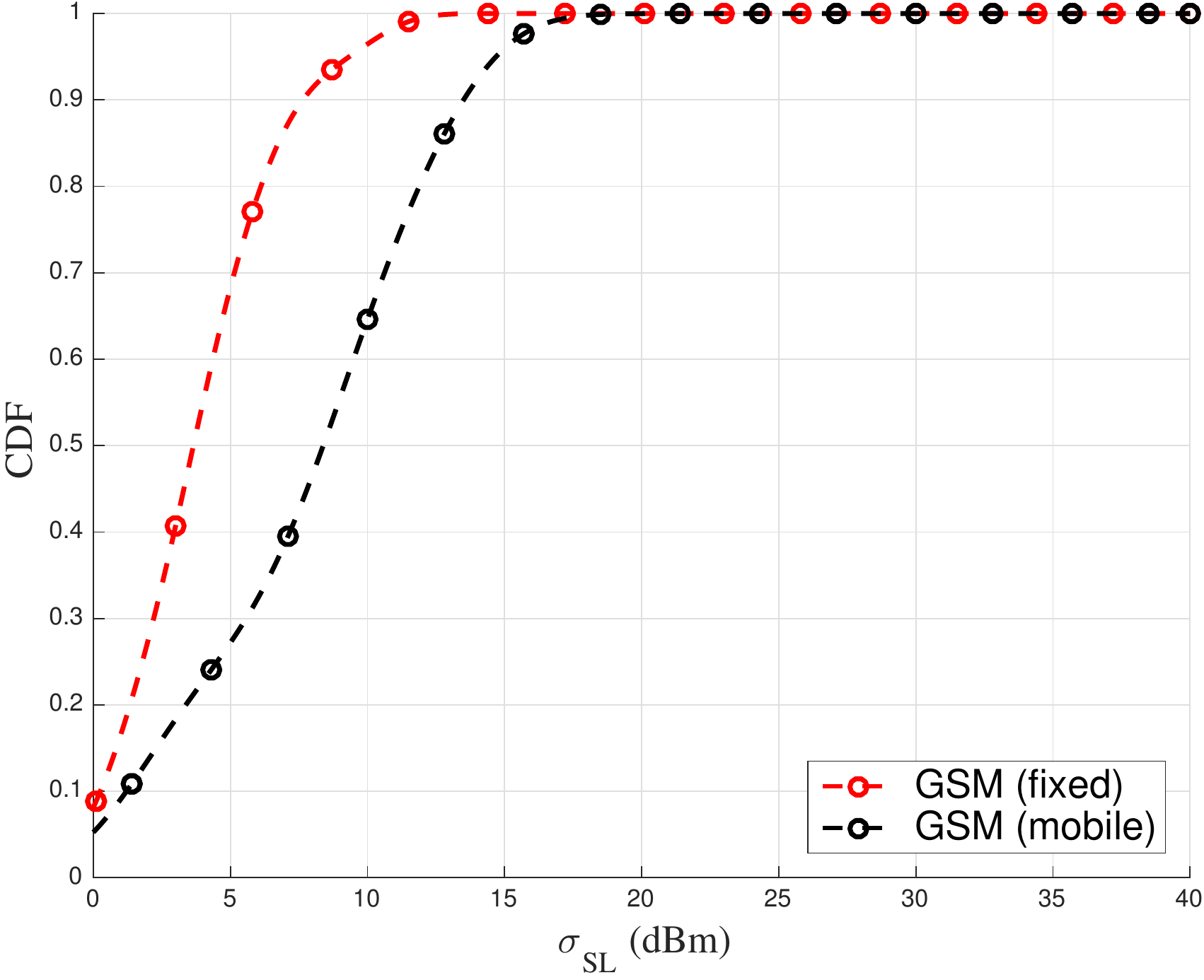}}  \hspace{2mm}
		\subfloat{\includegraphics[scale=0.25]{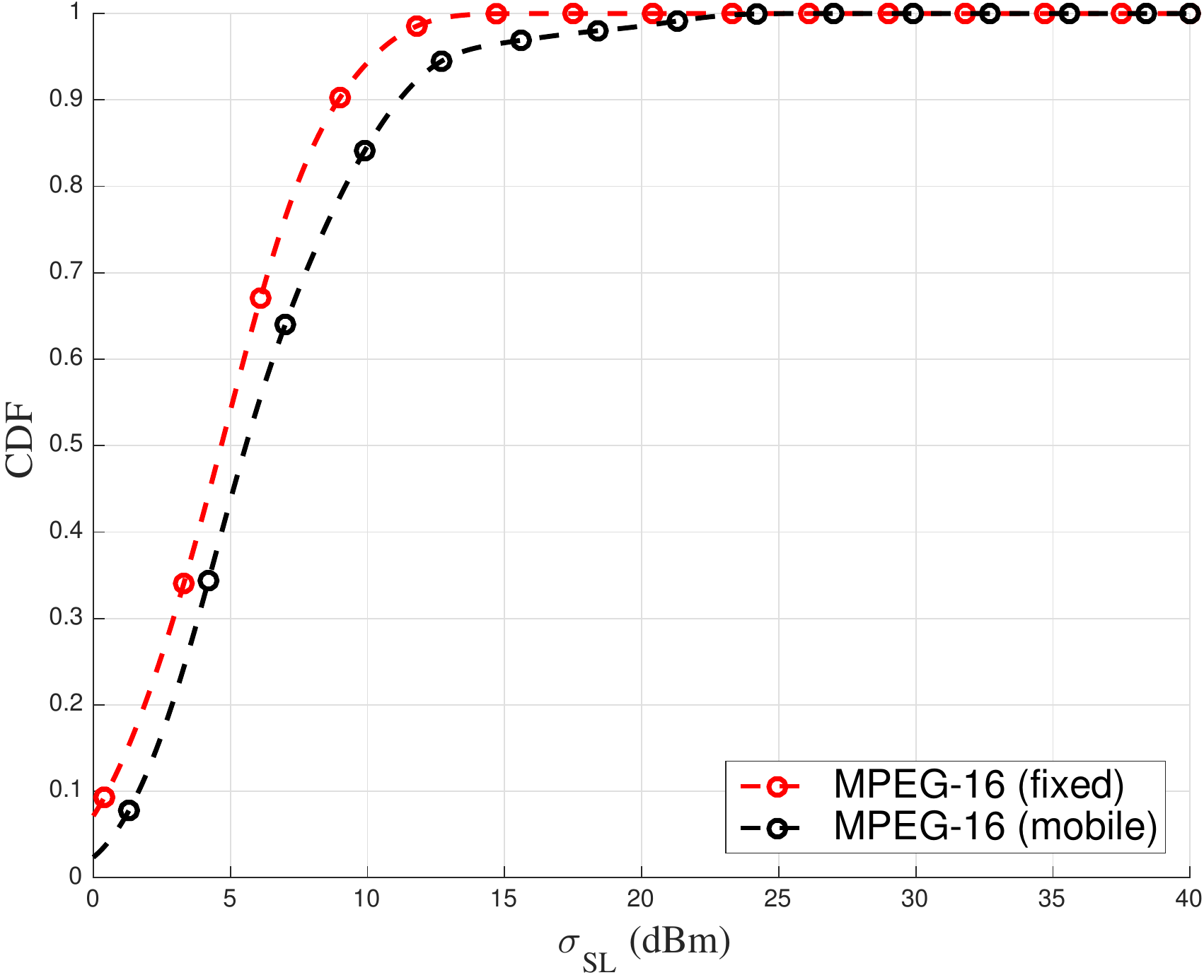}} \\
		\subfloat{\includegraphics[scale=0.25]{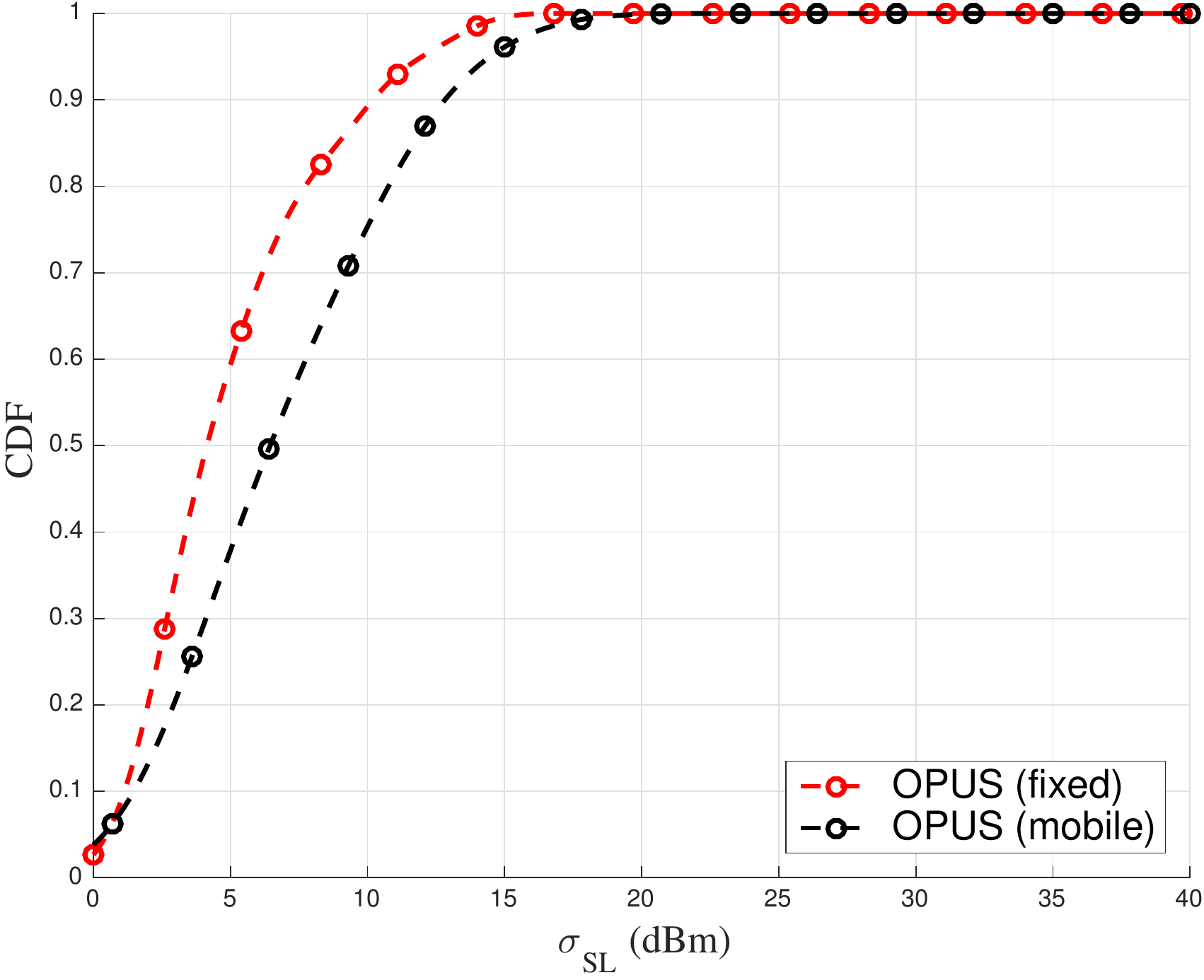}}  \hspace{2mm}
		\subfloat{\includegraphics[scale=0.25]{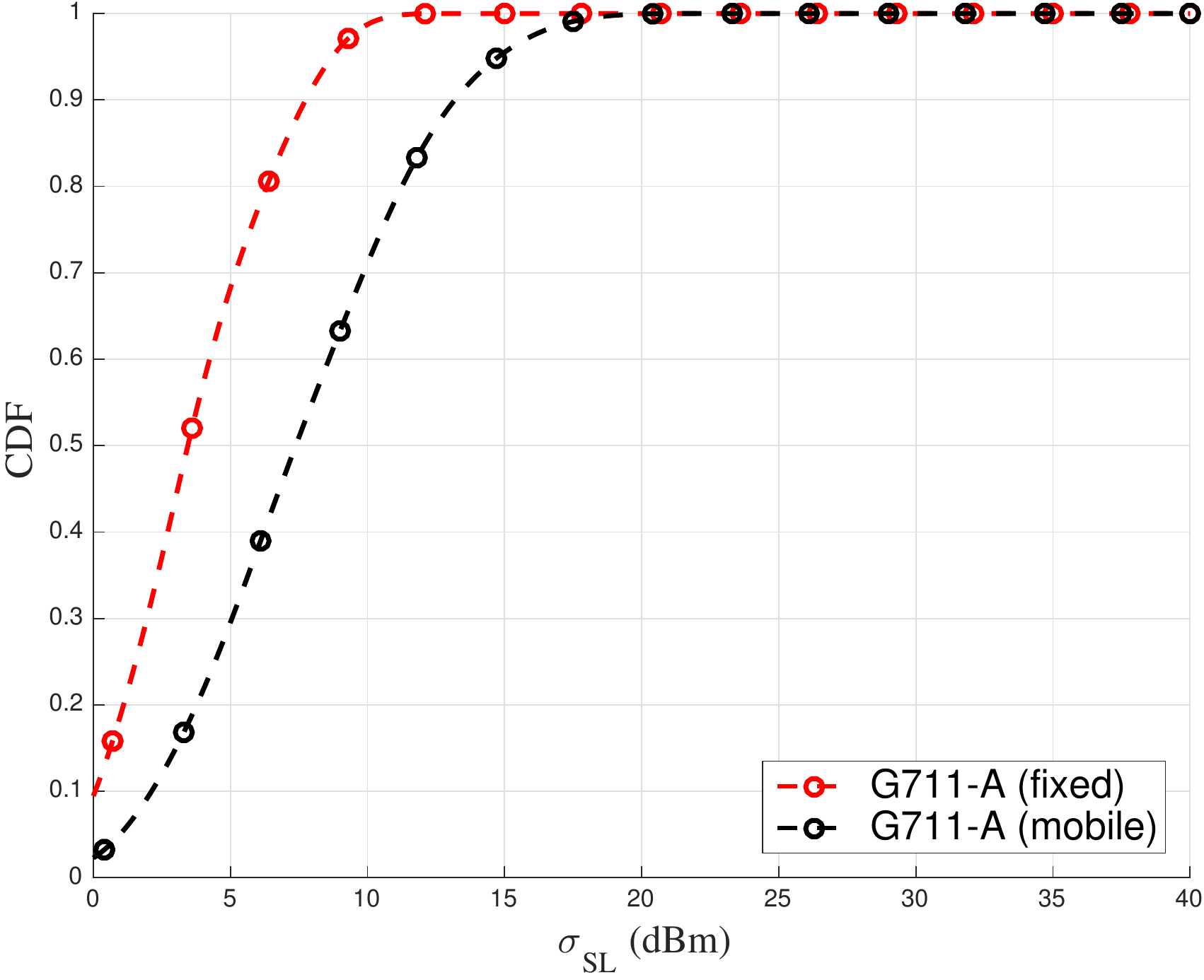}} \hspace{2mm}
		\subfloat{\includegraphics[scale=0.25]{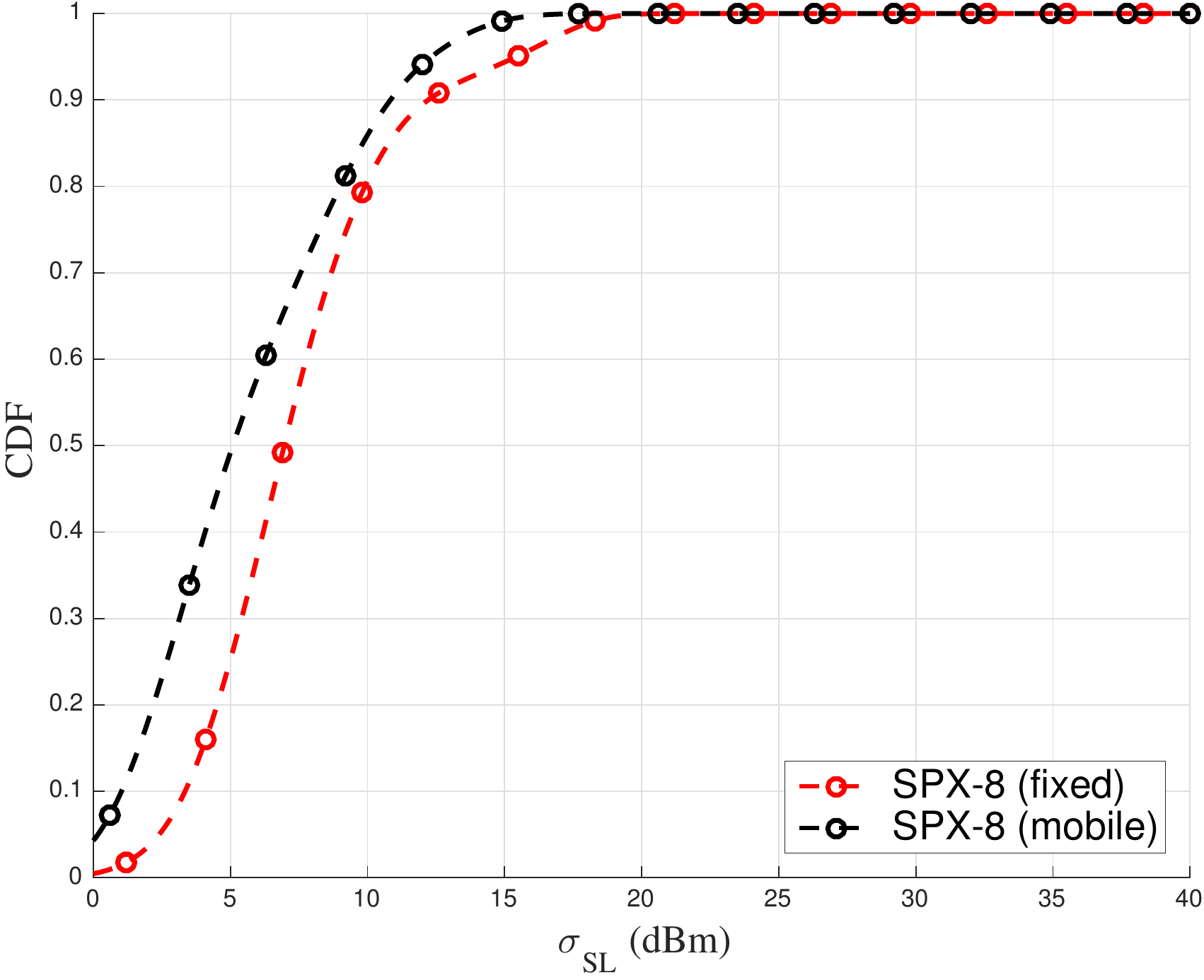}} \hspace{2mm}
		\subfloat{\includegraphics[scale=0.25]{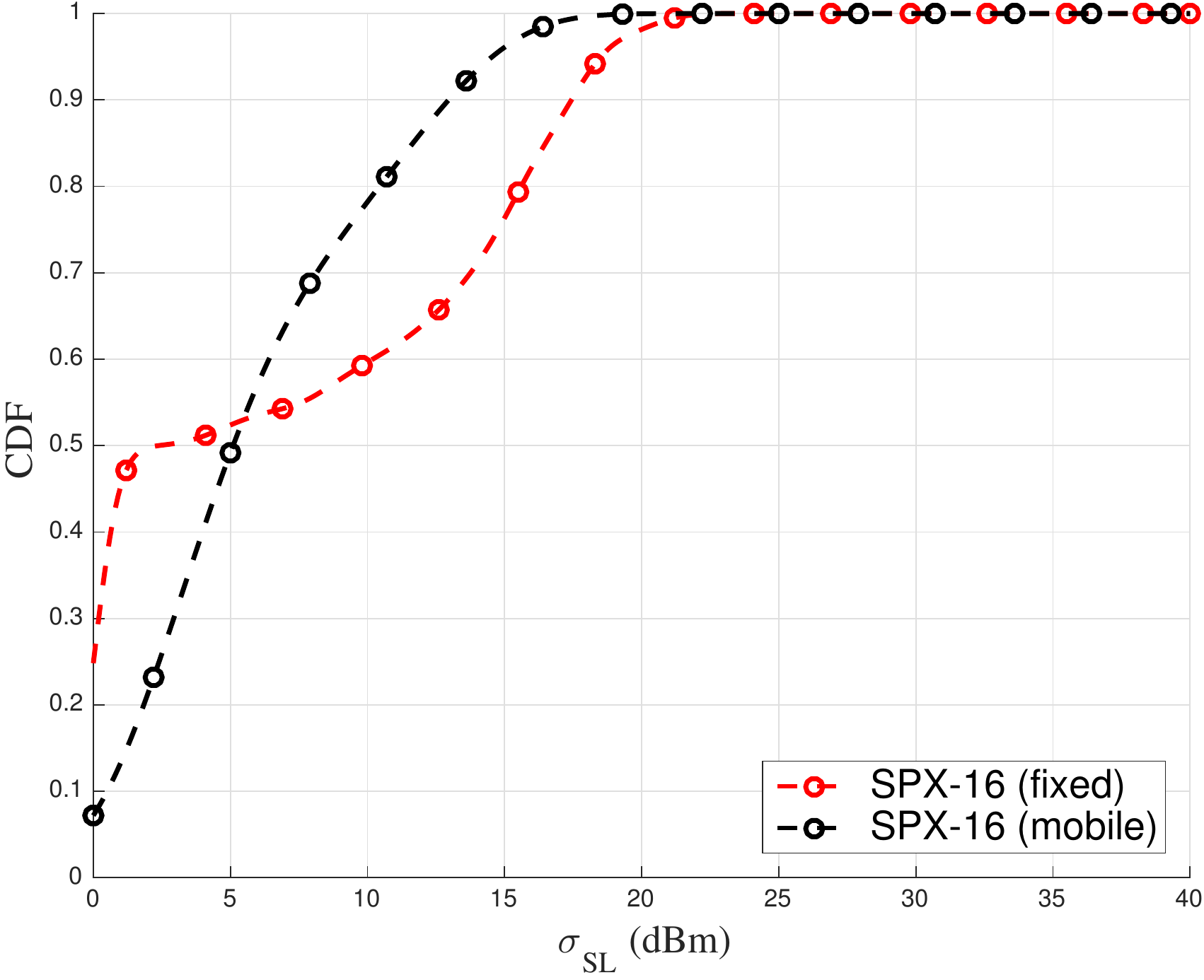}} 
	\end{tabular}
	\caption{Cumulative Distribution Functions (CDFs) of signal level standard deviation $\sigma_{SL}$ (dBm) in two cases: fixed scenario (red curve) and mobile scenario (black curve) for various VoIP sessions with different codecs.}
		\label{fig:cdfdevvoice}
\end{figure*}

\begin{figure*}[t!] 
	\centering
	\begin{tabular}{cccc}
		\subfloat{\includegraphics[scale=0.25]{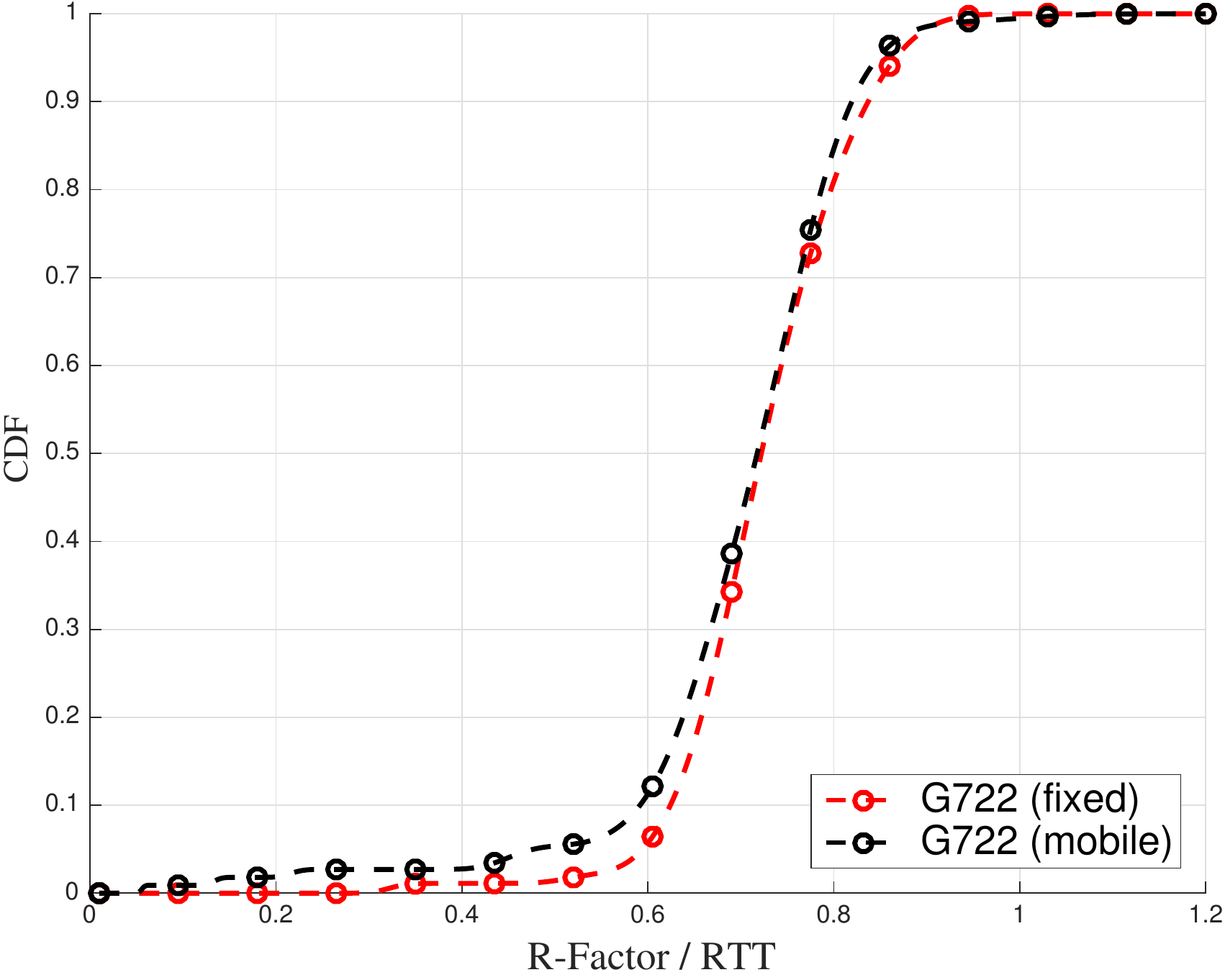}}  \hspace{2mm}
			\subfloat{\includegraphics[scale=0.25]{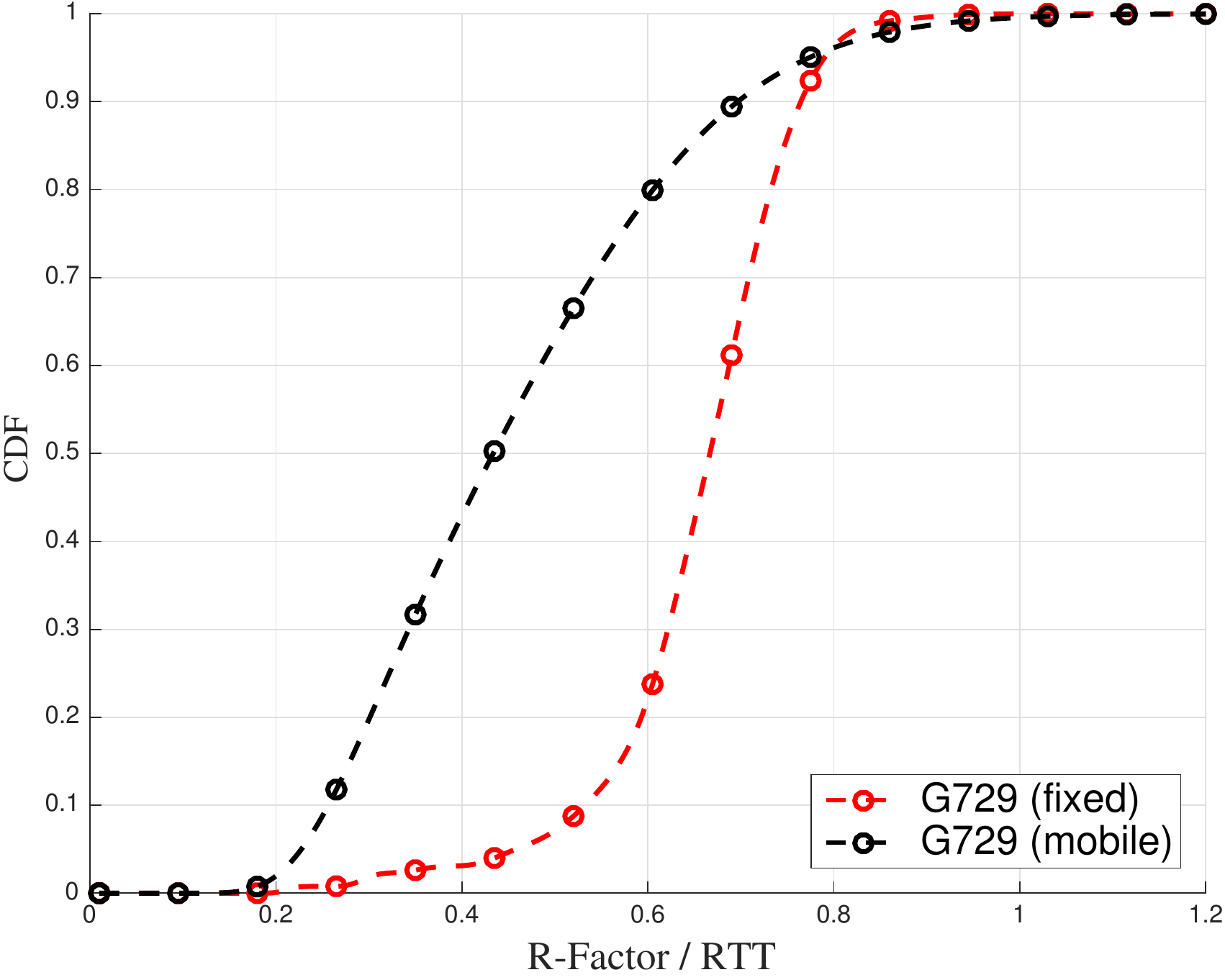}}  \hspace{2mm}
		\subfloat{\includegraphics[scale=0.25]{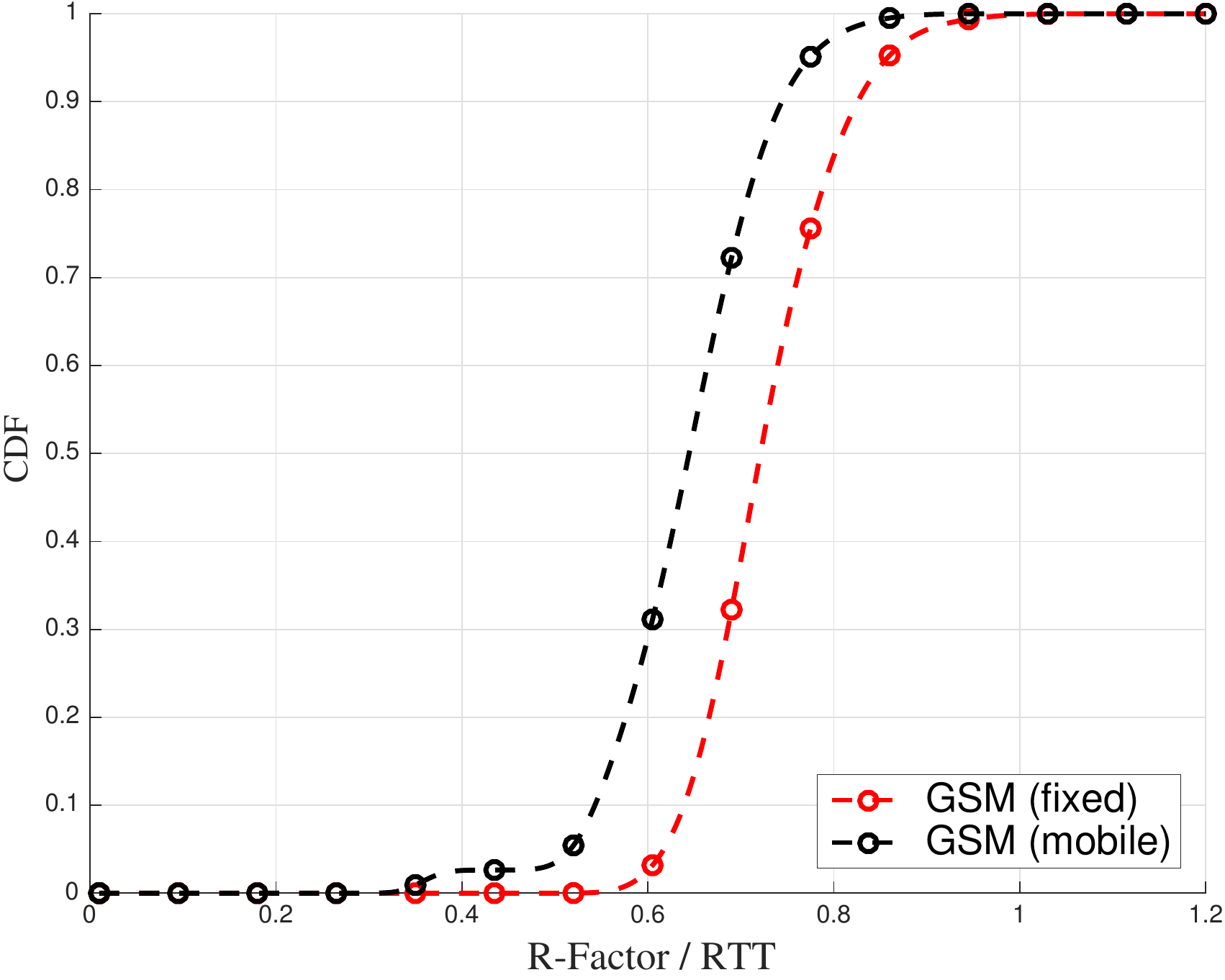}}  \hspace{2mm}
		\subfloat{\includegraphics[scale=0.25]{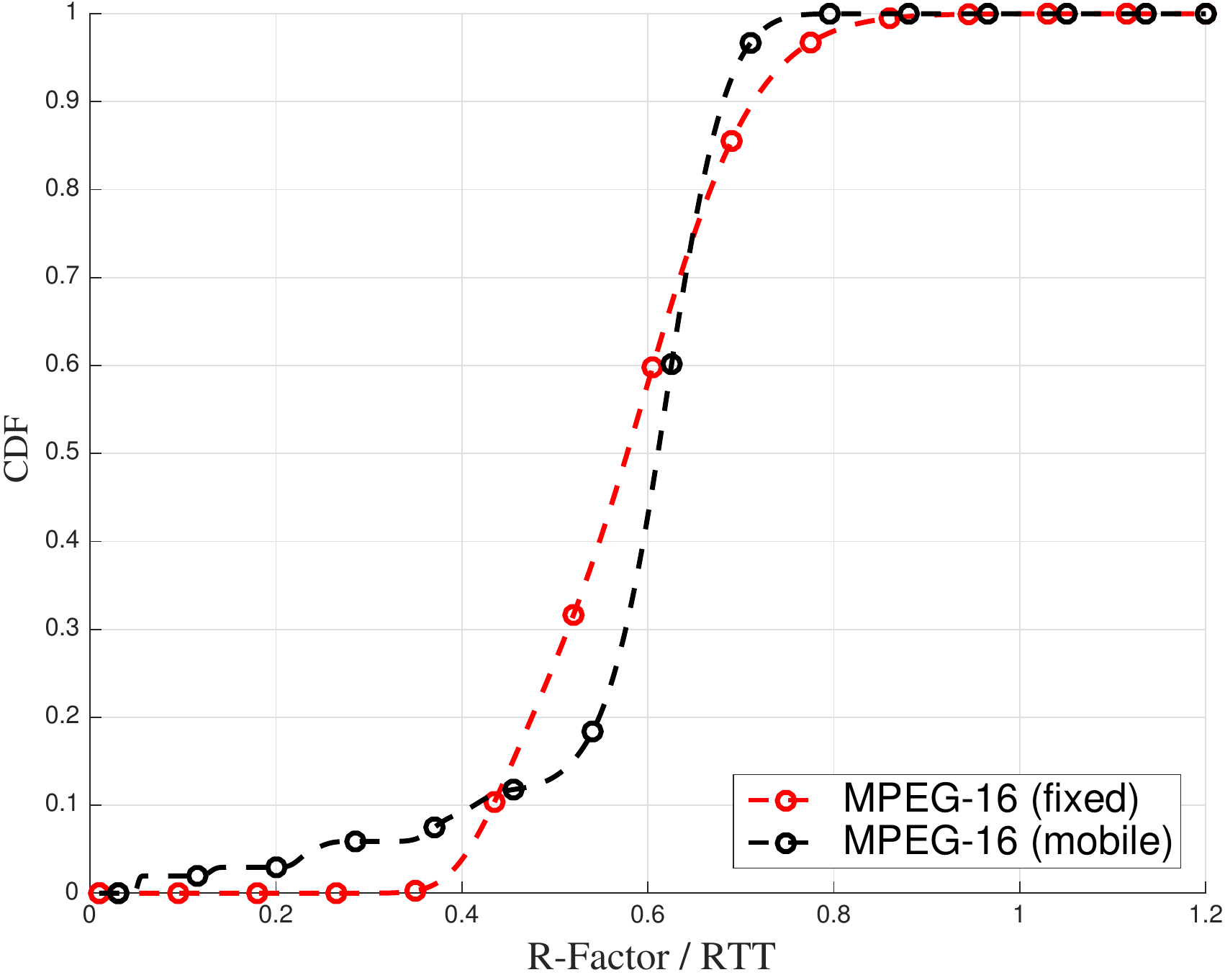}} \\
		\subfloat{\includegraphics[scale=0.25]{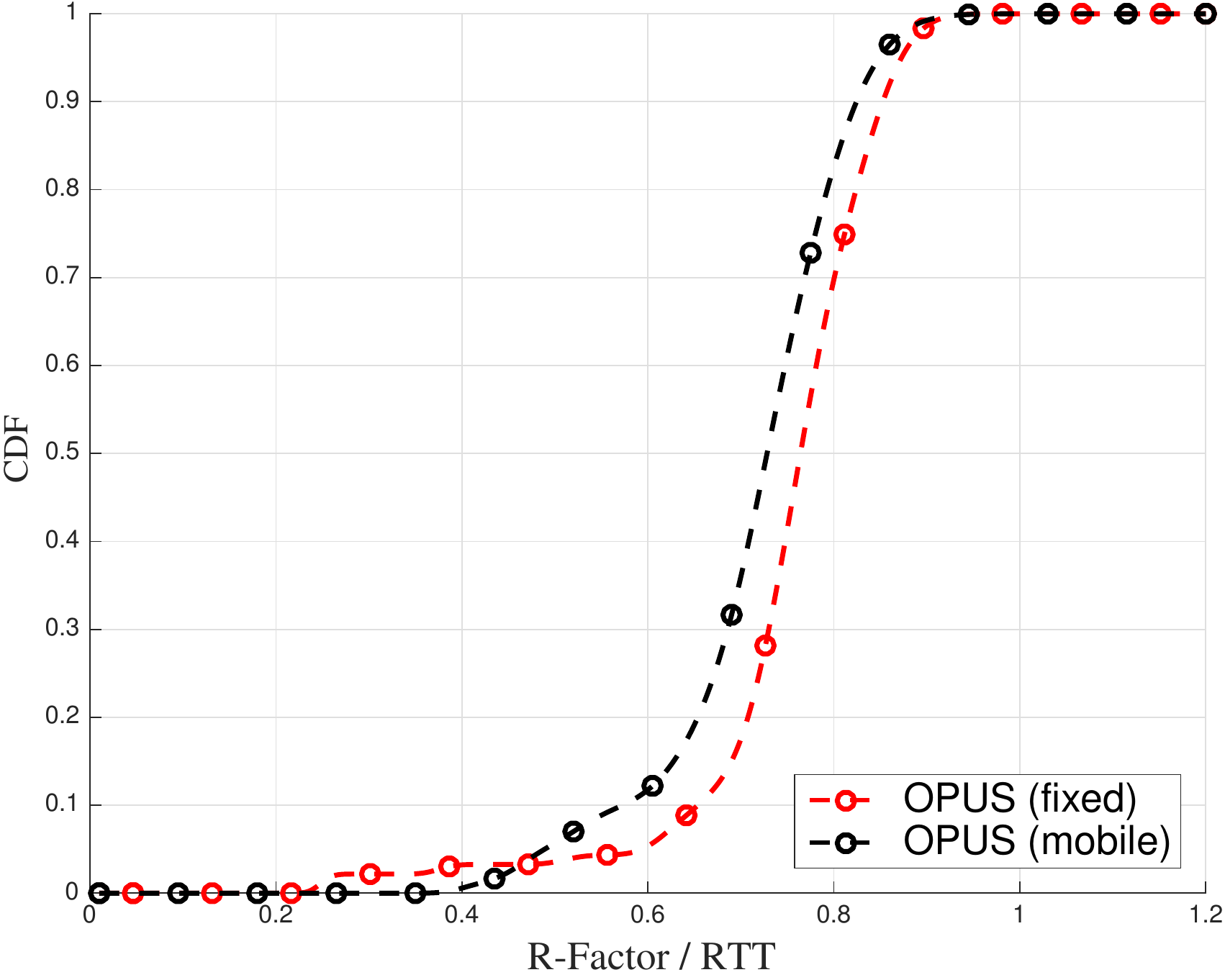}}  \hspace{2mm}
		\subfloat{\includegraphics[scale=0.25]{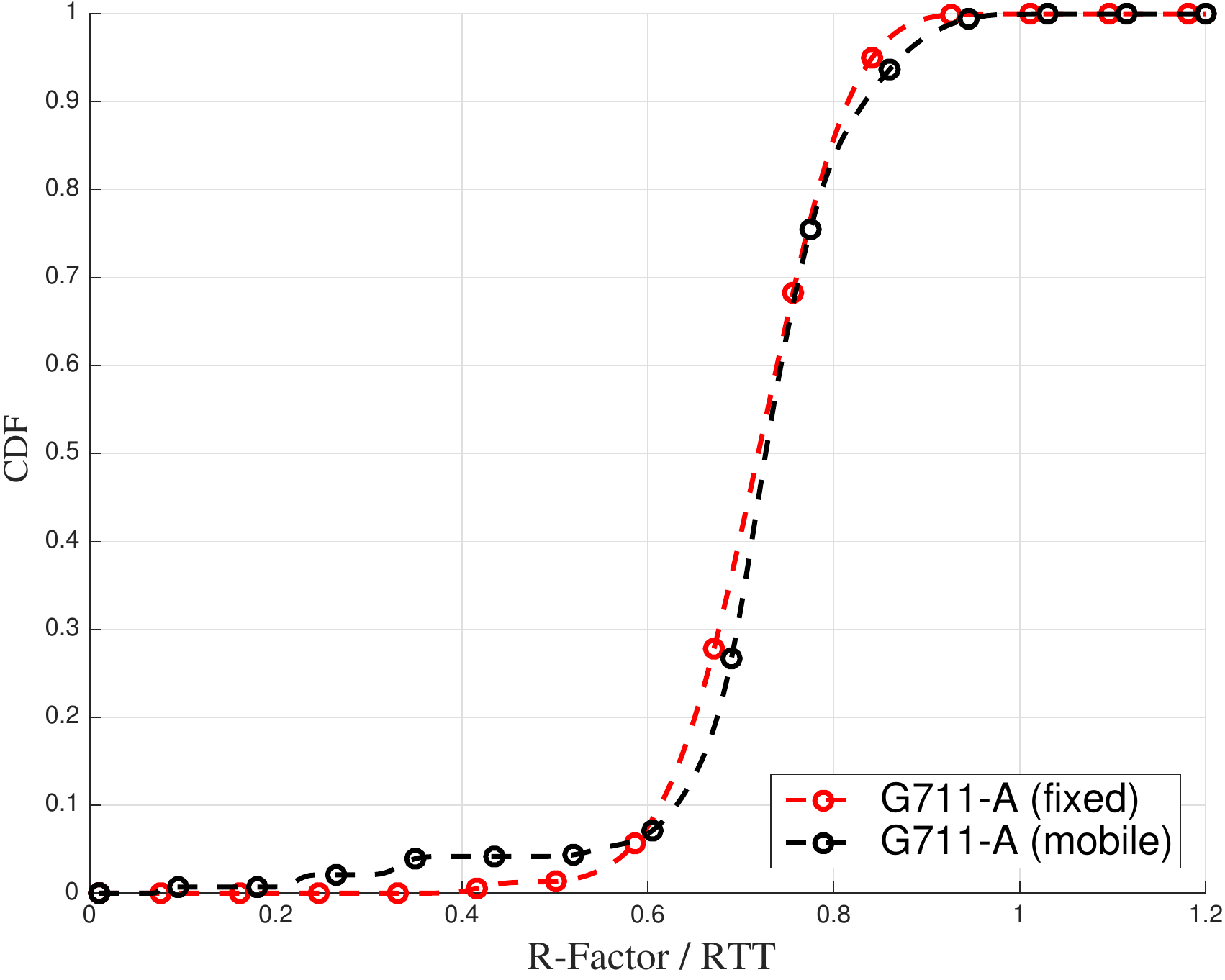}} \hspace{2mm}
		\subfloat{\includegraphics[scale=0.25]{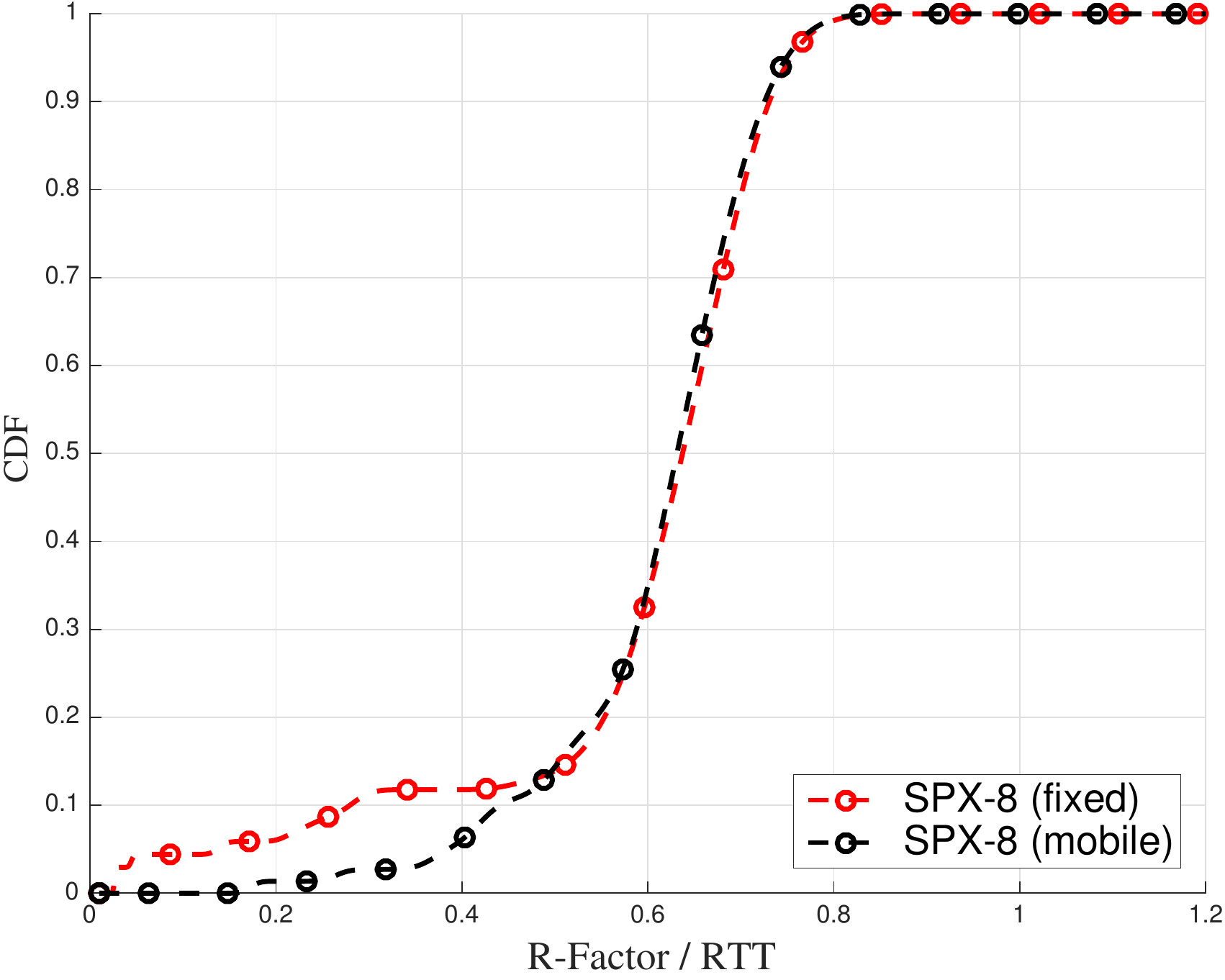}} \hspace{2mm}
		\subfloat{\includegraphics[scale=0.25]{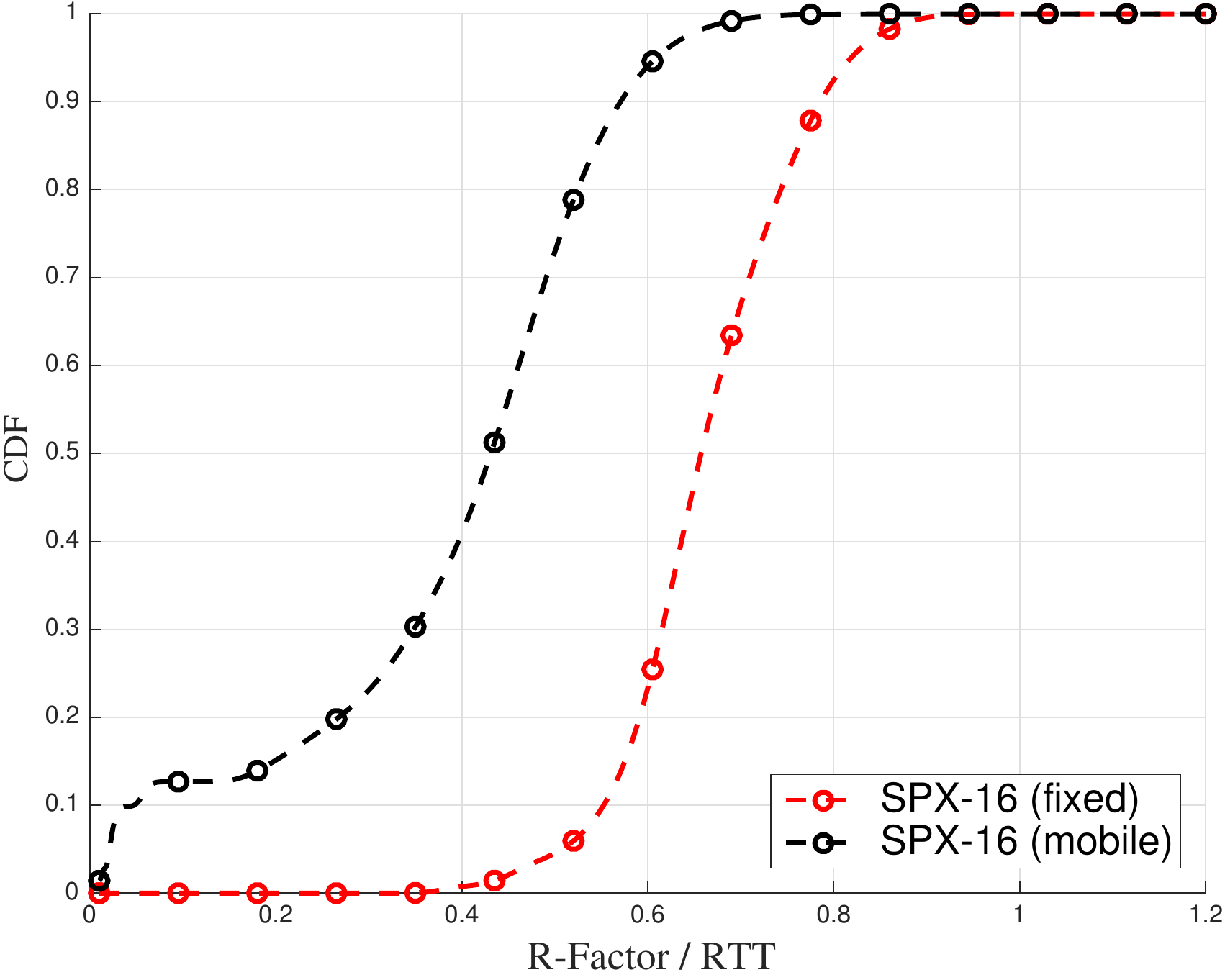}} 
	\end{tabular}
	\caption{Cumulative Distribution Functions (CDFs) of R-Factor normalized by RTT in two cases: fixed scenario (red curve) and mobile scenario (black curve) for various VoIP sessions with different codecs.}
	\label{fig:cdf_rfac_rtt}
\end{figure*}

\subsection{Impact of fixed/mobile scenario on VoIP metrics}

\begin{figure}[t]
	\centering
	\captionsetup{justification=centering}
	\includegraphics[scale=0.5]{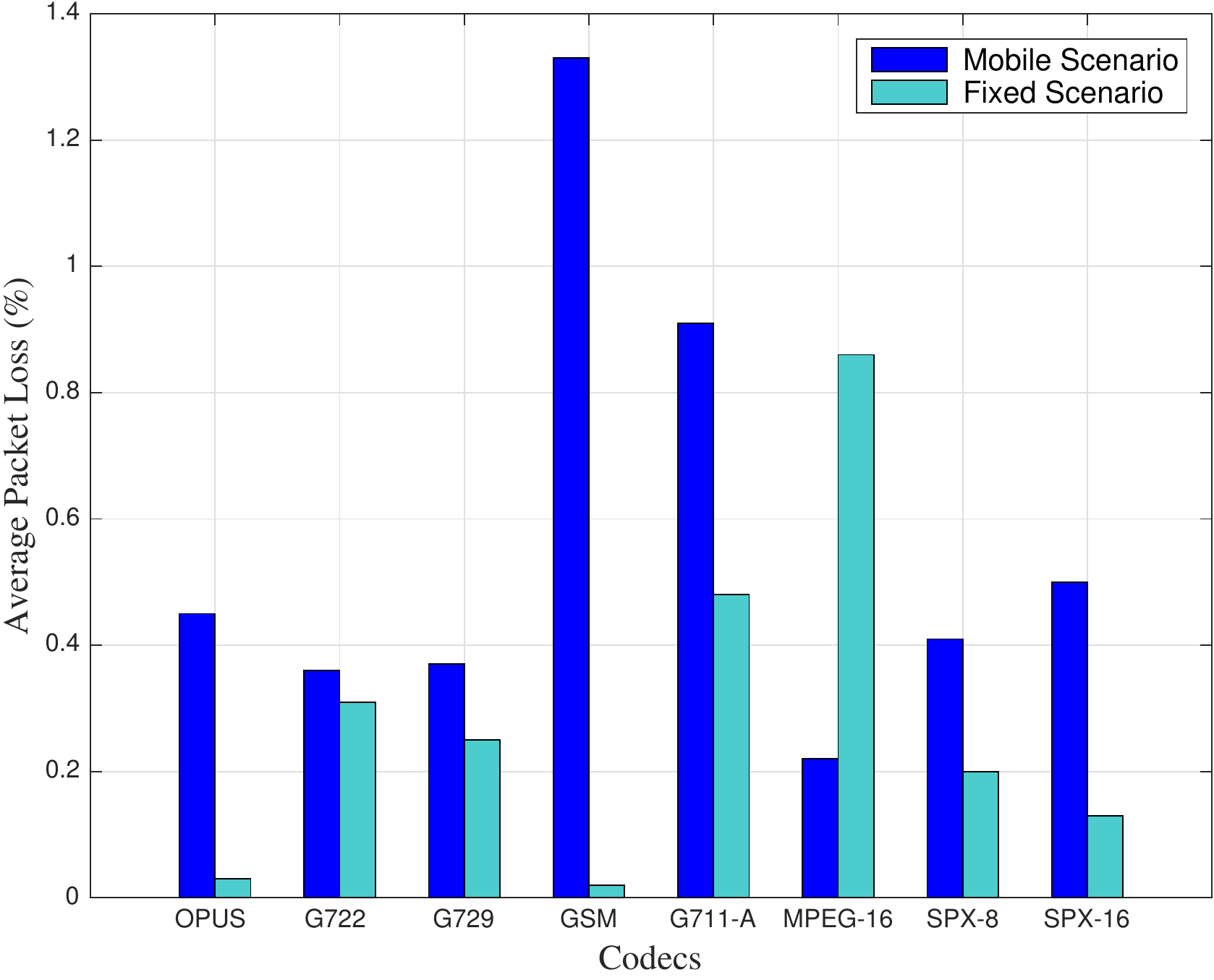}
	\caption{Percentage of packet loss (average) for mobile and fixed scenario and for various codecs.}
	\label{fig:loss}
\end{figure} 

Let's now analyze the effect of the external environment on some critical metrics. When comparing results across the two considered settings, namely fixed and mobility LTE scenarios, we have to account that interference phenomena occurring in mobile and fixed environments are reasonably comparable, since the trials were made in the same city environment.

In a first set of measurements, we compare the behavior of $\sigma_J$ across the two scenarios (fixed and mobile), and for the different codecs. More specifically, Fig. \ref{fig:cdfdevjit} reports the comparative results in terms of CDFs (red dashed curves for fixed scenario and black dashed curves for mobile scenario). Even if the jitter variation lies under acceptable thresholds for real-time communications \cite{sheluhin-book}, we observe a slightly increase in  the mobile case. This phenomenon, that is particularly evident for VoIP sessions with codecs G722, GSM and SPX-16, is due to the higher variability encountered in mobile scenarios, where hand-overs and scenery changing play a key role. We observe an exception for the case of the G729 codec, where the fixed environment seems to introduce a greater (although by just a small extent) jitter variability than mobile setting.

In a second set of measurements, we evaluate the variation of the voice signal level metric across the two reference scenarios. This metric is directly related to the quality of radio channel, noise on line and packet loss \cite{raake-book}. An excessive signal level change (e.g. too much attenuation or too much gain) affects the human perception. Thus, a network infrastructure with a balanced loss plan, should contain such fluctuations in a range of few dBms, up to a value of 20 dBm. Figure \ref{fig:cdfdevvoice} confirms such a trend, by showing that the most significant fluctuations occur during the mobile trials, due to the more unstable conditions of the radio channel. Acceptable deviations are represented by the case of codecs G729, SPX-8 and SPX-16, where a slightly more unstable behavior is exhibited in the case of fixed scenario, probably due to temporary decay of radio channel conditions.  

A last set of measurements is shown in Fig. \ref{fig:cdf_rfac_rtt}, where the ratio between R-Factor and RTT is analyzed, for the two scenarios and for different codecs. The expected behavior is that this ratio should increase in case of fixed scenario (i.e., greater R-Factor and/or smaller RTT). The results show that no dramatic differences exist across mobile and fixed scenarios (with the exception of G729 and SPX-16 flows), indicating that a mobility setting does not sensibly influence the combination of those two parameters.

Furthermore, we consider the impact of fixed and mobile scenarios on the average packet loss as reported in Fig. \ref{fig:loss}. Again, as a general (and expected) behavior, the mobile setting influences more consistently the packet loss phenomenon even if an exception occurs. This is the case of codec MPEG-16, where the percentage of packet loss in the fixed scenario overcomes that of the mobile case. As previously notices, some unexpected behaviors can be ascribed to uncontrolled phenomena such as overloaded eNBs, weather conditions, temporary interference.

\subsection{SIP metrics}

\begin{figure}[t]
	\centering
	\captionsetup{justification=centering}
	\includegraphics[scale=0.35,angle=90]{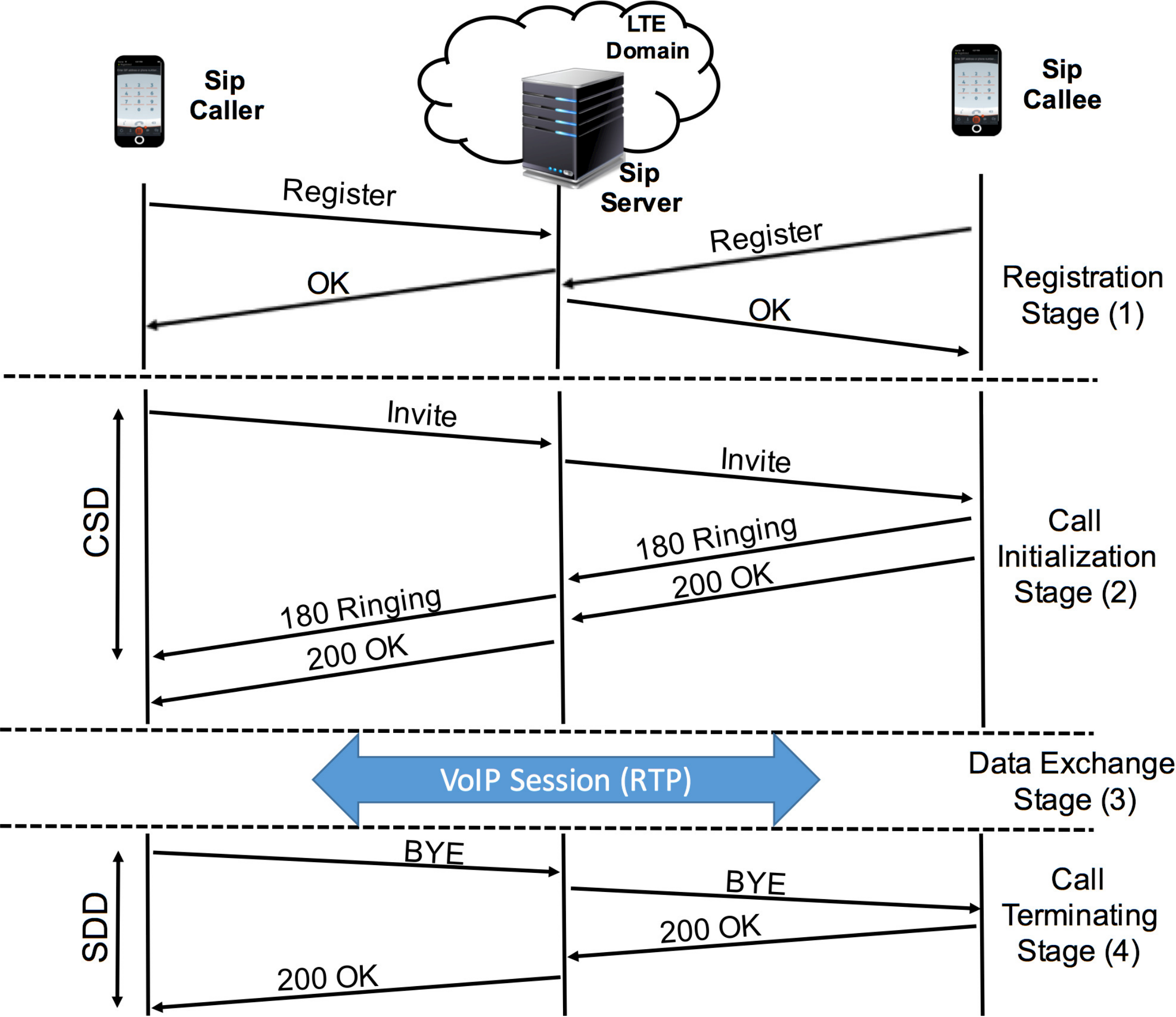}
	\caption{Snapshot of a SIP call with the two main delay metrics: Call Setup Delay (CSD) between Invite and $180$ Ringing messages and Session Disconnect Delay (SDD) between BYE and $200$ OK messages.}
	\label{fig:sipmetrics}
\end{figure} 


It is meaningful to evaluate also the performance of the SIP signaling protocol under different network conditions. In particular, we consider: \textit{i)} the Call Setup Delay (CSD), defined as the time interval between an Invite message sent from caller and the $180$ Ringing message received from the callee \cite{begain-book}; and \textit{ii)} the Session Disconnect Delay (SDD), a metric introduced to characterize the time needed to end a session managed by SIP protocol \cite{rfc6076}. In a sense, such metrics capture a macroscopic behavior of a voice call session quality, providing a coarse-grained association with the classic metrics such as jitter, RTT, and so forth.

Figure \ref{fig:sipmetrics} depicts a simplified scenario reproducing the four main stages characterizing a SIP-based session between two terminals, namely: registration stage, in which the two terminals attach to a SIP server and exchange information such as codecs to use; call initialization stage, representing the invite step in a voice call; data exchange stage, representing the voice data flow; call terminating stage that ends the call. More specifically, in our analysis we consider CSD and SSD metrics that intervene in call initialization stage (2) and call terminating stage (4), respectively.   

It is worth noting that, CSD and SDD can be considered high-level metrics (with respect to RTT or jitter for example). Thus, the codec dependency appears veiled. Accordingly, it is more useful to analyze the effect of mobile and fixed scenarios over the CSD and SDD metrics by including the effect of all used codecs. 

Results are shown in Fig. \ref{fig:csd_sdd}, and represented through boxplot diagrams that provide a concise representation of data distributions in terms of quartiles. On the left panel, the boxplot of CSD is reported. It is straightforward to notice that, in case of mobile scenario, CSD exhibits higher values due to more unstable conditions resulting in a longer initialization phase. The median value (horizontal red line inside the boxplot) of CSD in mobile setting amounts to $1.3920$ seconds, whereas it decreases to $0.8530$ seconds in fixed scenario. Another effect of unstable conditions in mobile setting is recognizable through the higher spread of values around the median. Such a behavior can be captured by means of the inter-quartile range (\textit{iqr}), defined as the difference between the third and the first quartiles. 
\begin{figure}[t]
	\centering
	\captionsetup{justification=centering}
	\includegraphics[scale=0.37]{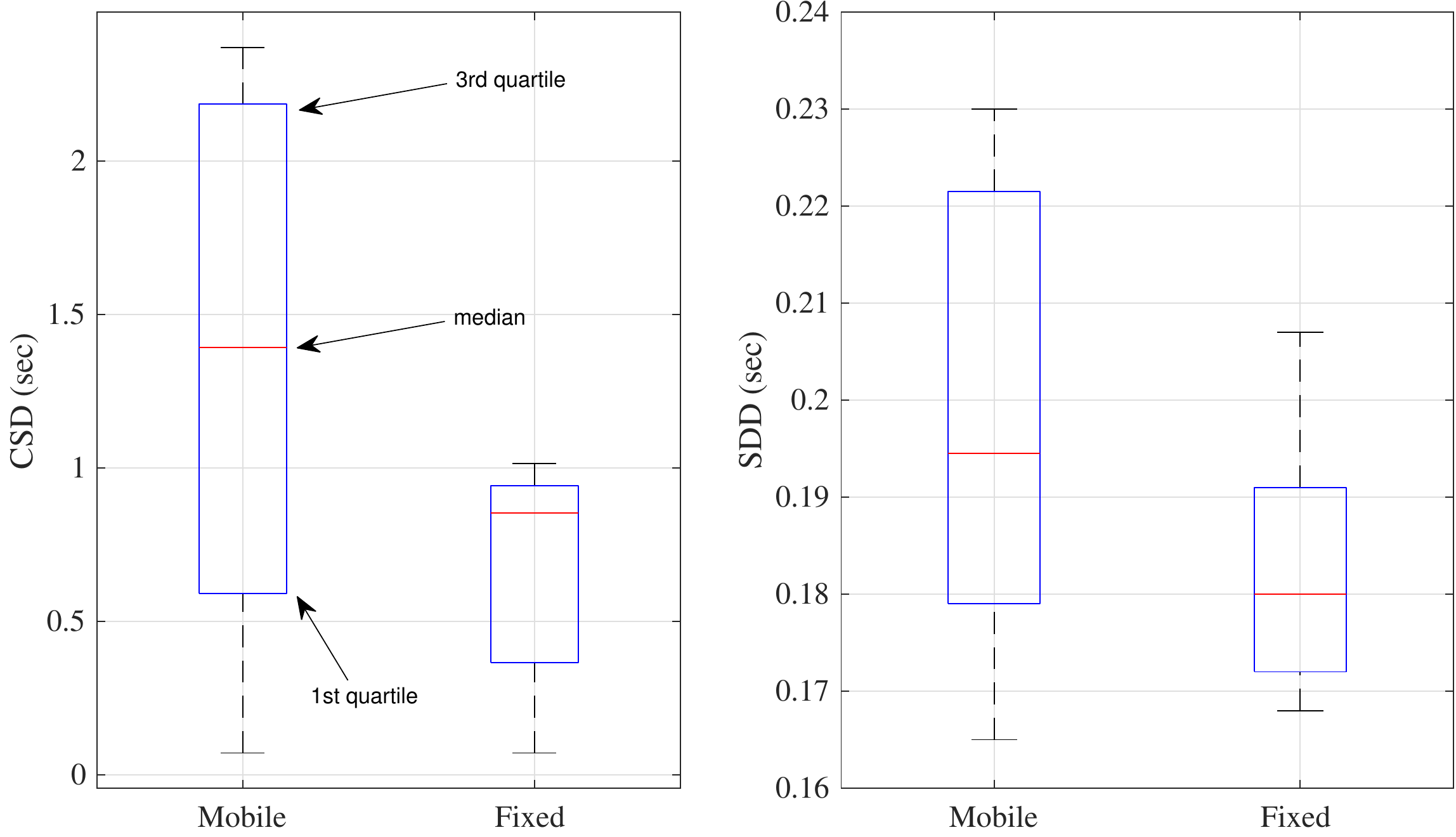}
	\caption{Boxplots representation for CSD (left panel) and SDD (right panel) in mobile and fixed scenarios.}
	\label{fig:csd_sdd}
\end{figure} 
The larger the \textit{iqr} is, the more dispersed will the values be. Accordingly, we measure \textit{iqr} values of $1.5960$ and $0.5770$ seconds, for mobile and fixed scenarios, respectively. It is worth highlighting that, the obtained values for CSD are in line with ITU-T (International Telecommunication Union - Telecommunication Standardization Bureau) specifications, being in the order of seconds \cite{itu1028}. 

Experiments about CSD measurements have been carried out also in \cite{curcio2002}, where the CSD is indicated by Post-Dialing Delay (PDD) and evaluated in a 3G-based environment. In this case, the CSD is in the order of hundred of milliseconds, but, the authors rely on a network simulator, by neglecting all the concurrent effects present in real networks. Similar under-estimated results are presented in \cite{awaludin2010} where NS-2 simulator equipped with a SIP module has been exploited.  

The right panel of Fig. \ref{fig:csd_sdd} shows the SDD boxplot representation for both the mobile and the fixed cases. With analogous considerations as those discussed for the CSD metric, the median SSD values for mobile and fixed cases amount to $0.1981$ seconds and $0.1638$ seconds, respectively. Similarly, being SSD values more spread around the median for the mobile case, we find \textit{iqr} values of $0.0425$ seconds and $0.0190$ seconds for mobile and fixed settings, respectively. In particular, for all cases, the SDD values never exceed 230 msec, and such values are in line with \cite{husic2014} (although authors operate in a simulated environment through NS-2 tool), and with \cite{brajdic2009}, where tests have been carried out within an emulated IP Multimedia Subsystem framework.

In summary, the performed assessment allows to discover: \textit{i)} how QoS (e.g. bandwidth consumption, jitter) and QoE (e.g. R-Factor) metrics of a call session are jointly impacted for different codecs and settings; \textit{ii)} how high level metrics such as CSD and SDD, related to the perceived call quality and typically neglected in classic literature, are influenced by fixed/mobile settings in a real LTE-A environment.

\subsection{Impact on Network Management}
Our evaluation results can be profitably exploited for the purposes of network management, specifically by network or telco providers, aiming at optimizing different network functions. Our measurements data will prove as invaluable \textit{ground truth} for other prediction-based approaches. Also it can be used as input to other optimization algorithms. 
Our findings will also impact the deployed architecture, where the most interested node involves the PCRF (Policy and Charging Resource Function). This is responsible for policy charging and QoS management, which takes decisions on how to handle services in terms of QoS Class Identifiers. }

Furthermore, QoS parameters can be mapped into Key Performance Indicators (KPI), whereas QoE parameters are mapped into Key Quality Indicators (KQI). The two defined indicators have an impact on  different  sub-infrastructures  of the LTE domain  \cite{Vaser2015}. 
Specifically, the E-UTRAN (Evolved UMTS Terrestrial Radio Access Network) infrastructure, representing the part of the domain involved in radio procedures management, is influenced by KPIs such as delay, jitter, packet loss, mobility success rate in terms of handover performance. 
These KPIs directly contribute to the Video/Audio quality KQI, which captures the ability of an end user to appreciate the video content and/or service audio. By contrast, the EPC (Evolved Packet Core) infrastructure, representing the  core  network  of the LTE domain, is influenced by KPIs such as accessibility success rate or EPC bearer utilization, which can  be mapped with KQIs responsible for measuring session setup time and session response time. 
The considered mappings will provide an operator with some useful indications about improving, for instance, E-UTRAN nodes (in case jitter or delay have to be taken under control), or optimizing EPC sub-domain functionalities (in case high-level metrics such as CSD or SSD must satisfy more challenging constraints). Latter considerations about CSD/SSD metrics (directly involving the performance of the SIP protocol) also hold true for 5G network scenarios, where the IMS framework (relying on SIP) represents a key element of the 5G core infrastructure.
Remarkably, LTE-A (and more generally 5G networks) can also benefit from the Self-Organizing-Networks (SON) paradigm, designed to automate Operation and Maintenance of cellular networks to help operators improving network efficiency and performance. In this direction, results derived in the proposed study can be first translated into KPIs/KQIs indicators, and then, embedded into SON routines. 
	

\section{Statistical Characterization of jitter and RTT in an LTE mobile environment}
\label{sec:statchar}

\begin{figure*}[t!] 
	\centering
	\begin{tabular}{cccc}
		\subfloat{\includegraphics[scale=0.25]{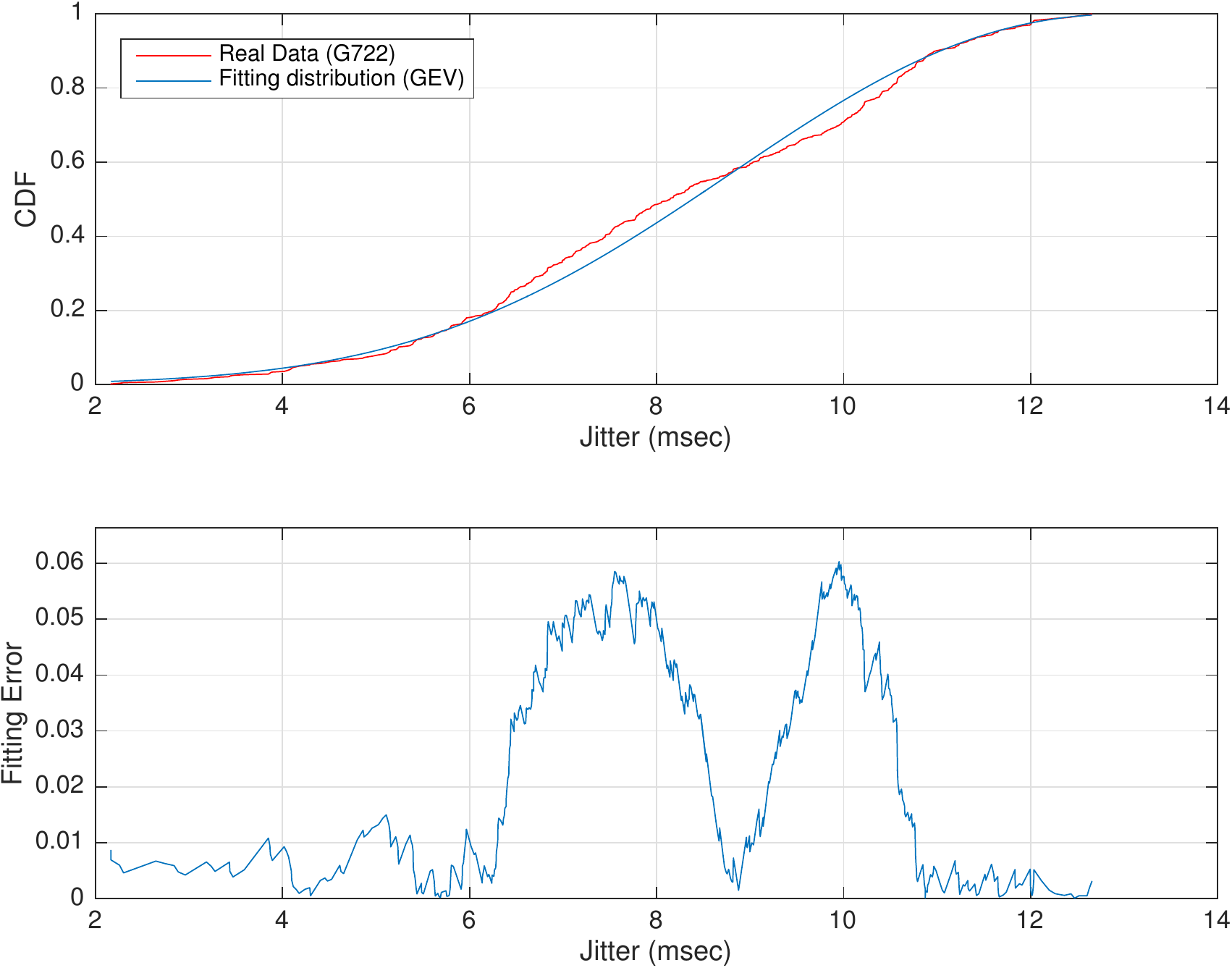}}  \hspace{2mm}
			\subfloat{\includegraphics[scale=0.25]{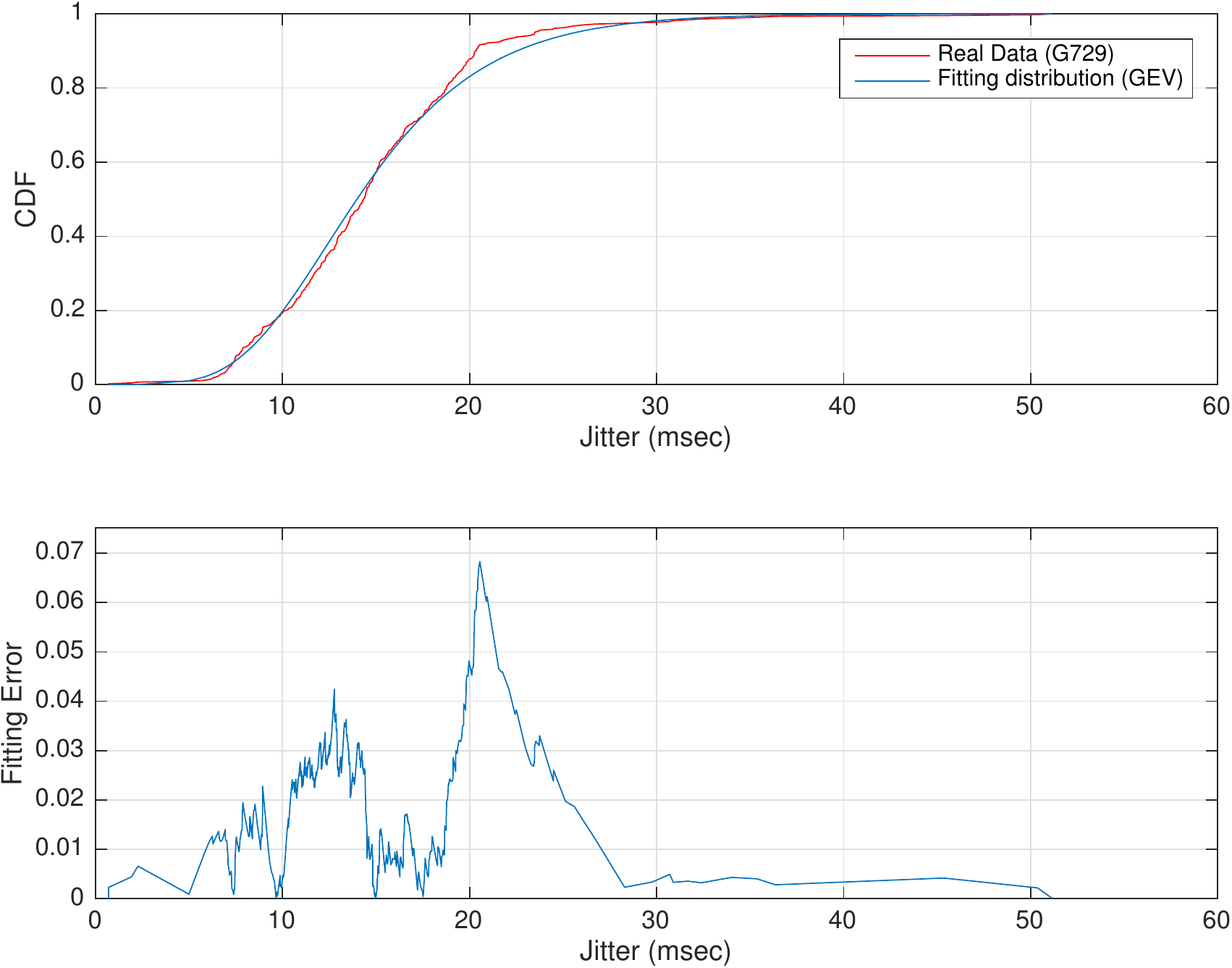}}  \hspace{2mm}
		\subfloat{\includegraphics[scale=0.25]{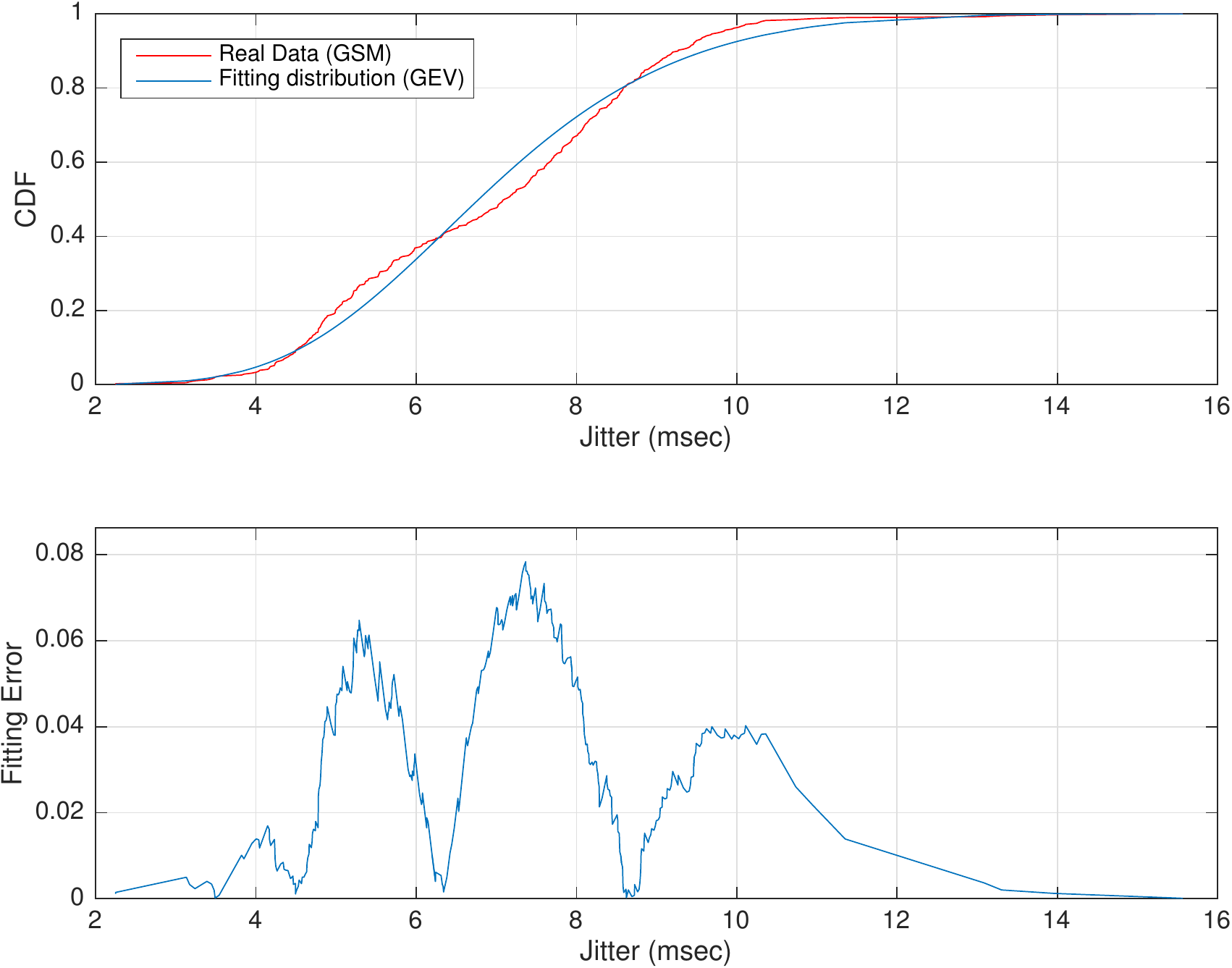}}  \hspace{2mm}
		\subfloat{\includegraphics[scale=0.25]{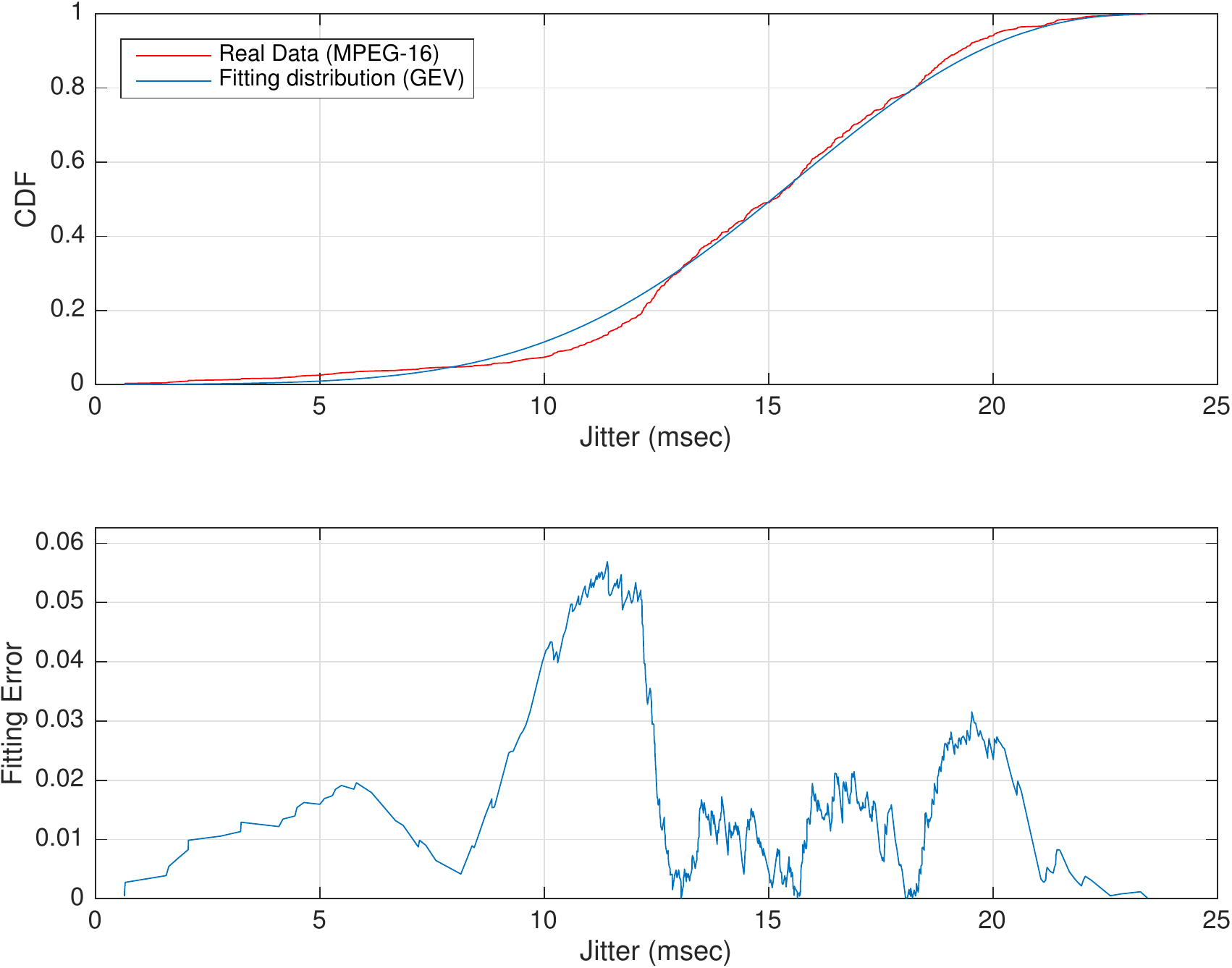}} \\
		\subfloat{\includegraphics[scale=0.25]{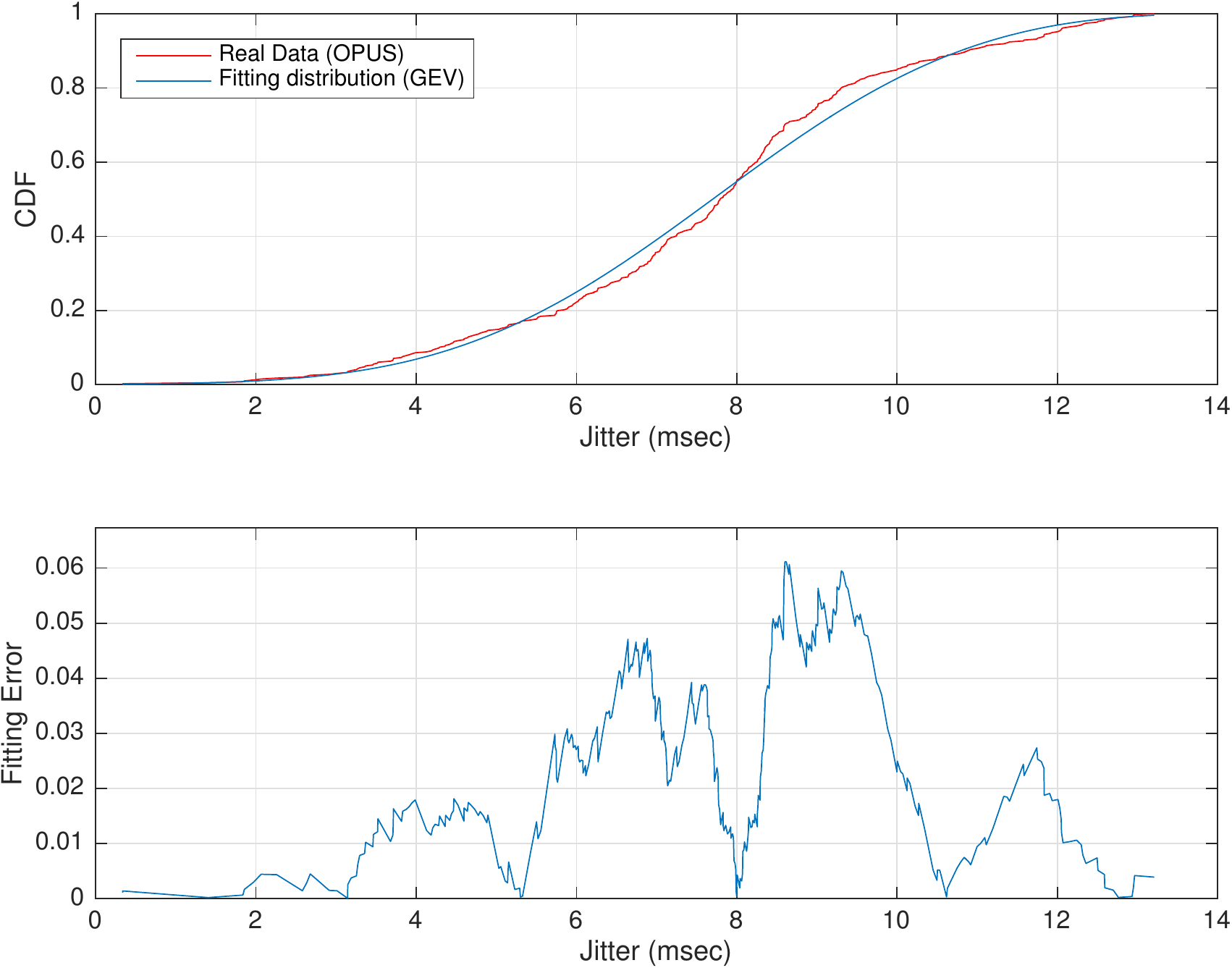}}  \hspace{2mm}
		\subfloat{\includegraphics[scale=0.25]{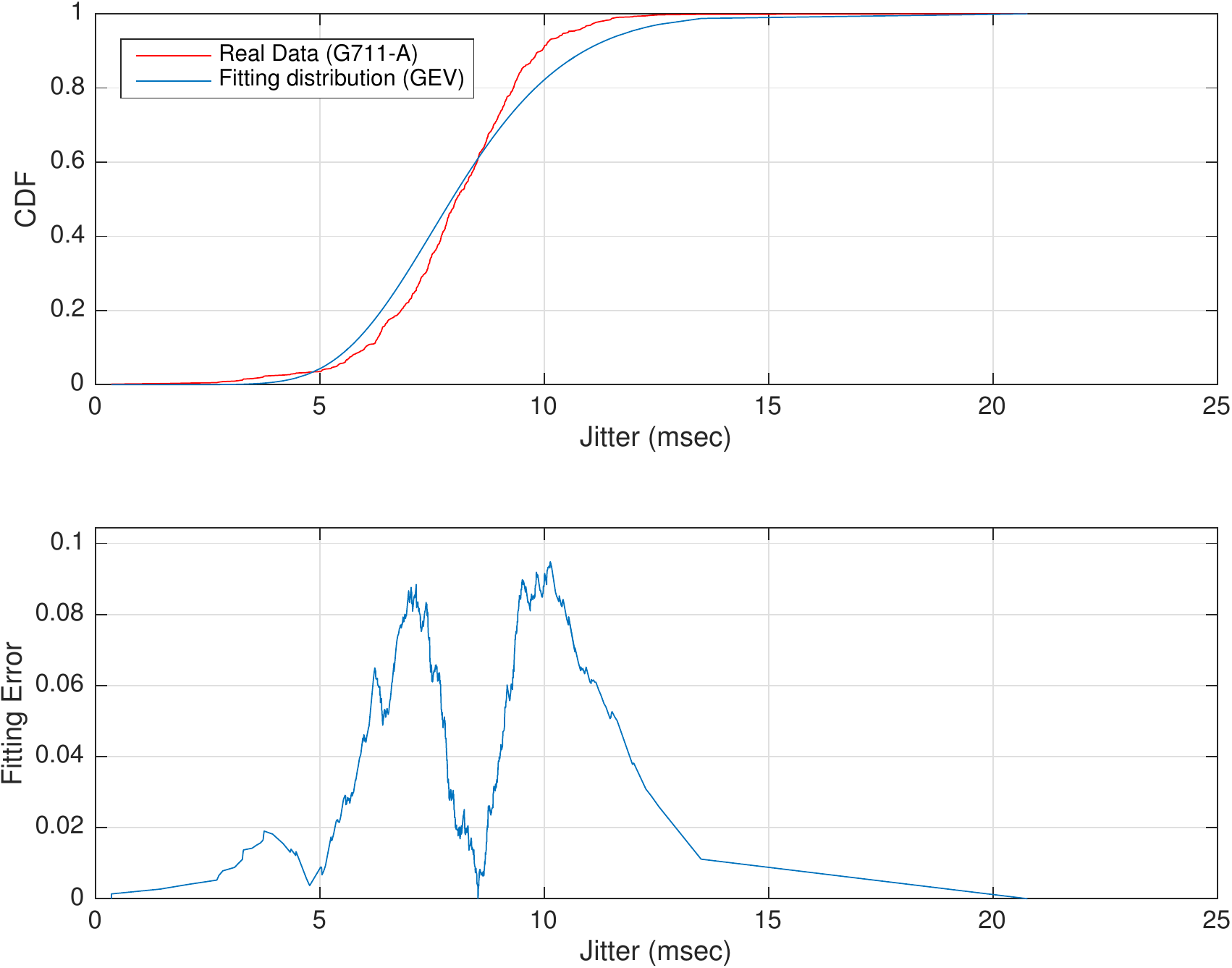}} \hspace{2mm}
		\subfloat{\includegraphics[scale=0.25]{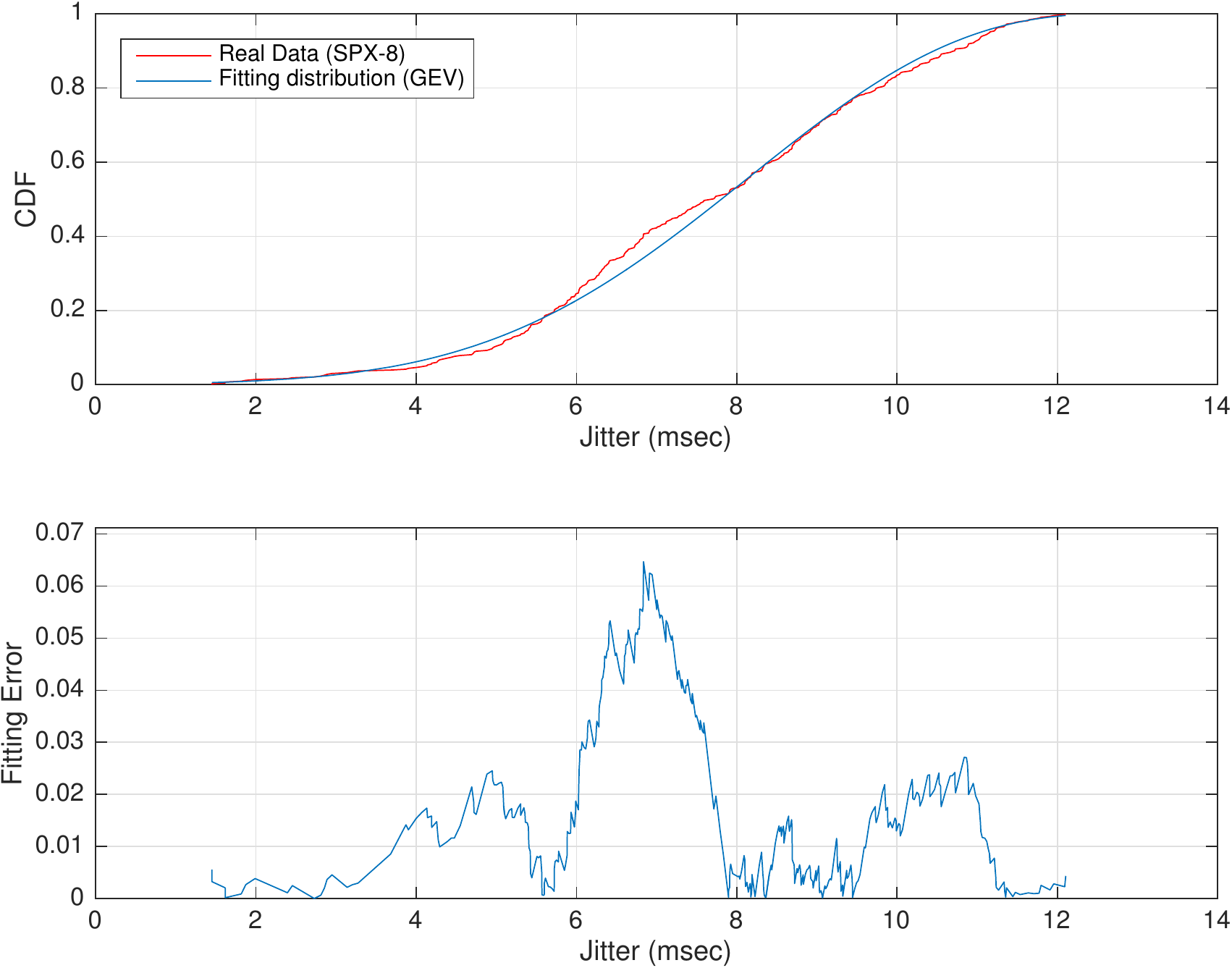}} \hspace{2mm}
		\subfloat{\includegraphics[scale=0.25]{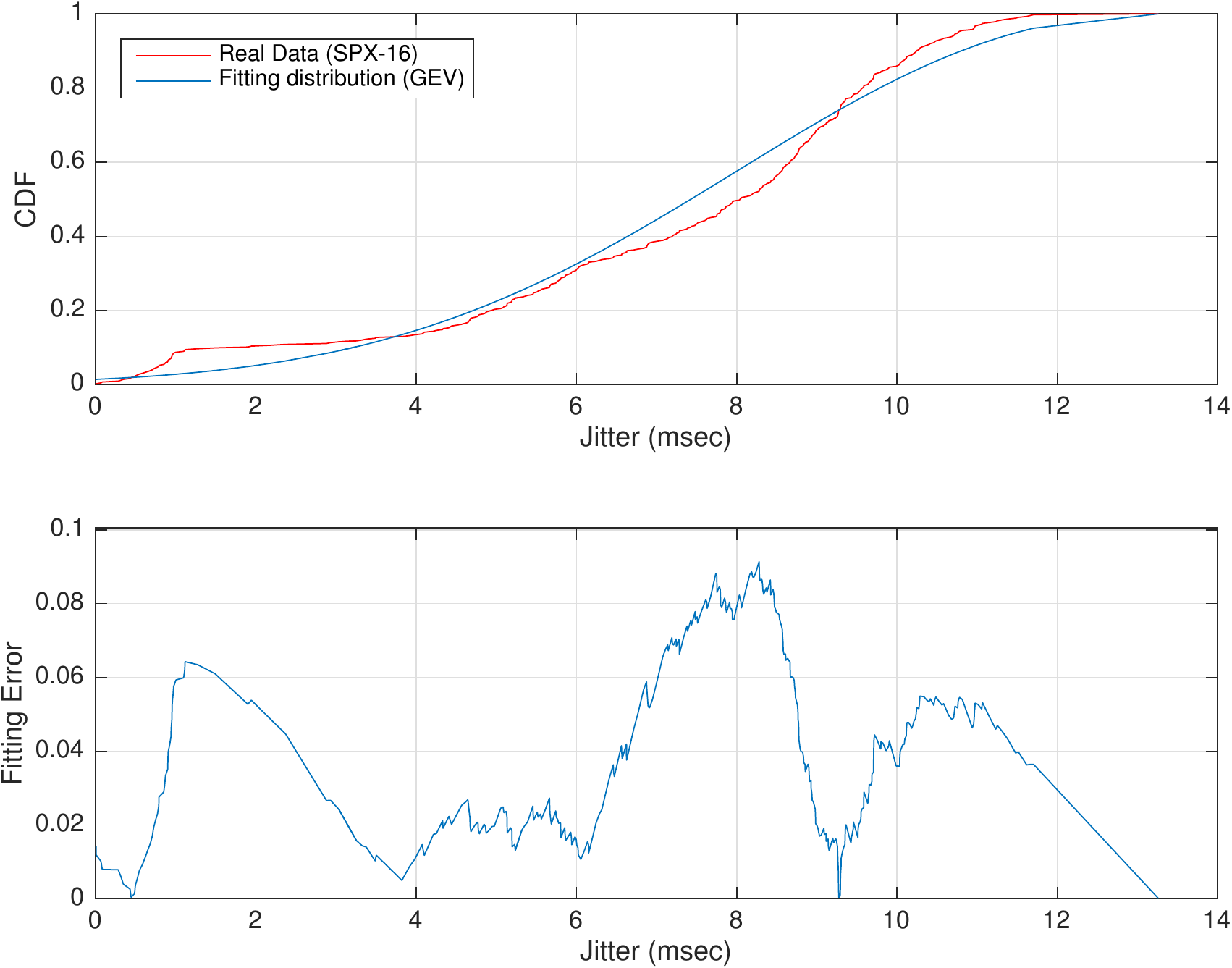}} 
	\end{tabular}
	\caption{Statistical modeling of jitter through GEV distribution. In the upper sub-panels, real data in red and fitting distribution in blue for various codecs. In the lower sub-panels, the pertinent fitting errors.} 
	\label{fig:fitting}
\end{figure*}

In this section, we analyze two QoS metrics that have a significant impact on communication, that are jitter and RTT \cite{dahmouni2012,kaup2016}. These are particularly important due to the best-effort nature of IP networks, whereby packet delivery time and order are not guaranteed. Hereafter, we introduce a statistical characterization of jitter and RTT (we just consider the case of mobile setting) based on the measurement campaign described in the previous sections. 

In the scientific literature, several work have been devoted at modeling these metrics, although most studies are based on simulations and have lead to approximate results. It is commonly accepted that many QoS-related metrics follow heavy-tailed distributions, as confirmed by relevant work reported in \cite{rizo2010_rtt}, \cite{acharya}, \cite{broido}, which specifically analyze RTT distributions.

Even more sophisticated statistical analyses have been devoted to jitter, leading to similar considerations. For instance, some analytical approximations are provided in \cite{matragi1997}, which characterize jitter as a function of background traffic in ATM networks. In \cite{huremovic2015}, the authors propose an analytic model for jitter, by simulating incoming traffic as an interrupted Poisson process. In \cite{hammad2016}, a mathematical model of jitter in a queuing system supported by a simulated environment has been proposed. Again, in \cite{voznak2012} the authors propose a model through a generalized Pareto distribution. Whereas in \cite{cruz2013}, jitter has been modeled by invoking self-similar structures and by evaluating the Hurst parameter to confirm its memoryless property.

By contrast to previous literature, our analysis is devoted to characterize RTT and jitter in a real scenario, with emphasis on a variety of VoIP flows and considering 8 different codecs. Among candidate heavy-tailed distributions, we choose the Generalized Extreme Value (GEV) distribution which offers the flexibility of embodying more distributions. Such a choice is confirmed by our empirical analysis where we perform selective goodness of fit tests (among a set of 10 distributions) based on the Schwarz criterion \cite{schwarz78}. 

In this way, we can find the optimal distribution among 10 different candidate ones, representing jitter and RTT as Generalized Extreme Value (GEV) distributions.

The GEV distribution was originally introduced by Jenkinson in \cite{jenkinson1955,jenkinson1969}, and has a CDF defined as:

\begin{align*}
\centering
F_{X}(x) =
\left\{
\begin{array}{l}
{\begin{array}{ll}
	\hspace{-0.2cm}  e^{{-(1+\xi \frac{x-\mu}{\sigma})^{-1/\xi}}}  \;\;\  \left\{\begin{array}{lr}  -\infty < x \leq \mu - \sigma/\xi \;\;\; (\xi < 0);
\nonumber \\
 \\
 \mu - \sigma/\xi \leq x < \infty \;\;\;\;\;  (\xi > 0);  \end{array}\right. 
\nonumber \\
	\end{array}}
\\
\\
e^{-e^{- \frac{x-\mu}{\sigma}}} \;\;\;\;\;\;\;\;\;\;\;\;\;\;\;\;\;\;\;\;\;\; -\infty < x < \infty \;\;\;\;\; (\xi = 0),
\end{array}
\right.
\tag{1}
\label{gevcdf}
\end{align*}   
where $\xi$, $\sigma$, and $\mu$ are shape, scale, and location parameters, respectively. Among these, shape is the most critical parameter since its value sign determines the type of distribution \cite{kotz-book}. In particular, for $\xi>0$ the distribution is called Fr\'echet-type distribution; for $\xi<0$ the distribution becomes a Weibull-type distribution; and the case $\xi=0$ corresponds to the Gumbel distribution. Location and scale parameters govern the shift of distribution and the shift deviation, respectively. The distributions of RTT and jitter concerning the mobile traffic have been completely characterized by estimating GEV parameters through the ML criterion \cite{coles-book}. 

We should note that, although the maximum likelihood estimate (MLE) has been successfully applied, when dealing with extreme distributions a possible issue could arise. The regularity conditions required for the asymptotic properties of MLE could be violated in some cases. Smith \cite{smith1985} faced in detail this issue by proving that: for $\xi>-0.5$, ML estimators exhibit the standard asymptotic properties; for $-1<\xi<-0.5$, ML estimators are generally achievable even if they do not have asymptotic properties; and for $\xi<-1$, it is unlikely to obtain ML estimators. In our regimes we always fall in the first two cases, so no critical issues occur.   

Let us now consider the PDF associated to (\ref{gevcdf}):
\begin{align*}
\centering
p_{X}(x) =
\left\{
\begin{array}{l}
{\begin{array}{ll}
	\hspace{-0.2cm}  e^{{-(1+\xi \frac{x-\mu}{\sigma})^{-1/\xi}}} \frac{1}{\sigma} \{ 1 + \xi (\frac{x-\mu}{\sigma})\}^{-\frac{1}{\xi}-1}  \;\;\;\;\;\;\;\;   
	\\
	\\
	\;\;\;\;\;\;\;\;\;\;\;\;\;\;\;\;\; \left\{\begin{array}{lr}  -\infty < x \leq \mu - \sigma/\xi \;\; (\xi < 0);
	\\
	\\
  \mu - \sigma/\xi \leq x < \infty \;\;\;\;  (\xi > 0); \end{array}\right.
	\end{array}}
\\
\\
e^{-e^{- \frac{x-\mu}{\sigma}}} \frac{1}{\sigma} e^{-\frac{x-\mu}{\sigma}} \;\;\;\;\;\;\;-\infty < x < \infty \;\; (\xi = 0).
\end{array}
\right.
\tag{2}
\label{gevpdf}
\end{align*} 

Let us neglect the particular case $\xi=0$, which implies some different considerations concerning the Gumbel limit of GEV distribution, and consider the case $\xi \neq 0$. We will show in the forthcoming analysis that this case is in accordance with estimated shape parameter. 

Let $\theta=(\xi,\sigma,\mu)$ be the triplet of unknown distribution parameters to be estimated; and assume that the empirical data (representing RTT and jitter realizations) $Z_1, \dots, Z_n$ form a sequence of i.i.d. random variables. We can express the likelihood function as follows:

\beqa
\setcounter{equation}{3}
L(\theta; Z)&=&\prod_{i=1}^{n} \frac{1}{\sigma}\left(1 + \xi \frac{z_i-\mu}{\sigma} \right) ^{-\frac{1}{\xi}-1} e^{-\left( 1+  \xi \frac{z_i-\mu}{\sigma}\right)^{-\frac{1}{\xi}}} \\ \nonumber
&=& 
\sigma^{-n} \left[ \prod_{i=1}^{n} \left( 1+ \xi \frac{z_i-\mu}{\sigma} \right) \right] e^{-\left( 1+  \xi \frac{z_i-\mu}{\sigma}\right)^{-\frac{1}{\xi}}},
\eeqa
for $1+\xi \frac{z_i - \mu}{\sigma}>0$ and for all $z_i$.

The corresponding log-likelihood has the following form:

\beq
l(\theta; Z)= -n \ln \sigma - (1+\xi) \sum_{i=1}^{n} \omega_i - \sum_{i=1}^{n} e^{-\omega_i}, 
\label{eq:loglike}
\eeq
where we set $\omega_i=\xi^{-1} \ln \left( 1+ \xi \frac{z_i - \mu}{\sigma} \right)$.
By definition, the MLE $\hat{\theta}=(\hat{\xi},\hat{\sigma},\hat{\mu})$ is defined as:

\beq
\hat{\theta}=\argmax_{\theta \in \Theta} l(\theta; Z),
\eeq
being $\Theta$ the set of variable bounds.
The solution of likelihood system is obtained by differentiating (\ref{eq:loglike}), which does not however admit a closed form, and is solved with the Newton-Raphson's method. 
\begin{figure*}[t!] 
	\centering
	\begin{tabular}{cccc}
		\subfloat{\includegraphics[scale=0.25]{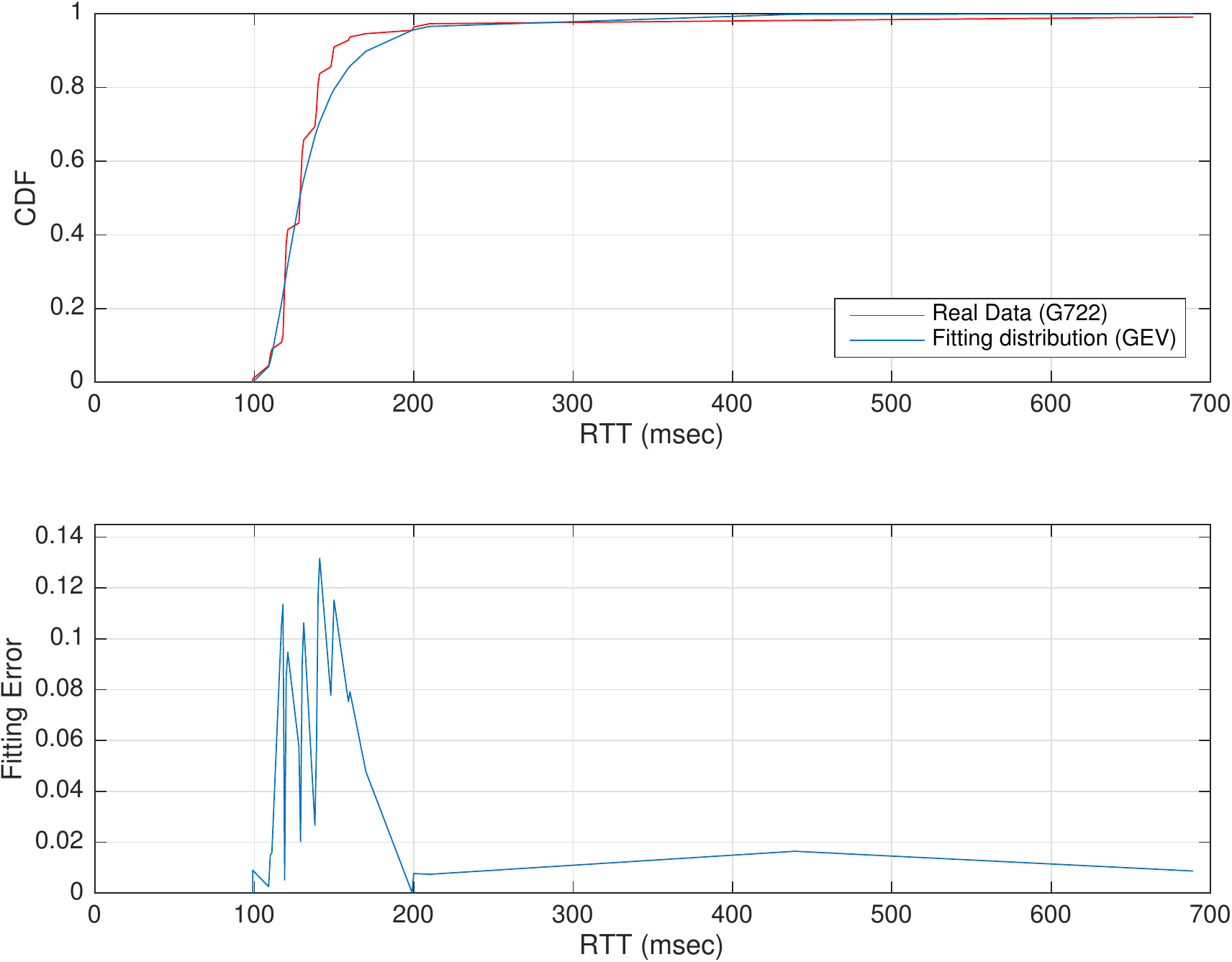}}  \hspace{2mm}
		\subfloat{\includegraphics[scale=0.25]{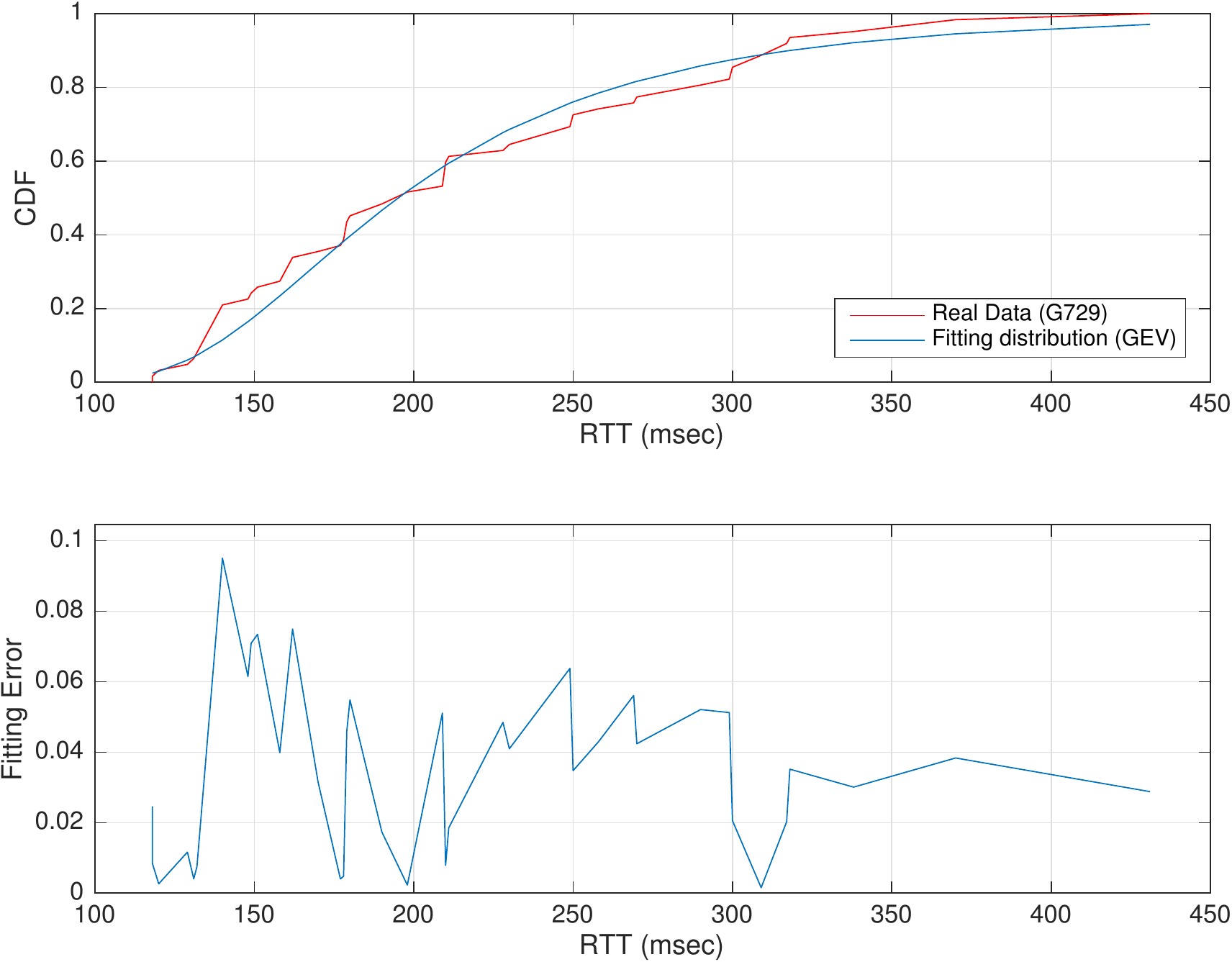}}  \hspace{2mm}
		\subfloat{\includegraphics[scale=0.25]{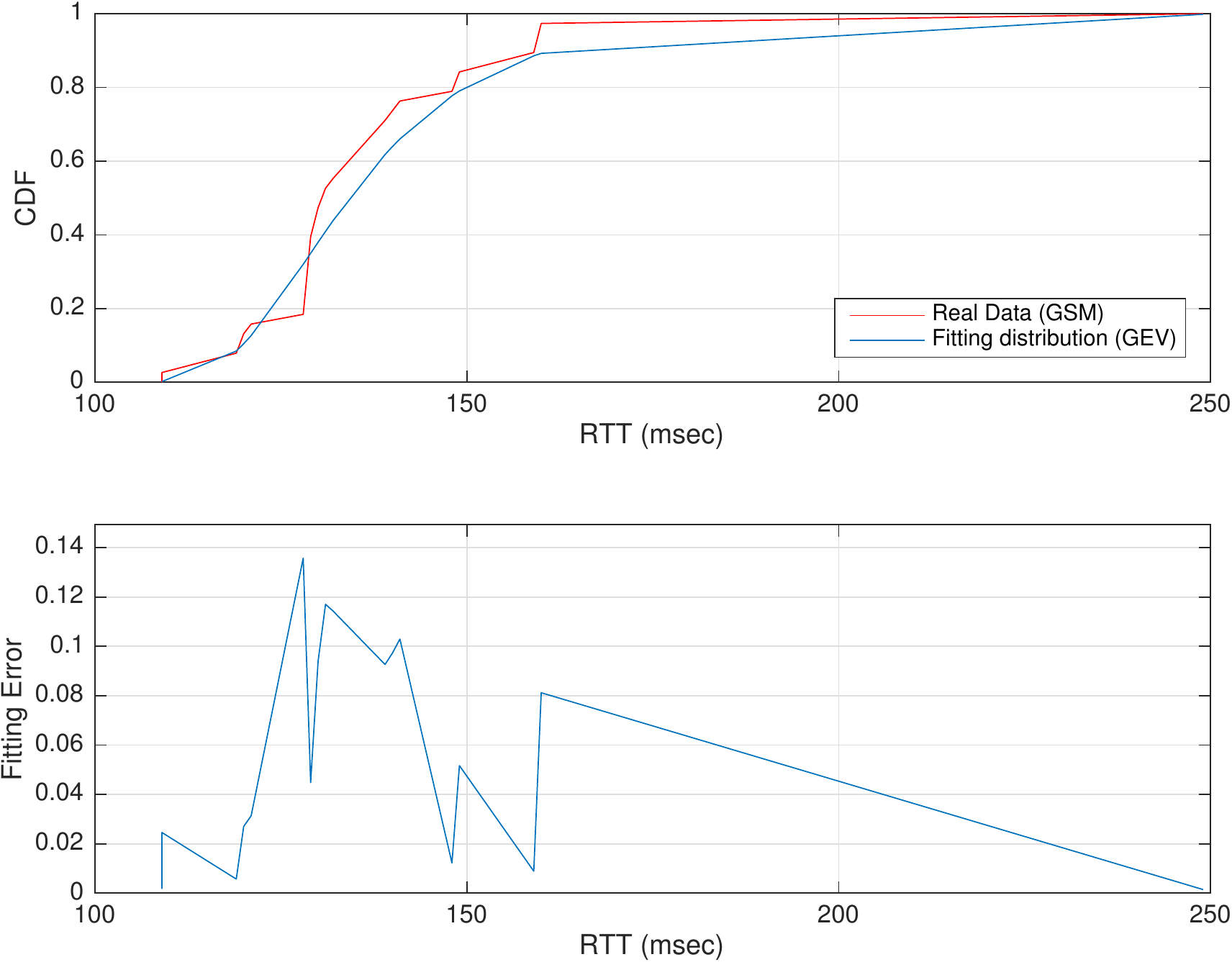}}  \hspace{2mm}
		\subfloat{\includegraphics[scale=0.25]{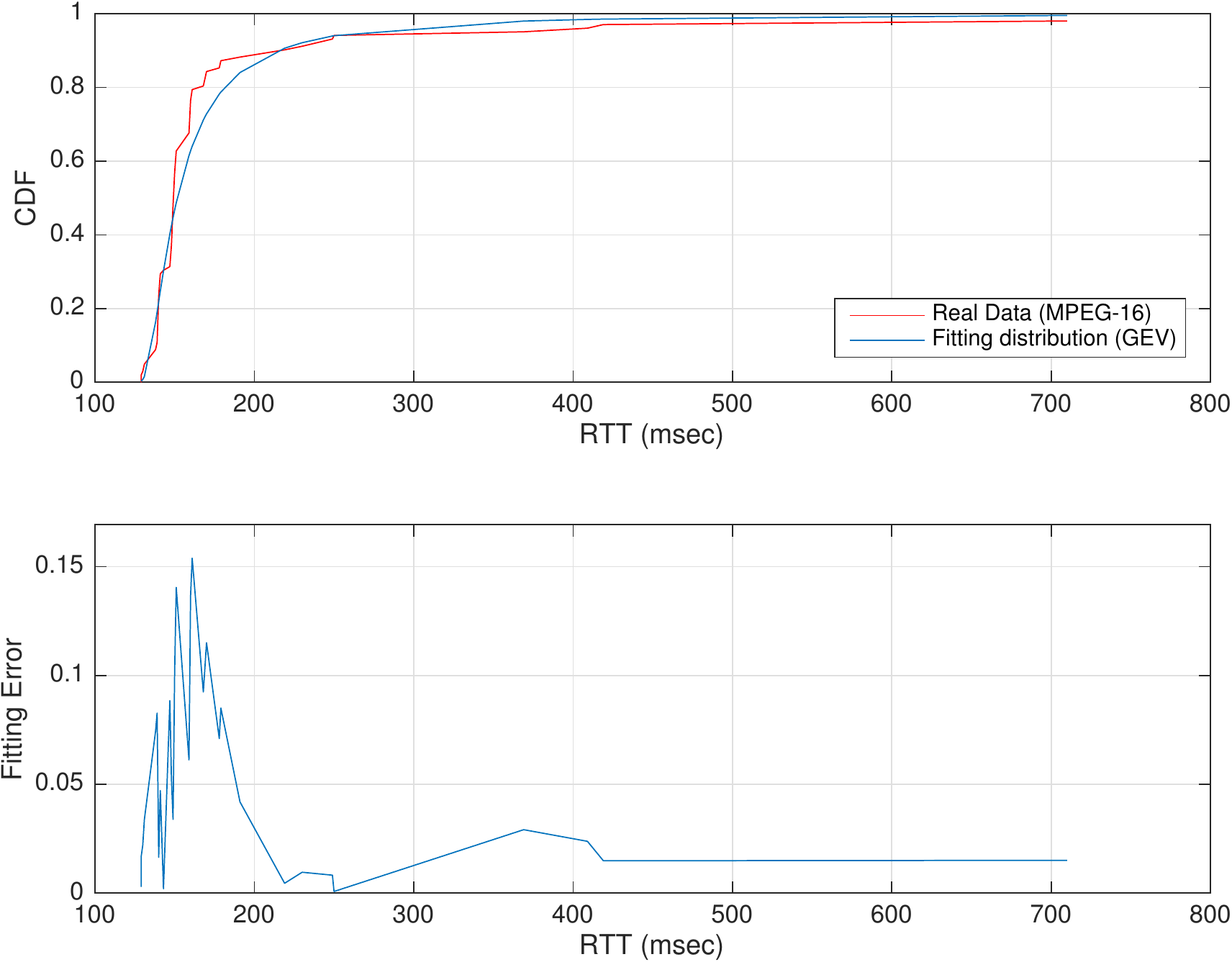}} \\
		\subfloat{\includegraphics[scale=0.25]{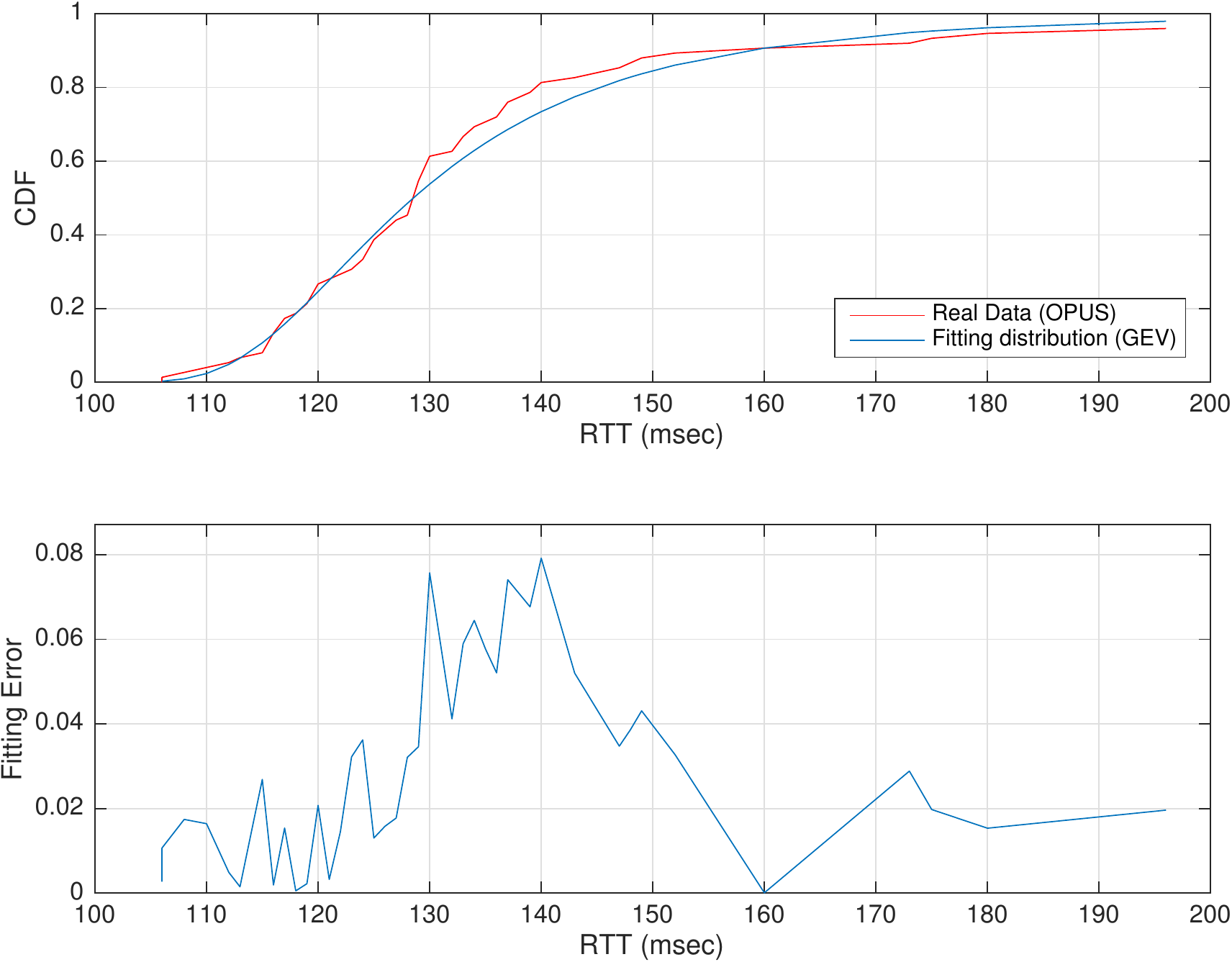}}  \hspace{2mm}
		\subfloat{\includegraphics[scale=0.25]{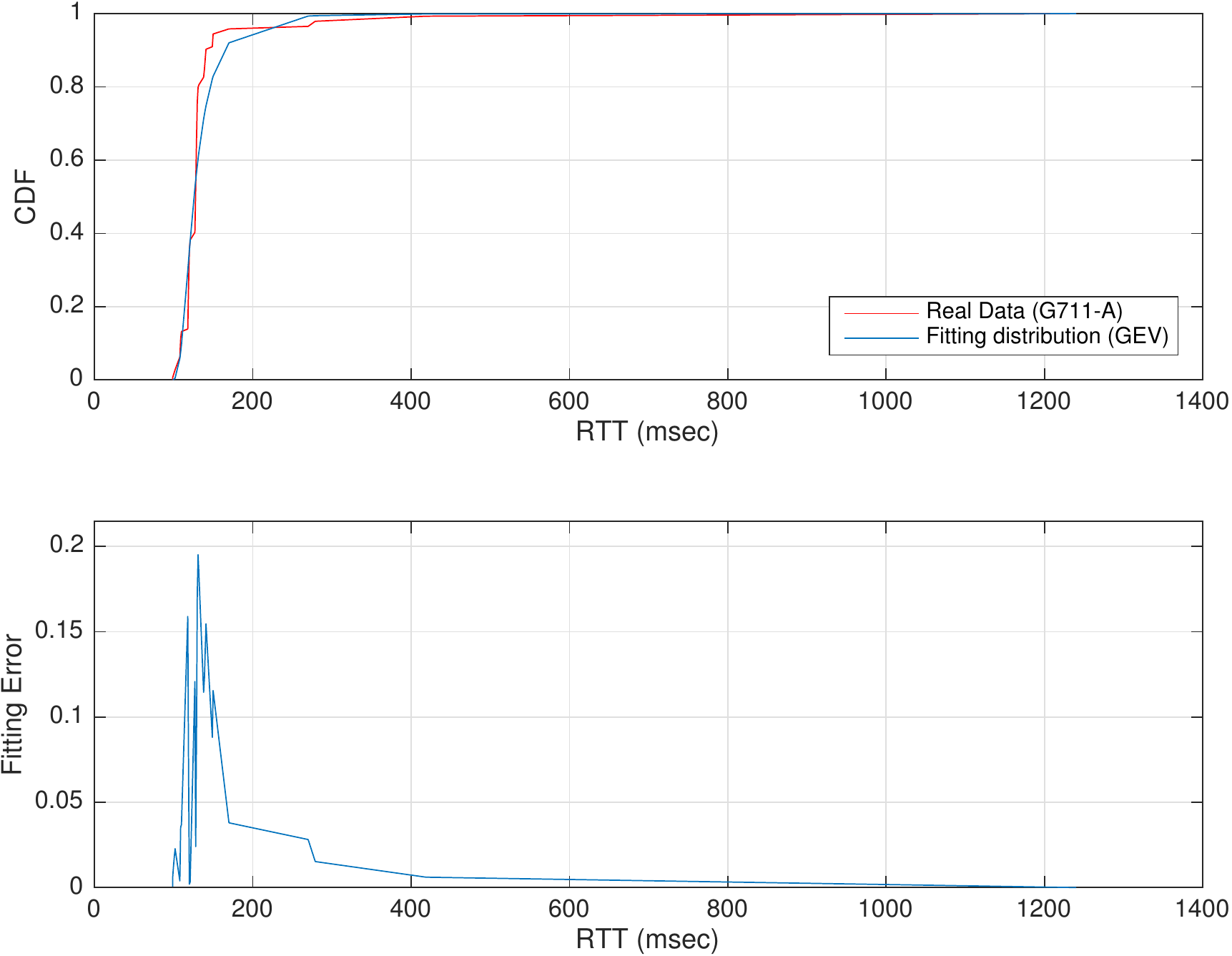}} \hspace{2mm}
		\subfloat{\includegraphics[scale=0.25]{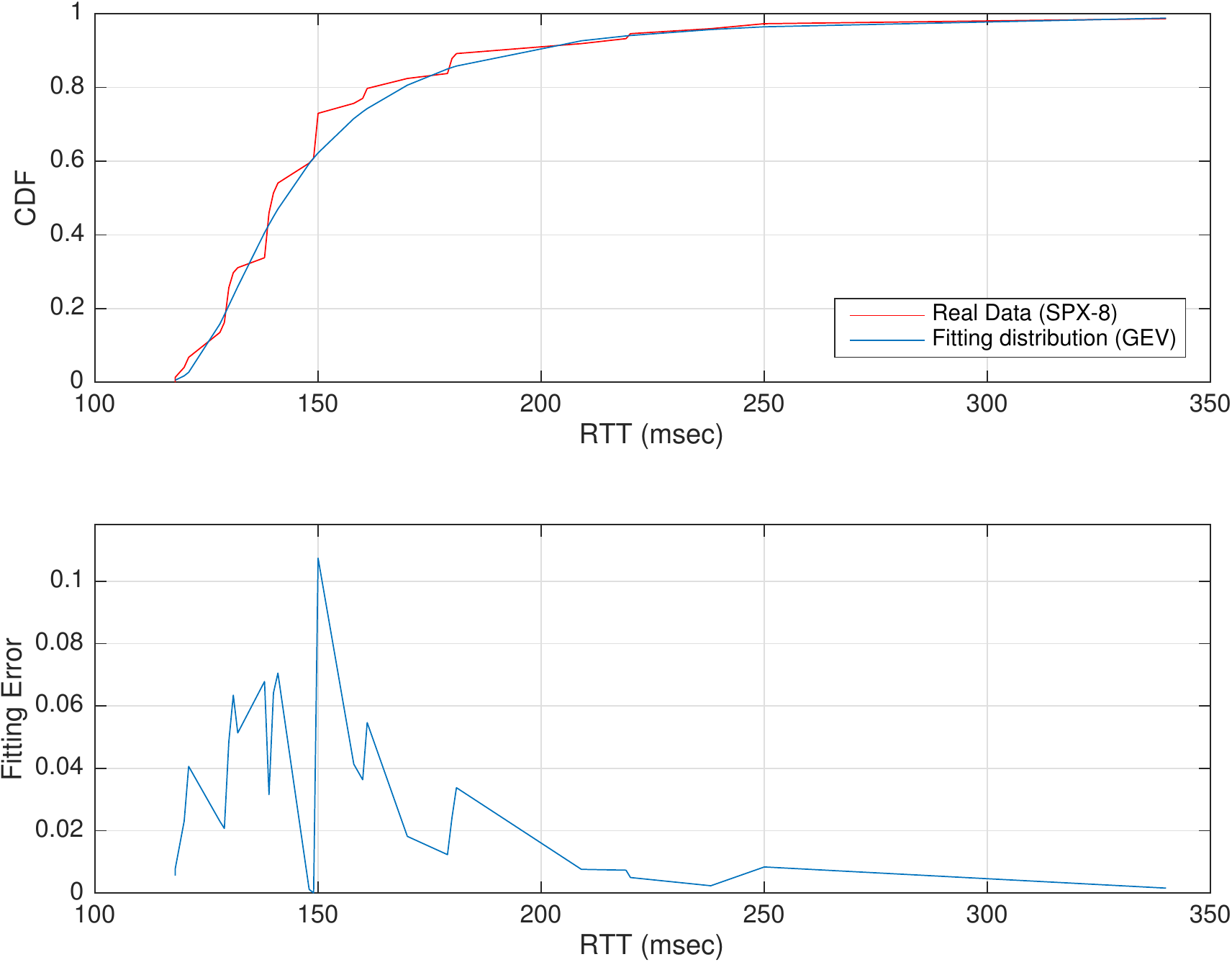}} \hspace{2mm}
		\subfloat{\includegraphics[scale=0.25]{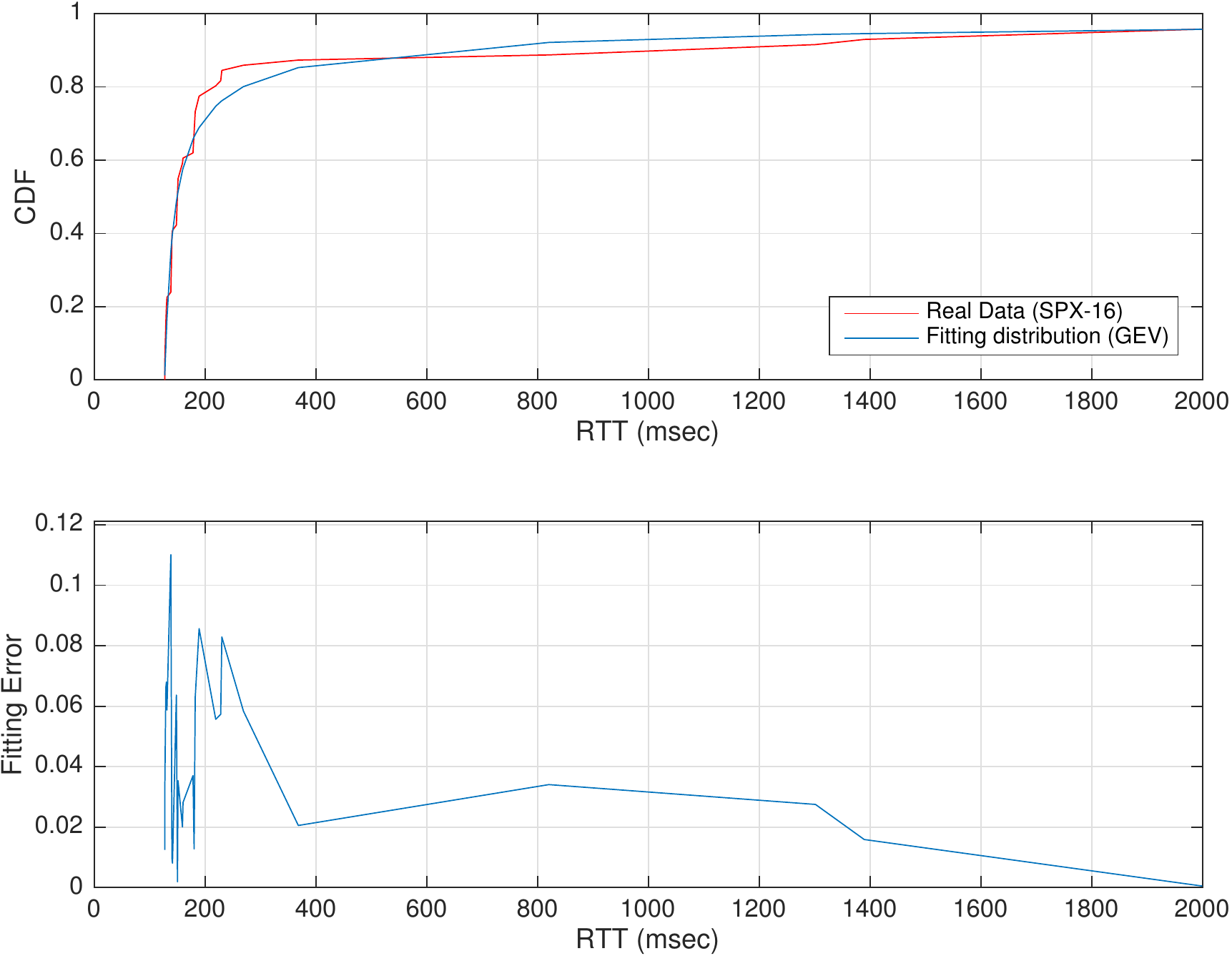}} 
	\end{tabular}
	\caption{Statistical modeling of RTT through GEV distribution. In the upper sub-panels, real data in red and fitting distribution in blue for various codecs. In the lower sub-panels, the pertinent fitting errors.} 
	\label{fig:fitting_rtt}
\end{figure*}
More precisely, in \cite{hosking1985}, the author proposes a modified version of Newton-Raphson's iteration method designed to improve the speed and the probability of convergence. Such a $4$-step procedure (called MLEGEV) allows to guarantee the convergence by the introduction of some ad-hoc constraints.

In short, the cited routine solves the likelihood equation $\partial l / \partial \theta$ through the following iteration:
\beqa
\theta_{k+1} &=& \theta_k + \delta \theta \\ \nonumber
\delta \theta &=& H^{-1} v \\ \nonumber
v= \frac{\partial l}{\partial \theta}\Bigr|_{\substack{\theta = \theta_k}} , 
H&=&\frac{-\partial^{2} l}{\partial \theta \partial \theta^{'}} \Bigr|_{\substack{\theta = \theta_k}},
\eeqa
where the derivatives $v$ and $H$ follow by numerical calculation as analyzed in \cite{prescott83}.

Figures \ref{fig:fitting} and \ref{fig:fitting_rtt} summarize the obtained results for jitter and RTT, respectively. More precisely, each subfigure consists of two panels: the upper one shows the real data distribution for each codec (red curve) and the GEV fitting distribution (blue curve), expressed through CDFs. The lower panel shows the absolute error obtained by considering the punctual difference between observed and fitted data.

Tables \ref{tab:gevvalues} and \ref{tab:gevvalues_rtt} report the estimated values $\hat{\xi}$, $\hat{\sigma}$ and $\hat{\mu}$ for GEV distributions, modeling the jitter and RTT, respectively. The last column ($E_{max}$) indicates the maximum error between real data and fitted distributions, expressed in terms of Kolmogorov-Smirnov distance, namely $E_{max}=\sup_{x} | D_n(x)-D(x)|$, where $D_n(x)$ and $D(x)$ are the empirical and hypothesized distributions, respectively. 

An interesting consideration may be drawn from the results, concerning the shape parameter $\xi$ that is strictly related to the tail behavior of distribution \cite{coles-book}. Considering jitter, $\xi$ has always a negative value, indicating a short-tailed behavior. In other words, the jitter probability distribution decays to zero very fast. This result is perfectly reasonable with the strict constraints (often indicated by Service Level Agreements - SLAs) that telco operators must obey to, since jitter is considered to be the primary source of quality degradation during a VoIP session. Thus, a de-jittering buffer on receiver side is often needed to mitigate this negative effect.

On the contrary, in case of RTT, $\xi$ always exhibits positive values, indicating a long-tailed behavior. This suggests that extreme values of RTT can occur more frequently than the case with jitter. This can be justified by considering that in complex (real-world) mobile scenarios, RTT is subjected to unpredictable variations due to phenomena that negatively interfere over the packet flows (e.g. multi-path fading). Thus, these more dispersed values slow down the tail's decay. 

These findings have interesting implications in terms of improving existing models and simulations of jitter and RTT, particularly in mobile communication scenarios. Furthermore, these results suggest that, regardless of the codec type, the probabilistic behavior of jitter and RTT may be well represented by a set of GEV distributions, which can capture both short- and long-tail structures at the same time. 

\begin{table}[t!]
	\caption {GEV($\hat{\xi}$,$\hat{\sigma,}$$\hat{\mu}$) estimated values for jitter distribution} \label{tab:gevvalues}
	\centering
	\resizebox{.47\textwidth}{!}{
		\begin{tabular}{c|c|c|c|c}
			\hline
			Codec & $\hat{\xi}$ (shape) & $\hat{\sigma}$ (scale) & $\hat{\mu}$ (location) & $E_{max}$ \\
			\hline
			\\[-8pt]
			G722 & -0.437297  & 2.42213 & 7.56518 & 0.060283 \\ \hline
			G729 &  -0.0244 & 4.6840 & 12.2693 & 0.0683 \\ \hline
			GSM & -0.13388 & 1.78116 & 6.14511 & 0.078412 \\ \hline
			MPEG-16 & -0.405995 & 4.06973 & 13.6804 & 0.056911 \\ \hline
			OPUS & -0.321161 & 2.45704 & 6.84896 & 0.061159 \\ \hline
			G711-A & -0.125761 & 1.84636 & 7.27644 & 0.094862 \\ \hline
			SPX-8 & -0.413113 & 2.35586 & 7.00671  & 0.064678 \\ \hline
			SPX-16 & -0.448086 & 4.06973 & 13.6804 & 0.091352 \\ \hline
			\hline
		\end{tabular}}
	\end{table}

	\begin{table}[t!]
		\caption {GEV($\hat{\xi}$,$\hat{\sigma,}$$\hat{\mu}$) estimated values for RTT distribution} \label{tab:gevvalues_rtt}
		\centering
		\resizebox{.45\textwidth}{!}{
			\begin{tabular}{c|c|c|c|c}
				\hline
				Codec & $\hat{\xi}$ (shape) & $\hat{\sigma}$ (scale) & $\hat{\mu}$ (location) & $E_{max}$ \\
				\hline
				\\[-8pt]
				G722 & 0.3230  & 14.4228 & 122.8554 & 0.1317 \\ \hline
				G729 &  0.1945  & 50.1967 & 176.0254 & 0.0951 \\ \hline
				GSM & 0.1131 & 12.3366 &  129.5999 & 0.1357 \\ \hline
				MPEG-16 & 0.5800 & 14.9812 & 145.5648 & 0.1539 \\ \hline
				OPUS &  0.2077 & 12.0708 & 123.9454 & 0.0791 \\ \hline
				G711-A & 0.2727 & 13.8707 & 120.5886 & 0.1952 \\ \hline
				SPX-8 & 0.4260 & 15.5273 & 136.3658 & 0.1074 \\ \hline
				SPX-16 & 1.5807 & 21.0076 & 139.0054 & 0.1101 \\ \hline
				\hline
			\end{tabular}}
		\end{table}

\section{Conclusion}
\label{sec:concl}

A plethora of network tools available today aim at simulating VoIP traffic, with great interest arising around LTE-based environments. Yet, at the current state of play, it is not yet possible to realistically deal with the complexity of end-to-end mobile VoIP across an LTE infrastructure. This is because it is so hard to reproduce in simulation the  phenomena that affect radio communications, such as time-variant interference, effect of weather, and state of load on nodes, to mention but a few. Due to these limitations, it is hard to accurately characterize QoS metrics. 

In this article, we address this challenge, designing a detailed measurement campaign of voice traffic across an urban LTE-A domain. Our trials take into account two operative settings: a first scenario involving a set of LTE-A voice sessions between two fixed users; and a second scenario that accounts for LTE-A voice sessions between a mobile user (a driver) and a fixed one. The field trials have been carried out by considering families of $8$ different codecs, and a total of around 750,000 voice RTP packets. In this way, we carry out performance tests aimed at evaluating the impact that external environmental factors have on voice communication. We also capture the non-linear influence that different types of voice codecs have on QoS. 

Based on the collected data, we then propose a statistical characterization of jitter and RTT, the two most critical quality metrics in VoIP. We derive a Generalized Extreme Value distribution for both metrics, estimating the pertinent parameters (shape, scale, and location) by means of Maximum Likelihood estimators. The resulting model gives a contrasting behaviour for jitter and RTT, which exhibit a short- and long-tail, respectively. 

Our work offers a novel viewpoint on voice traffic characterization, based on metrics that are typically neglected in classical studies, particularly the R-Factor and voice signal level. Experimental results show how environmental phenomena (such as interference) can generate unpredictable behaviour that may result in counter-intuitive results. On occasions, we noticed that some mobility settings were exhibiting more stability than fixed ones. Statistical characterization helps to better understanding non-linear behaviour and appreciating end-to-end effects on quality. It is also of paramount importance in the improvement of simulation modeling. 

Interestingly, our findings could be applied also to 5G networks, which have many aspects in common with the LTE-A architecture. This is the case of high-level metrics such as SDD/CSD that, derived in terms of SIP protocol performance, allow to easily extend the analysis to 5G enabling architectures such as the IP Multimedia Subsystem.

Finally, our study may be further extended, particularly in the scale of the measurement campaign and considering a broader range of parameters (e.g. user's speed and environmental conditions). It will be beneficial to gain greater insights on the specific effects of handover on QoS. Vertical and horizontal handover are expected to create different effects, which will become increasingly important in future hybrid networks. 
\vspace{-1pt}
\begin{appendices}
	\section{}
	\label{app:A}
\begin{figure}[ht]
	\centering
	\captionsetup{justification=centering}
	\includegraphics[scale=0.29,angle=90]{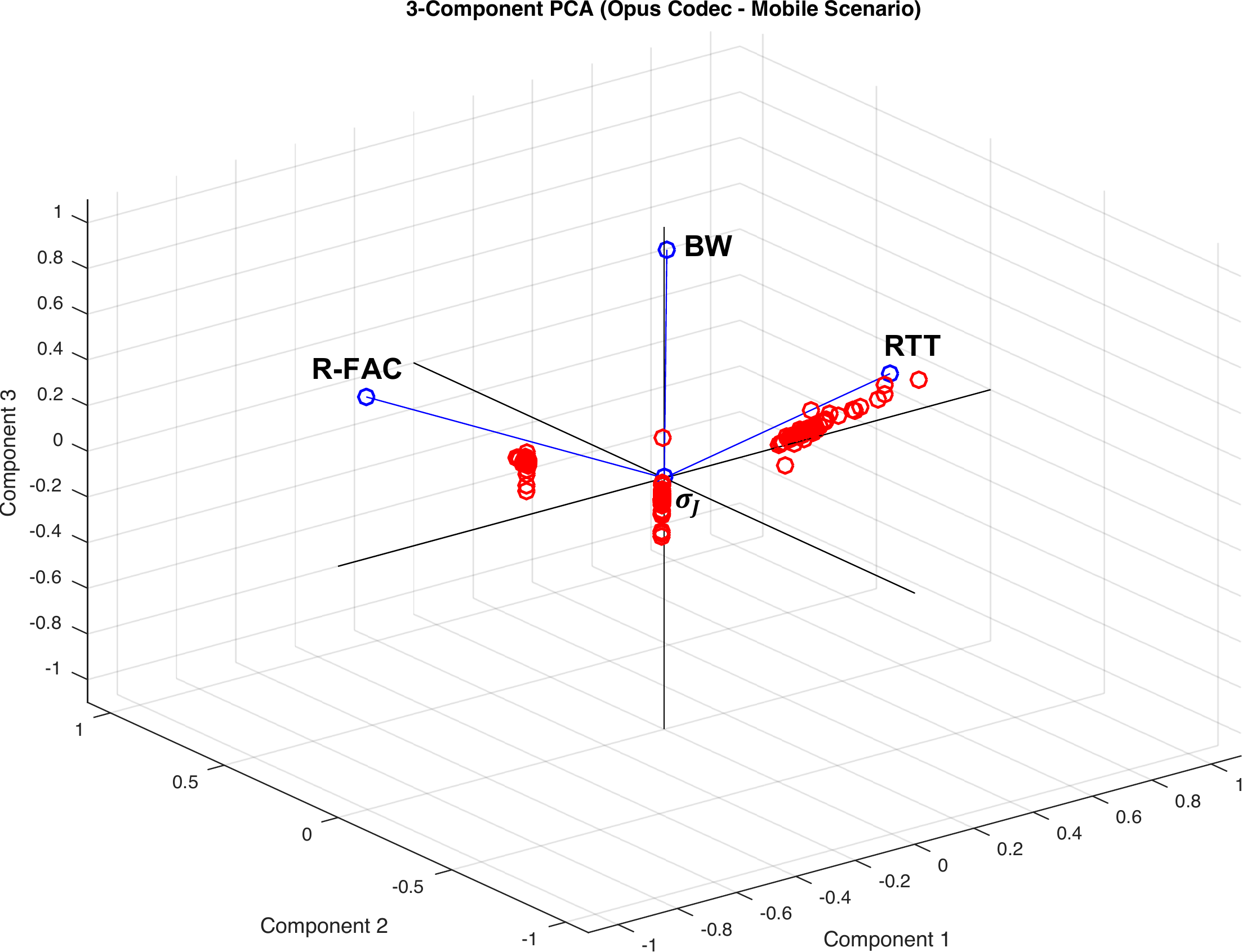}
	\caption{Exemplary $3$-component PCA with four variables (R-Factor, Bandwidth consumption, RTT, $\sigma_J$) for the OPUS codec case in mobile scenario.}
	\label{fig:PCA}
\end{figure}

An alternative representation to the bivariate distributions reported in Figs. $3-5$ might be offered by Principal Component Analysis (PCA), one (among various) method going in the direction of data dimensionality reduction. PCA allows to recast mutually correlated variables (representative of multidimensional data) into new uncorrelated variables (a.k.a. principal components) through linear combinations of the original variables, with minimal information loss. The first principal component takes into account as much of data variability as possible. Whereas, succeeding components take into account the remaining variability. 
One drawback of PCA is that, being a domain transformation allowing to find the best projections that maximize variances, a physical interpretation could be not so immediate. As an example, we consider the case in which the OPUS codec is used in mobile scenario. 
In Fig. \ref{fig:PCA}, we derive the representation of PCA with 3 principal components.
According to such a representation, PCA shows how to recast four observables (R-Factor, Bandwidth consumption, RTT, and $\sigma_J$) in terms of four vectors (in blue), whose directions and lengths indicate the contribution to the three considered components (red points representing the scores). Macroscopically, it can be seen that each variable contribute to various components differently. 
\end{appendices}

\begin{IEEEbiography}[{\includegraphics[width=1in,height=1.15in,clip,keepaspectratio]{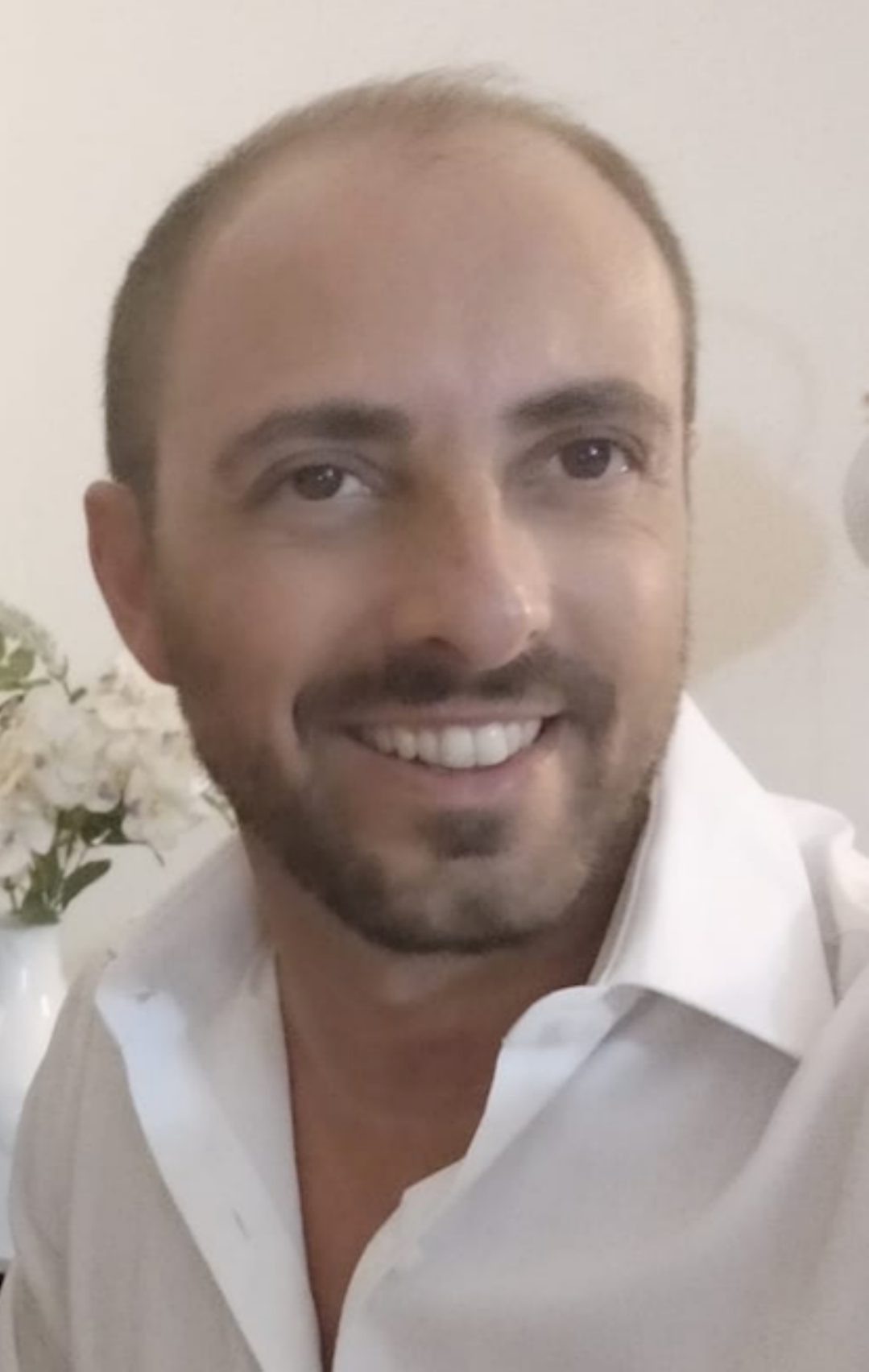}}] 
	{Mario Di Mauro} received the Laurea degree in electronic engineering from the University of Salerno (Italy), the M.S. degree in networking from the University of L'Aquila (Italy) jointly with the Telecom Italia Centre, and the PhD degree in information engineering from University of Salerno.
	He was a Research Engineer with CoRiTel (Research Consortium on Telecommunications, led by Ericsson Laboratory, Italy) and then a Research Fellow with University of Salerno. He has authored several scientific papers, and holds a patent on a telecommunication aid for impaired people. His main fields of interest include: network performance, network security and availability, data analysis for
	telecommunication infrastructures.
\end{IEEEbiography}

\begin{IEEEbiography}[{\includegraphics[width=1in,height=1.15in,clip,keepaspectratio]{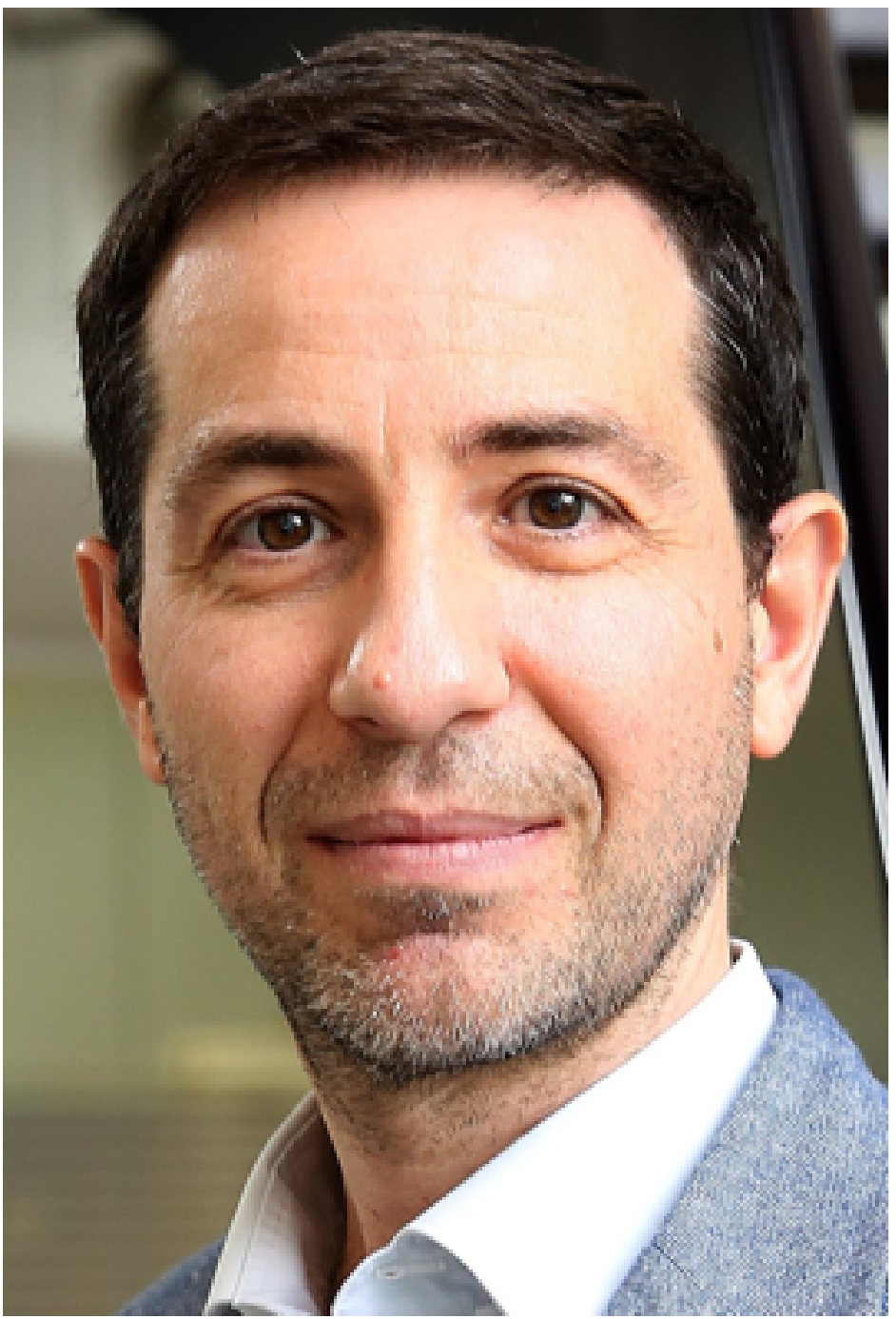}}] 
{Antonio Liotta} is Professor of Data Science and Intelligent Systems at Edinburgh Napier University (U.K.). Previously, he was Professor of Data Science and the founding director of the Data Science Research Centre, University of Derby, UK. His team is at the forefront of influential research in data science and artificial intelligence, specifically in the context of smart cities, Internet of Things, and smart sensing. He is renowned for his contributions to miniaturized machine learning, particularly in the context of the Internet of Things. He has led the international team that has recently made a breakthrough in artificial neural networks, using network science to accelerate the training process. Antonio is a Fellow of the U.K. Higher Education Academy. He has 6 patents and over 350 publications to his credit, and is the Editor-in-Chief of the Springer Internet of Things book series.
\end{IEEEbiography}



\begin{thebibliography}{1}

\bibitem{eri2017}
{Ericsson}, ``{Ericsson Mobility Report},''
2017 [Online].
\newblock Available: 
\url{https://www.ericsson.com/assets/local/mobility-report/documents/2017/ericsson-mobility-report-november-2017-central-and-eastern-europe.pdf}.

\bibitem{cisco}
{Cisco Systems}, ``{Quality of Service Design Overview},''
[Online].
\newblock Available: 
\url{https://www.cisco.com/c/en/us/td/docs/solutions/Enterprise/WAN_and_MAN/QoS_SRND/QoS-SRND-Book/QoSIntro.pdf}.



\bibitem{murroni}
M.~Murroni, R.~Rassool, L.~Song, R.~Sotelo ``Guest Editorial Special Issue on Quality of Experience for Advanced Broadcast Services,'' {\em IEEE Trans. Broadcast.}, vol.~64, no.~2, pp.~335--340, 2018.

\bibitem{atzori1}
A.~Floris, L.~Atzori, ``Quality of Experience in the Multimedia Internet of Things: Definition and practical use-cases,'' in  {\em 2015 IEEE International Conference on Communication Workshop (ICCW)}, pp 1747-1752, 2015.

\bibitem{atzori2}
A.~Ahmad, L.~Atzori, M.G.~Martini ,``Qualia: A multilayer solution for QoE passive monitoring at the user terminal,'' in  {\em 2017 IEEE International Conference on Communications (ICC)}, pp 1-16, 2017.


\bibitem{torres18}
M.~Torres Vega, C.~Perra, F.~De Turck, A.~Liotta, ``A Review of Predictive Quality of Experience Management in Video Streaming Services,'' {\em IEEE Trans. Broadcast.}, vol.~64, no.~2, pp.~432--445, 2018.

\bibitem{torres18_2}
M.~Torres, C.~Perra, and A.~Liotta, ``Resilience of Video Streaming Services to Network Impairments,'' {\em IEEE Trans. Broadcast.}, vol.~64, no.~2, pp.~220--234, 2018.

\bibitem{olaifa16}
J. O.~Olaifa, D.~Arifler, ``Using system-level simulation to evaluate downlink throughput performance in LTE-A networks with clustered user deployments,'' in {\em 1st International Workshop on Link and System Level Simulations}, pp.~1--6, 2016.

\bibitem{xu2013}
X.~Xu, F.~Schroeder, B.~Gevrekce, G.~Kutrolli, M.~Ni, R.~Mathar ``A physical layer simulator based on radio wave propagation for LTE cellular networks,'' in {\em 2013 7th European Conference on Antennas and Propagation (EuCAP)}, pp.~1007--1010, 2013.

\bibitem{sarker2014}
Z.~Sarker, V.~Singh, C.~Perkins, ``An evaluation of RTP circuit breaker performance on LTE networks,'' in  {\em 2014 IEEE Conference on Computer Communications Workshops (INFOCOM)}, pp.~251--256, 2014.

\bibitem{carullo16}
G.~Carullo, M.~Tambasco, M.~Di Mauro, M.~Longo, ``A performance evaluation of WebRTC over LTE,'' in {\em 12th Annual Conference on Wireless On-demand Network Systems and Services}, pp.~1--6, 2016.

\bibitem{ahmad15}
T.~Ahmad, A.M.~Abbas, ``Delay and throughput of Long Term Evolution under TCP traffic,'' in {\em Annual IEEE India Conference}, pp.~1--6, 2015.

\bibitem{abidi14}
K.~Abidi, E.~Hajlaoui, M.~Abdellaoui, ``Handover simulation of LTE and LTE-A standards,'' in {\em 6th International Conference of Soft Computing and Pattern Recognition}, pp.~157--162, 2014.

\bibitem{bermudez17}
H.F.~Bermudez, R.~Sanchez-Iborra, J.L.~Arciniegas, W.Y.~Campo, M.D.~Cano, ``Handover simulation of LTE and LTE-A standards,'' in {\em 6th International Conference of Soft Computing and Pattern Recognition}, pp.~157--162, 2014.

\bibitem{kassim17}
M.~Kassim, R. A.~Rahman, M.A.A.~Aziz, A.~Idris, M.I.~Yusof, ``Performance analysis of VoIP over 3G and 4G LTE network,'' in {\em International Conference on Electrical, Electronics and System Engineering}, pp.~37--41, 2017.

\bibitem{wylie14}
M.P.~Wylie, T.~Svensson, ``Throughput, Capacity, Handover and Latency Performance in a 3GPP LTE FDD Field Trial,'' in {\em  IEEE Global Telecommunications Conference GLOBECOM 2010}, pp.~1--6, 2010.

\bibitem{buenestado14}
V.~Buenestado, J.M.~Ruiz-Aviles, M.~Toril, S.~Luna-Ramirez, A.~Mendo ``Analysis of Throughput Performance Statistics for Benchmarking LTE Networks,'' {\em IEEE Commun. Lett.}, vol.~18, no.~9, pp.~1607--1610, 2014.

\bibitem{murroni}
C.~Desogus, M.~Anedda, M.~Murroni, G.~Muntean, ``A Traffic Type-Based Differentiated Reputation Algorithm for Radio Resource Allocation During Multi-Service Content Delivery in 5G Heterogeneous Scenarios,'' {\em IEEE Access}, vol.~7, pp.~27720--27735, 2019.

\bibitem{trivisonno2015}
R.~Trivisonno, R.~Guerzoni, I.~Vaishnavi, A.~Frimpong, ``Network Resource Management and QoS in SDN-Enabled 5G Systems,'' in {\em  IEEE Global Telecommunications Conference GLOBECOM 2015}, pp.~1--7, 2015.

\bibitem{5gqoe}
J.~Nightingale, P.~Salva-Garcia, J.~M.~A.~Calero, Q.~Wang, ``5G-QoE: QoE Modelling for Ultra-HD Video Streaming in 5G Networks,'' {\em IEEE Trans. Broadcast.}, vol.~64, no.~2, pp.~621--634, 2018.

\bibitem{cellmapper}
[Online].
\url{https://www.cellmapper.net}.

\bibitem{voip-handbook}
{\em VoIP Handbook: Applications, Technologies, Reliability, and Security}.
\newblock CRC Press, Ed. by S.A.~Ahson, M.~Ilyas, 2008.


\bibitem{chang2012}
C.L.~Chang, S.P.~Huang, ``The interleaved video frame distribution for P2P-based VoD system with VCR functionality,'' {\em Computer Networks}, vol.~56, no.~5, pp.~1525--1537, 2012.


\bibitem{jou2003}
J.S.~Jou, J.S.~Baras, ``Jitter analysis of CBR streams in multimedia networks,"  in {\em Proceedings of  IEEE Conference on Control Applications}, pp.~1339--1344, 2003.


\bibitem{icufn}
H.G.~Kim, S.H.~Ryu, ``Adaptive jitter estimation for voice-over IP networks,"  in {\em Proceedings of IEEE International Conference on Ubiquitous and Future Networks}, pp.~328--329, 2012.

\bibitem{rtcp}
{IETF}, ``{RTP: A Transport Protocol for Real-Time Applications},''
2003 [Online].
\newblock Available: 
\url{https://tools.ietf.org/html/rfc3550}.

\bibitem{rtcp-xr}
{IETF}, ``{RTP Control Protocol Extended Reports (RTCP XR)},''
2003 [Online].
\newblock Available: 
\url{https://tools.ietf.org/html/rfc3611}.

\bibitem{itutg107}
{ITU-T}, ``{The E-model: a computational model for use in transmission planning},''
2015 [Online].
\newblock Available: 
\url{https://www.itu.int/rec/T-REC-G.107-201506-I/en}.

\bibitem{eiatia2000}
{TIA/EIA-810-A}, ``{Transmission Requirements for Narrowband Voice over IP
	and Voice over PCM Digital Wireline Telephones},'' 2000.


\bibitem{garch}
	Y.~Zhang, D.~Fay, L.~Kilmartin, A.~W.~Moore, ``A Garch-based adaptive playout delay algorithm for VoIP,'' {\em Computer Networks}, vol.~54, no.~17, pp.~3108--3122, 2010.


\bibitem{cisco-codec}
{Cisco Systems}, ``{Voice Over IP - Per Call Bandwidth Consumption)},''
[Online].
\newblock Available: 
\url{https://www.cisco.com/c/en/us/support/docs/voice/voice-quality/7934-bwidth-consume.html}.

\bibitem{sheluhin-book}
O.I.~Sheluhin, S.M.~Smolskiy, A.V.~Osin {\em Self-Similar Processes in Telecommunications}.
\newblock West Sussex, John Wiley \& Sons, 2007.

\bibitem{raake-book}
A.~Raake, {\em Speech Quality of VoIP - Assessment and Prediction}.
\newblock West Sussex, John Wiley \& Sons, 2006.

\bibitem{begain-book}
K.~Al-Begain, A.~Ali, {\em Multimedia Services and Applications in Mission Critical Communication Systems}.
\newblock Hershey (PA), IGI Global, 2017.

\bibitem{rfc6076}
{IETF}, ``{Basic Telephony SIP End-to-End Performance Metrics},''
2011 [Online].
\newblock Available: 
\url{https://tools.ietf.org/html/rfc6076}.

\bibitem{itu1028}
{ITU-T}, ``{G.1028 : End-to-end quality of service for voice over 4G mobile networks},''
[Online].
\newblock Available: 
\url{https://www.itu.int/rec/T-REC-G.1028/en}.

\bibitem{curcio2002}
I.D.D.~Curcio, M.~Lundan, ``SIP call setup delay in 3G networks,'' in {\em Proceedings ISCC 2002 Seventh International Symposium on Computers and Communications}, pp.~835--840, 2002.

\bibitem{awaludin2010}
A.~Awaludin, M.~Fathurahman, R.~Primardiansyah, R.F.~Sari, ``Performance evaluation of datagram congestion control protocol for SIP signalling using NS-2 simulation,'' in {\em Proceeding of the 3rd International Conference on Information and Communication Technology for the Moslem World (ICT4M) 2010}, pp.~67--71, 2010.

\bibitem{husic2014}
J.B.~Husic, A.~Hidic, M.~Hadzialic, S.~Barakovic, ``Simulation-based optimization of signaling procedures in IP multimedia subsystem,'' in {\em Proceedings of 15th Conference of Open Innovations Association FRUCT}, pp.~9--14, 2014.

\bibitem{brajdic2009}
A.~Brajdic, M.~Suznjevic, M.~Matijasevic, ``Measurement of SIP signaling performance for advanced multimedia services,'' in {\em 10th International Conference on Telecommunications}, pp.~381--388, 2009.

\bibitem{Vaser2015}
M.~Vaser, S.~Forconi, ``QoS KPI and QoE KQI Relationship for LTE Video Streaming and VoLTE Services,'' in {\em 9th International Conference on Next Generation Mobile Applications, Services and Technologies}, pp.~318--323, 2015.



\bibitem{dahmouni2012}
H.~Dahmouni, A.~Girard, M.~Ouzineb, B.~Sanso, ``The Impact of Jitter on Traffic Flow Optimization in Communication Networks,'' {\em IEEE Trans. Netw. Service Manag.}, vol.~9, no.~3, pp.~279--292, 2012.

\bibitem{kaup2016}
F.~Kaoup, F.~Michelinakis, N.~Bui, J.~Widmer, K~Wac, D.~Hausheer ``Assessing the Implications of Cellular Network Performance on Mobile Content Access,'' {\em IEEE Trans. Netw. Service Manag.}, vol.~13, no.~2, pp.~168--180, 2016.

\bibitem{rizo2010_rtt}
L.~Rizo-Dominguez, D.~Torres-Roman, D.~Munoz-Rodriguez, C.~Vargas-Rosales, ``Packet variation delay distribution discrimination based on Kullback-Leibler divergence,'' {\em IEEE Latin-American Conference on Communications}, pp.~1--4, 2010.

\bibitem{acharya}
A.~Acharya, J.~Saltz, ``A Study of Internet Round-trip Delay,'' {\em Univ. of Maryland, Tech. Rep.}, 1996

\bibitem{broido}
A.~Broido, E.~Basic, K.C.~Klaffy ``Invariance of the Internet RTT spectrum.
Global RTT analysis,'' 2002
[Online].
\newblock Available: 
\url{http://www.caida.org/ broido/rtt/rtt.html,}.

\bibitem{matragi1997}
W.~Matragi, K.~Sohraby, C.~Bisdikian, ``Jitter calculus in ATM networks: multiple nodes,'' {\em IEEE/ACM Trans. Netw.}, vol.~5, no.~1, pp.~122--133, 1997.

\bibitem{huremovic2015}
A.~Huremovic, M.~Hadzialic, ``Novel approach to analytical jitter modeling,'' {\em IEEE Journal of Communications and Networks}, vol.~17, no.~5, pp.~534--540, 2015.

\bibitem{hammad2016}
K.~Hammad, A.~Moubayed, A.~Shami, S.~Primak, ``Analytical Approximation of Packet Delay Jitter in Simple Queues,'' {\em IEEE Wireless Communications Letters}, vol.~5, no.~6, pp.~564--567, 2016.

\bibitem{voznak2012}
M.~Voznak, A.~Kovac, M.~Halas, ``Effective Packet Loss Estimation on VoIP Jitter Buffer,'' in {\em NETWORKING 2012 Workshops}, pp.~157--162, 2012.

\bibitem{cruz2013}
H.~Toral-Cruz, A.K.~Pathan, and J.C.~Ramirez Pacheco, ``Accurate modeling of VoIP traffic QoS parameters in current and future networks with multifractal and Markov models,'' {\em Mathematical and Computer Modelling}, Elsevier vol.~57, no.~11-12, pp.~2832--2845, 2013.

\bibitem{schwarz78}
G.E.~Schwarz, ``Estimating the dimension of a model,'' {\em Annals of Statistics}, vol.~6, no.~2, pp.~461--464, 1978.

\bibitem{jenkinson1955}
A.F.~Jenkinson, ``The frequency distribution of the annual maximum (or minimum) values of meteorological elements,'' {\em Quarterly Journal of Royal Meteorological Society}, vol.~81, pp.~58--171, 1955.

\bibitem{jenkinson1969}
A.F.~Jenkinson, ``Estimation of maximum floods,'' {\em World Meteorological Organization Technical Note no.~98}, pp.~183--227, 1969.

\bibitem{kotz-book}
S.~Kotz, S.~Nadarajah, {\em Extreme Value Distributions. Theory and Applications}.
\newblock London, Imperial College Press, 2000.

\bibitem{coles-book}
S.~Coles,  {\em An introduction to statistical modeling of extreme values}.
\newblock London, Springer Verlag, 2001.

\bibitem{smith1985}
R.L.~Smith, ``Maximum likelihood estimation in a class of non-
regular cases'' {\em Biometrika}, no.~72, pp.~67--90, 1985.

\bibitem{hosking1985}
J.R.M.~Hosking, ``Algorithm AS 215: Maximum-likelihood estimation
of the parameter of the generalized extreme-value distribution,'' {\em Applied Statistics}, no.~34, pp.~301--310, 1985.

\bibitem{prescott83}
P.~Prescott, A.T.~Walden, ``Maximum likelihood estimation of the parameters of the three-parameter generalized extreme-value distribution from censored samples,'' 
{\em Journal of Statistical Computation and Simulation}, vol.~16, no.~3-4, pp.~241--250, 1983.





\end{thebibliography}
\end{document}